\documentclass[12pt,reqno]{amsart}

\usepackage{amsmath,amssymb,amsfonts}
\usepackage{geometry}
\usepackage{booktabs}
\usepackage{graphicx}
\usepackage{hyperref}

\graphicspath{{figures/}{./}}

\theoremstyle{remark}
\newtheorem{remark}{Remark}[section]

\begin{document}

\title[Singular layer neural networks for the Navier--Stokes equations]
{Singular layer neural networks for the Navier--Stokes equations at small
viscosity}

\author[T.-Y. Chang, 
G.-M. Gie, 
Y. Hong, and 
C.-Y.  Jung]
{
Teng-Yuan Chang$^{1}$, 
Gung-Min Gie$^{2,*}$, 
Youngjoon Hong$^{3,4}$, and   
Chang-Yeol  Jung$^5$}
\address{$^1$ Delta Electronics, Taipei, Taiwan}
\address{$^2$ Department of Mathematics, University of Louisville, Louisville,
KY 40292, USA}
\address{$^3$ Research Institute of Mathematics, Seoul National University,
Seoul, Korea}
\address{$^4$ Department of Mathematical Sciences, Seoul National University,
Seoul, Korea}
\address{$^5$ Department of Mathematical Sciences, Ulsan National Institute of
Science and Technology,
Ulsan 44919, Korea}
\email{ty.chang9408@gmail.com}
\email{gungmin.gie@louisville.edu}
\thanks{$^*$Corresponding author.}
\email{hongyj@snu.ac.kr}
\email{cjung@unist.ac.kr}

\begin{abstract}
We develop neural network methods for the two-dimensional incompressible
Navier--Stokes equations at a small viscosity $\varepsilon > 0$, whose
accuracy is uniform in $\varepsilon$ and whose residual is evaluated on a grid
that does not resolve the vorticity boundary layer.
In the vorticity--stream function formulation with the slip boundary
conditions, a vorticity layer of width $\sqrt{\varepsilon}$ forms along the
walls.
The trial function of the singular layer Physics-Informed Neural Network
(sl-PINN) contains the corrector of the asymptotic analysis, together with the
law that fixes its amplitude at the wall. The amplitude is tied to the wall
trace of the network itself, so that the method requires no a priori
knowledge of the limit solution.
The sl-PINN keeps the error uniform in $\varepsilon$ over six decades,
for the Stokes and the Navier--Stokes equations alike, although in
the latter the wall trace travels with the flow and is not known in advance.
An enrichment that uses the thin scale but leaves the profile and the
amplitude of the layer free deteriorates as $\varepsilon\to0$.
Since the corrector carries the layer analytically, the Laplacian in the
residual can be discretized by a finite volume flux difference on a coarse
grid. The resulting singular layer Finite
Volume Neural Network (sl-FVNN) attains the same uniform accuracy as the
sl-PINN, to within the scatter of the random seed, at about half the cost of
a training step. The combination of the asymptotic corrector with the
coarse-grid finite volume discretization is the main result of this article.
\end{abstract}


\maketitle

\section{Introduction}
\label{sec:intro}

We develop neural network methods for singularly perturbed problems whose
solutions carry boundary layers, with an accuracy that is uniform in the small
parameter. Across a boundary layer the solution varies on a scale much smaller
than the scale of the domain. An approximation that neither resolves that
scale nor accounts for it otherwise fails to predict the solution near the
wall. We embed the corrector of the asymptotic analysis in the trial function.
The amplitude of the corrector is tied to the wall trace of the network itself
rather than to the limit solution, so that no a priori knowledge of the limit
solution is required. Our main target is the vorticity layer of the two-dimensional incompressible
Navier--Stokes equations at a small viscosity $\varepsilon$. This is the most
demanding case, since the problem is nonlinear and time dependent, the wall
trace of the vorticity travels with the flow, and the amplitude of the layer
is not known in advance.
Compared with our earlier singular layer PINNs
\cite{GieHongJung2024,GieHongJungMunkhjin2024,GieHongJungPolygonal,GieHongJungLee2026,ChangGieHongJungBurgers,ChangGieHongJung2024,MoonGieHongJungPipe}
and finite volume neural networks
\cite{GieKim2025,GieJungKimFVNN,GieJungKimBurgers}, the present article is new
in three respects. First, the closest earlier work treats the plane-parallel
flow, whose corrector is an explicit exponential in the wall-normal variable
\cite{ChangGieHongJung2024}. Here the corrector is the solution of the
transport--diffusion equation of \cite{GieJung2013}, which depends on time and
travels along the wall with the flow, and the layer carries the
nonlinearity. We show that this corrector, derived from the linear analysis,
transfers to the nonlinear Navier--Stokes system at a cost which does not
depend on the viscosity. Second, the earlier singular layer FVNN combines the
corrector with the finite volume residual on the stationary viscous Burgers
equations \cite{GieJungKimBurgers}. We show that the combination remains
consistent, on a grid that does not resolve the layer, for the present
time-dependent nonlinear system. The resulting sl-FVNN attains the
$\varepsilon$-uniform accuracy of the sl-PINN at half the cost of a training
step. Third, we compare the corrector-informed enrichment with an enrichment
that uses the scale $\sqrt{\varepsilon}$ only, and we show that the scale
alone does not suffice.

We consider, throughout the article, the two-dimensional incompressible
Navier--Stokes equations, 
\begin{equation}
\begin{cases}
\displaystyle
\frac{\partial \mathbf{u}^{\varepsilon}}{\partial t}
-\varepsilon \Delta \mathbf{u}^{\varepsilon}
+(\mathbf{u}^{\varepsilon}\cdot\nabla)\mathbf{u}^{\varepsilon}
+\nabla p^{\varepsilon}
=\mathbf{F},
& \text{in }\Omega\times(0,T),\\[1ex]
\nabla\cdot\mathbf{u}^{\varepsilon}=0,
& \text{in }\Omega\times(0,T),
\end{cases}
\label{eq:NS}
\end{equation}
supplemented with the slip boundary conditions
\begin{equation}
\mathbf{u}^{\varepsilon}\cdot\mathbf{n}=0,
\qquad
\omega^{\varepsilon}:=\operatorname{curl}\mathbf{u}^{\varepsilon}=0
\qquad
\text{on }\Gamma .
\label{eq:slip}
\end{equation}

We introduce the stream function $\psi^\varepsilon$ and the vorticity
$\omega^\varepsilon$ in the usual manner, by setting
\begin{equation}
\mathbf{u}^{\varepsilon}
=
(u^\varepsilon,v^\varepsilon)
=
\left(
\frac{\partial\psi^\varepsilon}{\partial y},
-\frac{\partial\psi^\varepsilon}{\partial x}
\right),
\qquad
\omega^\varepsilon
=
\frac{\partial v^\varepsilon}{\partial x}
-
\frac{\partial u^\varepsilon}{\partial y},
\label{eq:stream-vort-def}
\end{equation}
so that
\begin{equation}
\omega^\varepsilon=-\Delta\psi^\varepsilon.
\label{eq:vort-poisson}
\end{equation}

With the convention \eqref{eq:stream-vort-def}, the Navier--Stokes equations
take the vorticity--stream function form
\begin{equation}
\frac{\partial\omega^\varepsilon}{\partial t}
-\varepsilon\Delta\omega^\varepsilon
+
\frac{\partial\psi^\varepsilon}{\partial y}
\frac{\partial\omega^\varepsilon}{\partial x}
-
\frac{\partial\psi^\varepsilon}{\partial x}
\frac{\partial\omega^\varepsilon}{\partial y}
=
f,
\label{eq:vorticity-NS}
\end{equation}
where $f=\operatorname{curl}\mathbf{F}$, coupled with the Poisson equation, 
\begin{equation}
-\Delta\psi^\varepsilon=\omega^\varepsilon.
\label{eq:stream-poisson}
\end{equation}

For the Stokes problem, which we consider as well, the nonlinear convection term
in \eqref{eq:vorticity-NS} is absent, and the system reads
\begin{equation}
\frac{\partial\omega^\varepsilon}{\partial t}
-\varepsilon\Delta\omega^\varepsilon
=
f,
\qquad
-\Delta\psi^\varepsilon=\omega^\varepsilon.
\label{eq:stokes-vs}
\end{equation}

The numerical experiments of this article lead to three main conclusions:

First, and this is the main conclusion, the enrichment carries the layer
analytically, so that the part left to the network varies on the
$\mathcal{O}(1)$ scale only. The Laplacian in the residual can therefore be
replaced by a finite volume flux difference computed on a coarse grid that
does not resolve $\sqrt{\varepsilon}$ (Section~\ref{sec:exp10}).
Applied to the whole field, this substitution fails, since the discrete
Laplacian of the exact solution is then in error by the factor
$2.5\times10^{4}$ at $\varepsilon=10^{-10}$. Applied after the splitting, it
is accurate to $4.2\times10^{-4}$ uniformly in
$\varepsilon$. The trained approximation then attains the same
$\varepsilon$-uniform error, to within a factor of $1.5$ over three random
seeds, at $0.51$ times the cost of a training step. That
cost is $1.08$ times the cost of an unenriched step.

Second, the corrector transfers to the nonlinear system at a constant cost.
For the Navier--Stokes problem the wall trace travels with the flow, and the
trial function contains no information about this transport, so that the
network has to recover it from the residual (Section~\ref{subsec:slpinn-nse}).
At the fixed random seed used throughout the article, the error is again
uniform in $\varepsilon$, and it is $1.30$, $1.32$, $1.31$ and $1.29$ times
the Stokes
error at the four viscosities considered. Over three random seeds at
$\varepsilon=10^{-8}$ the same ratio takes the
values $1.31$, $1.63$ and $1.33$. The cost of the nonlinearity should
therefore be read as a factor of about $1.4$ with a scatter of about $20\%$.
The four viscosities establish that this factor does not grow as the layer
becomes thinner.

Third, the analysis supplies the corrector, namely the profile of the layer
together with its amplitude law $\Theta|_\Gamma=-\omega^0|_\Gamma$, and the
scale $\sqrt{\varepsilon}$ alone does not suffice. In Section~\ref{sec:slpinn}
we compare three enrichments which all use the scale $\sqrt{\varepsilon}$ and
all satisfy the boundary conditions exactly. The corrector-informed enrichment inserts the analytic profile and ties the
amplitude of the layer to the wall trace of the network, as the amplitude law
requires, and the limit solution $\omega^0$ itself enters nowhere. The
scale-only enrichment leaves the profile and the amplitude to a sub-network of
the stretched variable, shared by the two walls or one per wall. The corrector-informed enrichment gives
an error that is uniform in
$\varepsilon$ over six decades, while the scale-only enrichment degrades as
$\varepsilon\to0$. Its error grows by a factor of three over the same
decades, and its maximum pointwise error inside the layer reaches the
amplitude of the layer. Giving each wall its own sub-network reduces the error of the
scale-only enrichment by a factor of two to seven, and leaves the growth. At
$\varepsilon\le10^{-8}$ the corrector-informed enrichment is $1.9$ to $7.5$
times more accurate than either scale-only enrichment. Adding collocation points inside the layer does not remove
this defect.

We also emphasize one point concerning the error measures used in this
article. A global relative $L^2$ error is a poor measure for a boundary layer
problem. The layer occupies a few Fourier modes, and those modes carry little
of the energy of the solution. A uniform evaluation grid does not resolve the
layer either. We therefore decompose the error by the Fourier mode in $x$,
since the layer
is built into one or two of the modes. We also evaluate the error on a grid
graded toward the walls, which resolves the layer and extends far beyond it,
since the error of a network which misses the layer is not confined
to the layer. The ratio between the error
on the uniform grid and the error on the wall-graded grid is $3.2$ for the
unenriched network, $3.8$ for its finite volume counterpart, and at
most $1.05$ for
the enriched one.

Physics-Informed Neural Networks (PINNs) train a network by minimizing the
residual of
the differential equation at a set of collocation points. They go back to
\cite{LagarisLikasFotiadis1998}, and they have been developed extensively
since \cite{RaissiPerdikarisKarniadakis2019}. We refer to the survey
\cite{KarniadakisEtAl2021}, to the software of
\cite{LuMengMaoKarniadakis2021}, and to the hard enforcement of the
constraints of \cite{LuPestourieEtAl2021}. It is well-understood, however,
that a network trained by the loss alone
struggles once the solution carries a thin layer. The reasons have been
examined from several directions. A fully connected network is biased toward
the low frequencies \cite{RahamanEtAl2019,CaoFangWuZhouGu2021}. The terms of
the composite loss are unbalanced, and that imbalance produces pathologies of
the gradient flow \cite{WangTengPerdikaris2021}. Failures to predict stiff or
sharply varying solutions are reported for transport in porous media
\cite{FuksTchelepi2020}, for stiff chemical kinetics
\cite{DeFlorioSchiassiFurfaro2022}, and more generally in
\cite{FernandezEtAl2023}. For the boundary layers in particular, the
difficulty is documented in
\cite{ArzaniCasselDSouza2023,BararniaEsmaeilpour2022,CaoGaoGuoYuan2023}, and
it is precisely the difficulty which the present article measures and then
removes.

The method that we study in this article continues a line of work of the authors.
Semi-analytic and singular layer PINNs (sl-PINNs) were developed earlier for
singularly
perturbed boundary value problems \cite{GieHongJung2024}, in a rectangular
domain \cite{GieHongJungMunkhjin2024}, and in a convex polygonal domain, where
the layers meet the corner singularities \cite{GieHongJungPolygonal}. They
were developed as well for convection-dominated problems in two dimensions
\cite{GieHongJungLee2026}, and for Burgers' equation
\cite{ChangGieHongJungBurgers}. An overview of these
methods is given in \cite{GieJungKimChapter}. The closest to the present
article is the
plane-parallel flow at small viscosity \cite{ChangGieHongJung2024}. There the
corrector, or its curl, is embedded in the structure of the network.  The same
construction has since been carried to the periodic parallel pipe flow, where
the curvature of the wall enters the corrector \cite{MoonGieHongJungPipe}.  The
finite-volume residual of \S\ref{subsec:fv-residual} possesses antecedents as
well, both in
the classical numerics, where enriched finite volume approximations were
constructed for the plane-parallel flow \cite{GieJungLee2020}, and in the neural
setting \cite{GieKim2025,GieJungKimFVNN}.

The main novelty of the present article lies in the way the corrector
structure changes both the trial function and the residual discretization that
is admissible with it. The problem is also different from the earlier ones.
The plane-parallel flow is one-dimensional in the wall-normal variable, and
its corrector is an explicit exponential. The problem that we consider in this
article is the two-dimensional Navier--Stokes system in the vorticity--stream
function
form. The corrector of this system is the solution of the transport--diffusion
equation
\eqref{eq:gj-corrector} of \cite{GieJung2013}, and the nonlinearity is present
inside the layer.
This difference makes the second and the third conclusions above possible.
The question whether the thin scale or the corrector itself is the essential
ingredient requires a corrector which depends on time and is transported along
the wall.
The question whether a corrector derived by a linear analysis survives
requires a nonlinear residual.

The wider literature on the boundary layers and the neural networks divides,
roughly speaking, into the methods which resolve the thin scale and the methods
which encode it.  The methods of the first kind stretch or decompose the
domain. They include
the coordinate transformation for high Reynolds number boundary layer flows
\cite{ZhangSuYuan2025}, the multi-scale networks built on the matched
asymptotic expansions \cite{HuangQiuWangWang2024}, and the zonal coupling of a
Blasius inner model to an outer Navier--Stokes model \cite{PangXiaoYang2026}.
The methods of the second kind put the asymptotic structure into the trial
space. They include the theory-guided networks for singularly perturbed
boundary layer problems \cite{ArzaniCasselDSouza2023}, a layer prior in the
architecture of the network \cite{WangZhaoMaSong2024}, and the composite
expansion of Chien \cite{WangZhangHe2024}. In \cite{LiuZhangLiHe2026},
composite-expansion networks are trained once per asymptotic term and then
assembled for any small $\varepsilon$. In \cite{ShenJuZhu2026}, transferable
neural networks approximate the inner and the outer solutions of the matched
asymptotic expansions, and a rescaling of the inner variable lets the same
parameters serve layers of different thicknesses.

The present article belongs to the second group, and it differs from the earlier
work of that group in two respects.  First, the trial function encodes the
corrector of the boundary layer analysis
of \cite{GieJung2013}, together with the amplitude law of the corrector,
rather than a generic composite expansion or a stretched variable. The comparison of Section~\ref{sec:slpinn} shows that the corrector, and not
the scale of the layer only, has to be built into the trial function.
Second, and to the best of our knowledge not addressed elsewhere, encoding the
corrector changes what the residual itself may be discretized with.  The
methods that resolve the thin scale must retain that scale in the
discretization. Removing the scale from the learned remainder removes it from
the discretization as well. The first conclusion above measures that effect.

Three strands of our earlier work meet in the present article. The singular
layer PINNs of
\cite{GieHongJung2024,GieHongJungMunkhjin2024,GieHongJungPolygonal,GieHongJungLee2026,ChangGieHongJungBurgers,ChangGieHongJung2024,MoonGieHongJungPipe}
put the corrector into the trial space.
The Finite Volume Neural Networks (FVNNs) of \cite{GieKim2025,GieJungKimFVNN}
replace
the pointwise residual by a conservative one, built from the first derivatives
of the network on a grid. The singular layer Finite Volume Neural Networks
(sl-FVNNs) of
\cite{GieJungKimBurgers} combine the two on the stationary viscous Burgers
equations, where the layers sit along the outflow sides of a rectangle. Those
works leave open whether the combination survives a time-dependent
system in which the corrector is transported along the wall. In such a system
the corrector is given by the solution of a
transport--diffusion equation rather than by an explicit exponential, and the
layer carries the nonlinearity. The present article settles that question on
the Navier--Stokes equations.
We embed the corrector of the boundary layer analysis of \cite{GieJung2013}
rather than a profile of the right width. This choice separates the thin scale
from the corrector, and makes the
conservative residual usable on a grid that does not resolve the layer.

All of the experiments in this article are posed on the flat channel
$\Omega=[0,2\pi]\times[0,1]$,
periodic in $x$, with the slip boundary conditions \eqref{eq:slip}, and the
limit flow of the problems of \S\ref{sec:exp7}--\S\ref{sec:exp10} has a
constant slip velocity along each wall.  Those three
choices make the boundary layer corrector of \S\ref{sec:corrector}
available in a closed form. The curvature term of the general analysis
vanishes. The walls are streamlines, so that the normal transport term
vanishes identically. The slip speed is constant, so that the corrector
equation reduces to the heat equation in a travelling frame.  The layer of the
exact solutions of \S\ref{sec:exp7} onwards is that
corrector, up to the order $\mathcal{O}(\varepsilon)$. An enrichment derived
from the corrector is then measured against a
problem that the analysis describes, and not against one that merely has a
thin layer.

The price that we pay is a limitation on what the experiments support.
A curved wall, a slip speed that varies along the wall, and the no-slip
boundary condition each restore a term that the present construction removes.
None of the three is tested in this article. We state in \S\ref{subsec:exp9-summary}
which of the claims of the article depend on the constancy of the slip speed
and which do not.  The no-slip Prandtl layer, in particular, is a different
object from the
$\sqrt{\varepsilon}$ vorticity layer that we study, and nothing in the present
article should be read as evidence about it.

This paper is organized as follows. In Section~\ref{sec:corrector}, we recall
the boundary layer corrector of \cite{GieJung2013} and reduce it to the flat
channel considered in this article. In Section~\ref{sec:framework}, we
introduce the PINN
formulation, the finite volume residual, and the error measures, and we verify
on smooth test problems that a small viscosity alone does not degrade the
approximations. In Section~\ref{sec:failure}, we introduce manufactured
solutions with a $\sqrt{\varepsilon}$ vorticity layer, for the Stokes and for
the Navier--Stokes equations, and we show that the vanilla PINN fails on both
equations. In Section~\ref{sec:exp7}, we construct exact solutions consistent
with the vanishing viscosity limit. In Section~\ref{sec:slpinn}, we introduce the
three singular layer PINN approximations, two scale-only ones and the
corrector-informed one, and we compare them on the Stokes and on the
Navier--Stokes problems. In Section~\ref{sec:exp10}, we replace the Laplacian
of the residual by a finite volume flux difference and introduce the singular
layer finite volume neural network. Section~\ref{sec:conclusion} contains the
conclusion.

\section{Boundary layer corrector of the vorticity equations}
\label{sec:corrector}

In \S\ref{sec:slpinn} we embed in the trial function the corrector of the
boundary layer analysis of Gie and Jung~\cite{GieJung2013}, rather than a
heuristic profile of the right width. In the present section we recall that
corrector in the notation of~\cite{GieJung2013}, and in
\S\ref{subsec:corrector-flat} we reduce it to the flat channel of our
experiments. The corrector is one instance of a broader theory. For that
theory, and for
the vanishing viscosity problem under the generalized Navier conditions of
which \eqref{eq:slip} is the special case $\alpha=2\gamma$, we refer the
reader to the monograph \cite{GieHamoudaJungTemam2018}, the survey
\cite{GieJungTemam2016}, and \cite{GieKelliher2012}.

\subsection[The corrector of Gie and Jung]{The corrector of Gie and
Jung~\protect\cite{GieJung2013}}
\label{subsec:corrector-gj}

As $\varepsilon$ tends to zero, the solution $(\omega^\varepsilon,\psi^\varepsilon)$
of \eqref{eq:vorticity-NS}--\eqref{eq:stream-poisson} is expected to converge to
the solution $(\omega^0,\psi^0)$ of the limit problem obtained by setting
$\varepsilon=0$,
\begin{equation}
\frac{\partial\omega^{0}}{\partial t}
+\frac{\partial\psi^{0}}{\partial y}\frac{\partial\omega^{0}}{\partial x}
-\frac{\partial\psi^{0}}{\partial x}\frac{\partial\omega^{0}}{\partial y}
=f,
\qquad
-\Delta\psi^{0}=\omega^{0},
\qquad
\mathbf{u}^{0}\cdot\mathbf{n}=0 \ \text{ on }\Gamma,
\label{eq:limit-problem}
\end{equation}
where $\mathbf{u}^0=(\partial_y\psi^0,-\partial_x\psi^0)$.  The limit problem
\eqref{eq:limit-problem} is of first order in the convection
and carries no diffusion. The limit problem accepts the impermeability condition
$\mathbf{u}^0\cdot\mathbf{n}=0$ only. The second condition $\omega=0$ of
\eqref{eq:slip} is in general lost in the limit.  The discrepancy
$\omega^{0}|_\Gamma\neq0$ is what the corrector has to remove, and it is removed
in a thin neighborhood of $\Gamma$.

Let $\boldsymbol\xi=(\xi_1,\xi_2)$ be the orthogonal curvilinear system adapted
to $\Gamma$, where $\xi_1$ is tangential with period $\ell$ and $\xi_2$ is the
distance to the boundary, and let $h(\xi_1,\xi_2)=1-\gamma(\xi_1)\xi_2$ be the
Jacobian,
$\gamma$ denoting the curvature.  Let $\Omega_{2\delta}$ be the tubular
neighborhood of width $2\delta$, and let $\sigma$ be the cut-off function of
\cite[(4.8)]{GieJung2013},
\begin{equation}
\sigma(\xi_2)=1 \ \ (0\le\xi_2\le\delta),
\qquad
\sigma(\xi_2)=0 \ \ (\xi_2\ge2\delta).
\label{eq:gj-cutoff}
\end{equation}
The width of that neighborhood follows from a balance of two terms.  We write
$\omega^\varepsilon\simeq\omega^0+\Theta$ and insert this into
\eqref{eq:vorticity-NS}. At the leading order, $\Theta$ has to satisfy the
difference between the viscous and the limit equations. In that difference,
the only term that can balance $\varepsilon\Delta\Theta$ is the one carrying
the largest normal derivative.  With the stretched normal variable
\begin{equation}
\bar\xi_2=\frac{\xi_2}{\sqrt{\varepsilon}},
\label{eq:stretched}
\end{equation}
we have $\partial_{\xi_2}=\varepsilon^{-1/2}\partial_{\bar\xi_2}$, so that
$\varepsilon\,\partial^2_{\xi_2}\Theta=\partial^2_{\bar\xi_2}\Theta$ is of the
order $\mathcal{O}(1)$, while the tangential diffusion
$\varepsilon\,\partial^2_{\xi_1}\Theta$ is of the order
$\mathcal{O}(\varepsilon)$ and is therefore dropped.  The thickness of the
layer is $\sqrt{\varepsilon}$, and the diffusion that
survives acts in the normal direction only. The transport coefficients are
evaluated at $\xi_2=0$. The variation of these coefficients across a layer of width
$\sqrt{\varepsilon}$ contributes at the order
$\mathcal{O}(\sqrt{\varepsilon})$.  The corrector is then given by
$\Theta=\sigma(\xi_2)\overline\Theta$ in $\Omega_{2\delta}$ and by $0$ outside
\cite[(4.9)]{GieJung2013}, where $\overline\Theta$ solves
\cite[(4.10)]{GieJung2013}
\begin{equation}
\frac{\partial\overline\Theta}{\partial t}
-\varepsilon\frac{\partial^2\overline\Theta}{\partial\xi_2^2}
+\frac{1}{h}\Big|_{\xi_2=0}\frac{\partial\psi^0}{\partial\xi_1}
   \frac{\partial\overline\Theta}{\partial\xi_2}
-\frac{1}{h}\Big|_{\xi_2=0}\frac{\partial\psi^0}{\partial\xi_2}
   \frac{\partial\overline\Theta}{\partial\xi_1}
=0,
\label{eq:gj-corrector}
\end{equation}
with $\overline\Theta=-\omega^0$ at $\xi_2=0$, $\overline\Theta=0$ at
$\xi_2=2\delta$, the periodicity in $\xi_1$, and $\overline\Theta|_{t=0}=0$.
Three features of \eqref{eq:gj-corrector} are used repeatedly in this article.
The diffusion acts in the normal direction only, and this sets the scale
$\sqrt{\varepsilon}$ of the layer.  The transport coefficients are frozen at the
wall, $\xi_2=0$, and hence they are data of the limit problem and not of the
layer.  Finally, the inhomogeneity enters only through the boundary value
$-\omega^0|_\Gamma$, so that \emph{the amplitude of the layer is the wall trace of
the outer solution}, and \S\ref{subsec:exp8-ansatz} builds that property into
the network.

The estimates of \cite[Lemma 4.2, (4.15)]{GieJung2013} fix the sizes, namely
$\partial^n\Theta/\partial\xi_2^n=\mathcal{O}(\varepsilon^{1/(2p)-n/2})$ in
$L^\infty(0,T;L^p)$, so that each normal derivative costs a factor
$\varepsilon^{-1/2}$ while the tangential derivatives are free.  That anisotropy
is the entire difficulty for a PINN, because the residual contains
$\varepsilon\,\partial_{\xi_2}^2\Theta=\mathcal{O}(1)$, which is built from a
function whose second derivative is of the size $\mathcal{O}(\varepsilon^{-1})$.

\begin{remark}[Orientation of the Jacobian]
\label{rem:sign}
We write the vorticity equation with the material derivative, as in
\eqref{eq:vorticity-NS}, so that, with
$\mathbf{u}=(\partial_y\psi,-\partial_x\psi)$, the
convection term is
$(\mathbf{u}\cdot\nabla)\omega=\partial_y\psi\,\partial_x\omega-\partial_x\psi\,\partial_y\omega$,
as in \cite[Remark 2.1]{GieJung2013}.  The definition
$J(v,w)=\partial_x v\,\partial_y w-\partial_y v\,\partial_x w$ of
\cite[(2.3)]{GieJung2013} carries the opposite
sign, and that sign propagates to the two convection terms of
\eqref{eq:gj-corrector}.  The tangential transport in
\eqref{eq:corrector-eq} below therefore appears as $+U\,\partial_x\Theta$, whereas
\eqref{eq:gj-corrector} as printed would give $-U\,\partial_x\Theta$.
\end{remark}

\subsection{Reduction to the flat channel}
\label{subsec:corrector-flat}

All of the experiments in this article are posed on
$\Omega=[0,2\pi]\times[0,1]$, with
$\xi_1=x$ and $\xi_2=y$ at the lower wall.  Two of the terms in
\eqref{eq:gj-corrector} disappear, for two different
reasons. We state the reasons separately, because a curved or a permeable
problem would restore each of these terms.

\emph{Curvature.}  The channel is flat, so that $\gamma\equiv0$ and $h\equiv1$.
In particular the operator $\mathcal{L}$ of \cite[(3.4)]{GieJung2013}, which
\cite[Remark 3.1]{GieJung2013} identifies as
$-\gamma(\xi_1)\partial_{\xi_2}+\mathcal{O}(\varepsilon^{1/2})\partial_{\xi_2}$
and singles out as the source of the critical $\varepsilon^{3/4}$ error in a
curved domain, is absent here. For boundary layers of the Navier--Stokes
equations in a curved domain, see also \cite{GieHamoudaTemam2012}.

\emph{Impermeability.}  The walls are streamlines, that is,
$\psi^0=\text{const}$ on $\Gamma$, so that
$\partial\psi^0/\partial\xi_1|_{\xi_2=0}\equiv0$ and the normal transport term of
\eqref{eq:gj-corrector} vanishes identically, not to the leading order only
but exactly.

The remaining term is the tangential transport, with the sign of
Remark~\ref{rem:sign}:
\begin{equation}
\partial_t\Theta + U(x,t)\,\partial_x\Theta = \varepsilon\,\partial_y^2\Theta,
\qquad
\Theta\big|_{\rm wall} = -\omega^0\big|_{\rm wall},
\qquad
\Theta\big|_{t=0}=0,
\label{eq:corrector-eq}
\end{equation}
where
\begin{equation}
U := \frac{\partial\psi^0}{\partial\xi_2}\bigg|_{\xi_2=0}
   = \partial_y\psi^0\big|_{\Gamma}
\label{eq:U-def}
\end{equation}
is the tangential slip velocity of the limit solution at the wall.  For the
Stokes problem the convection term is absent from the equation itself,
so that the transport term $U\,\partial_x\Theta$ is absent from
\eqref{eq:corrector-eq} as well, regardless of the slip velocity of the limit
solution. Equation \eqref{eq:corrector-eq} then reduces to the heat equation. For the
Navier--Stokes problem $U$ is determined by the limit solution $\psi^0$
through \eqref{eq:U-def}.

Two consequences of \eqref{eq:corrector-eq} shape the remainder of the
article. First, the layer of the exact solutions constructed in \S\ref{sec:exp7}
satisfies the equation in \eqref{eq:corrector-eq} up to the order
$\mathcal{O}(\varepsilon)$, with the initial condition replaced as described
in \S\ref{subsec:exp7-design}, and \S\ref{subsec:exp7-valid} verifies this
residual numerically. Second, when $U$ is constant in $x$,
\eqref{eq:corrector-eq} is the heat equation in the frame $\zeta=x-Ut$, and
its solution is a single complementary error function. The construction of
\S\ref{subsec:exp7-design} uses this fact, and \S\ref{subsec:exp9-summary}
states which of the conclusions depend on the constant slip speed.

\section{PINN formulation and smooth test problems}
\label{sec:framework}

In the present section we set up the network, the trial function, the
collocation sets, the residuals and the schedule of the optimizer that all of
the experiments of this article share. We also compute two baselines, a smooth
Stokes problem and a smooth Navier--Stokes problem, against which the boundary
layer results of \S\ref{sec:failure}--\S\ref{sec:exp10} are compared. The
exact solutions of the baselines are smooth on the entire domain, with
spatial length scales independent of $\varepsilon$, so that decreasing
$\varepsilon$ does not create a thin layer. The two baselines establish that a
small viscosity by itself does not degrade the method. We then replace the Laplacian
of the residual by a finite-volume flux difference and measure what that
substitution costs on the same smooth problems.

\subsection{Smooth Stokes test problem and the PINN formulation}
\label{subsec:smooth-stokes}

We first consider a smooth Stokes problem, in order to verify the basic PINN
implementation before any singular-layer structure is introduced.

Let
\[
\Omega=[0,2\pi]\times[0,1],
\qquad
t\in[0,1],
\]
and we take first \[ \varepsilon=1. \]

The solution is periodic in the $x$-direction and satisfies
\begin{equation}
\omega(x,0,t)=\omega(x,1,t)=0,
\qquad
\psi(x,0,t)=\psi(x,1,t)=0.
\label{eq:wall-bc-exp1}
\end{equation}

\subsubsection{Exact solution}

Let
\[
\lambda=1+\pi^2
\]
and define
\begin{equation}
\psi(x,y,t)
=
e^{-\varepsilon\lambda t}
\sin(x)\sin(\pi y).
\label{eq:psi-exp1}
\end{equation}
Then
\[
-\Delta\psi=\lambda\psi,
\]
and therefore
\begin{equation}
\omega(x,y,t)
=
\lambda e^{-\varepsilon\lambda t}
\sin(x)\sin(\pi y).
\label{eq:omega-exp1}
\end{equation}

Since
\[
\partial_t\omega=-\varepsilon\lambda\omega,
\qquad
\Delta\omega=-\lambda\omega,
\]
we have
\[
\partial_t\omega-\varepsilon\Delta\omega=0.
\]
Thus, in this experiment,
\begin{equation}
f=\operatorname{curl}\mathbf{F}=0.
\label{eq:f-exp1}
\end{equation}

\subsubsection{PINN ansatz and training setup}
\label{subsubsec:settings}

The trial functions are produced by a feed-forward neural network
\begin{equation}
\mathcal{N}_{\theta}(x,y,t)
=
\begin{pmatrix}
\widehat{\omega}_{\theta}(x,y,t)\\
\widehat{\psi}_{\theta}(x,y,t)
\end{pmatrix}
\in\mathbb{R}^{2},
\label{eq:raw-network-exp1}
\end{equation}
where $\theta$ denotes all of the trainable weights and biases.  The physical input
$\mathbf{z}^{(0)}=(x,y,t)^T\in[0,2\pi]\times[0,1]\times[0,1]$ is mapped
componentwise by the affine transformation $z\mapsto
2(z-z_{\min})/(z_{\max}-z_{\min})-1$ onto $[-1,1]^3$ before it is passed
through the network, and the hidden states are defined
recursively, a final linear layer producing the two outputs:

\[
\mathbf{z}^{(\ell)}=\sigma\bigl(W^{(\ell)}\mathbf{z}^{(\ell-1)}+\mathbf{b}^{(\ell)}\bigr),
\quad \ell=1,\ldots,L,
\qquad
\begin{pmatrix}\widehat{\omega}_{\theta}\\ \widehat{\psi}_{\theta}\end{pmatrix}
=W^{(L+1)}\mathbf{z}^{(L)}+\mathbf{b}^{(L+1)}.
\]

In all of the experiments of this article the trunk of the network has the
architecture
$3\longrightarrow50\longrightarrow50\longrightarrow50\longrightarrow50\longrightarrow2$,
that is, $L=4$ hidden layers with $50$ neurons in each layer. The activation
function is $\sigma(s)=\tanh(s)$. All of the computations are carried out in
double precision.

Since both $\omega$ and $\psi$ satisfy the homogeneous wall conditions
\eqref{eq:wall-bc-exp1}, we impose those conditions exactly through the ansatz
\begin{equation}
\omega_{\theta}(x,y,t)=y(1-y)\,\widehat{\omega}_{\theta}(x,y,t),
\qquad
\psi_{\theta}(x,y,t)=y(1-y)\,\widehat{\psi}_{\theta}(x,y,t),
\label{eq:hard-bc-exp1}
\end{equation}
so that no wall boundary loss is required.

We use three sets of training points, all of them drawn uniformly:

\[
\mathcal{T}_{\Omega}=\{(x_i,y_i,t_i)\}_{i=1}^{N_{\Omega}}\subset\Omega\times(0,1),
\qquad
\mathcal{T}_{\mathrm{per}}=\{(y_i^b,t_i^b)\}_{i=1}^{N_{\mathrm{per}}},
\qquad
\mathcal{T}_{\mathrm{IC}}=\{(x_i^0,y_i^0,0)\}_{i=1}^{N_{\mathrm{IC}}},
\]
the interior collocation set, the matched periodic set on the two ends of the
period, and the initial set, with $N_{\Omega}=10{,}000$,
$N_{\mathrm{per}}=1{,}000$ and $N_{\mathrm{IC}}=2{,}000$.

For the interior points, the Stokes residuals are
\begin{equation}
R_{\omega}
=
\frac{\partial\omega_{\theta}}{\partial t}
-\varepsilon
\left(
\frac{\partial^2\omega_{\theta}}{\partial x^2}
+
\frac{\partial^2\omega_{\theta}}{\partial y^2}
\right),
\label{eq:Rw-exp1}
\end{equation}
and
\begin{equation}
R_{\psi}
=
-\left(
\frac{\partial^2\psi_{\theta}}{\partial x^2}
+
\frac{\partial^2\psi_{\theta}}{\partial y^2}
\right)
-\omega_{\theta}.
\label{eq:Rpsi-exp1}
\end{equation}
All derivatives are evaluated by automatic differentiation.

The corresponding residual losses are the mean squares over
$\mathcal{T}_{\Omega}$,
\begin{equation}
L_{\omega}
=
\frac{1}{N_{\Omega}}
\sum_{i=1}^{N_{\Omega}}
\bigl|R_{\omega}(x_i,y_i,t_i;\theta)\bigr|^2,
\qquad
L_{\psi}
=
\frac{1}{N_{\Omega}}
\sum_{i=1}^{N_{\Omega}}
\bigl|R_{\psi}(x_i,y_i,t_i;\theta)\bigr|^2.
\label{eq:Lw-exp1}
\end{equation}

The periodic condition is imposed softly by matching the two boundaries at the
same $(y_i^b,t_i^b)$:
\begin{align}
L_{\mathrm{per}}
={}&
\frac{1}{N_{\mathrm{per}}}
\sum_{i=1}^{N_{\mathrm{per}}}
\Big[
\left|
\omega_{\theta}(0,y_i^b,t_i^b)
-
\omega_{\theta}(2\pi,y_i^b,t_i^b)
\right|^2
\nonumber\\
&+
\left|
\psi_{\theta}(0,y_i^b,t_i^b)
-
\psi_{\theta}(2\pi,y_i^b,t_i^b)
\right|^2
\nonumber\\
&+
\left|
\partial_x\omega_{\theta}(0,y_i^b,t_i^b)
-
\partial_x\omega_{\theta}(2\pi,y_i^b,t_i^b)
\right|^2
\nonumber\\
&+
\left|
\partial_x\psi_{\theta}(0,y_i^b,t_i^b)
-
\partial_x\psi_{\theta}(2\pi,y_i^b,t_i^b)
\right|^2
\Big].
\label{eq:Lper-exp1}
\end{align}

The initial condition is imposed on the vorticity,
\begin{equation}
L_{\mathrm{IC}}
=
\frac{1}{N_{\mathrm{IC}}}
\sum_{i=1}^{N_{\mathrm{IC}}}
\bigl|
\omega_{\theta}(x_i^0,y_i^0,0)-\omega(x_i^0,y_i^0,0)
\bigr|^2,
\label{eq:LIC-exp1}
\end{equation}
and the total loss is
\begin{equation}
L(\theta)=L_{\omega}+L_{\psi}+10L_{\mathrm{per}}+10L_{\mathrm{IC}}.
\label{eq:loss-exp1}
\end{equation}

The network is trained with the Adam optimizer at the learning rate
$10^{-3}$ for $5{,}000$ epochs, followed by an L-BFGS refinement of at most
$500$ iterations.

Unless stated otherwise, all of the experiments in this article use the
following settings.
\begin{enumerate}
\item[(i)] \emph{Architecture and initialization.} The trunk is
$3\longrightarrow50\longrightarrow50\longrightarrow50\longrightarrow50\longrightarrow2$,
that is, $L=4$ hidden layers of $50$ neurons, with the activation $\sigma(s)=\tanh(s)$. The weights are initialized by the
Xavier normal initialization and the biases by zero. The enrichments of
\S\ref{sec:slpinn} add the sub-networks described there, whose output layers
are initialized by zero, so that the enrichment contributes nothing at the
initialization.
\item[(ii)] \emph{Boundary conditions.} The wall conditions are imposed exactly
through the ansatz \eqref{eq:hard-bc-exp1}, so that no wall term enters the
loss.
\item[(iii)] \emph{Training points.} $N_{\Omega}=10{,}000$ interior collocation
points, $N_{\mathrm{per}}=1{,}000$ matched periodic pairs and
$N_{\mathrm{IC}}=2{,}000$ initial points, all drawn uniformly once per run and
kept fixed during the training.
\item[(iv)] \emph{Loss.} The mean-square residual loss \eqref{eq:loss-exp1},
with the weights $1$, $1$, $10$ and $10$ on $L_{\omega}$, $L_{\psi}$,
$L_{\mathrm{per}}$ and $L_{\mathrm{IC}}$ respectively.  All derivatives are
evaluated by automatic differentiation, except where a finite-volume residual
is used instead, as stated in \S\ref{subsec:fv-residual} and
\S\ref{sec:exp10}.
\item[(v)] \emph{Optimizer.} $5{,}000$ Adam epochs at the learning rate
$10^{-3}$, followed by at most $500$ L-BFGS iterations, with the learning rate
$1$, the strong Wolfe line search, the history size $100$ and the tolerances
$10^{-12}$ on the gradient and $10^{-14}$ on the change of the loss.
\item[(vi)] \emph{Precision, software and hardware.} All computations are
carried out in double precision with PyTorch 2.14 and Python 3.13 on the CPU
of a laptop (MacBook Pro with an Apple M3 Pro chip and 18 GB of memory), with
up to three runs executing concurrently.  The timings are collected in
Table~\ref{tab:fv-cost}.
\item[(vii)] \emph{Random seed.} The seed is $1234$ throughout, except in
\S\ref{subsec:exp9-seed}, where two further seeds are used.
\end{enumerate}

\subsubsection{Numerical results}

For $\varepsilon=1$ the relative $L^2$ errors are approximately
\[
0.045\% \quad\text{for }\omega,
\qquad
0.371\% \quad\text{for }\psi,
\]
and
\[
0.427\% \quad\text{for }u,
\qquad
0.897\% \quad\text{for }v.
\]

For comparison, when the wall boundary conditions are imposed softly through a
boundary loss instead, the relative $L^2$ errors are
\[
2.267100\times10^{-3},
\quad
1.395662\times10^{-2},
\quad
1.004514\times10^{-2},
\quad
1.808601\times10^{-2}
\]
for $\omega$, $\psi$, $u$, and $v$, respectively. We observe that the hard
enforcement of the homogeneous wall boundary conditions improves the accuracy
substantially, the error in the vorticity being reduced by a factor of about
five.

The vorticity--stream function formulation with the wall boundary conditions
imposed exactly therefore solves the smooth Stokes problem accurately at
$\varepsilon=1$.

In order to see whether a small viscosity by itself causes the vanilla PINN to
deteriorate, we repeat the same boundary-layer-free experiment for
\[
\varepsilon=1,\qquad 10^{-2},\qquad 10^{-4},
\]
without changing the architecture of the network, the training parameters, or
the spatial structure of the exact solution. Since the exact solution remains
a fixed smooth Fourier mode for all of the three values of $\varepsilon$, no
boundary layer is present in these computations.

The global space--time relative $L^2$ errors on
$\Omega\times[0,1]$ are summarized in Table~\ref{tab:stokes-eps-rel}.

\begin{table}[h!]
\centering
\begin{tabular}{ccccc}
\toprule
$\varepsilon$ & $E_\omega$ & $E_\psi$ & $E_u$ & $E_v$ \\
\midrule
$1$
& $4.508575\times10^{-4}$
& $3.705345\times10^{-3}$
& $4.269009\times10^{-3}$
& $8.965611\times10^{-3}$ \\
$10^{-2}$
& $1.675088\times10^{-3}$
& $9.945179\times10^{-4}$
& $1.624944\times10^{-3}$
& $3.663889\times10^{-3}$ \\
$10^{-4}$
& $1.789485\times10^{-3}$
& $1.261701\times10^{-3}$
& $1.727749\times10^{-3}$
& $3.751198\times10^{-3}$ \\
\bottomrule
\end{tabular}
\caption{Global space--time relative $L^2$ errors for the smooth,
boundary-layer-free Stokes problem.}
\label{tab:stokes-eps-rel}
\end{table}

The maximum absolute errors behave in the same way, the largest vorticity
error growing only from $5.86\times10^{-3}$ at $\varepsilon=1$ to
$4.15\times10^{-2}$ at $\varepsilon=10^{-4}$.

At $\varepsilon=10^{-4}$ the selected-time spatial relative $L^2$ errors
remain small over the entire time interval as well.

We observe from these results that the vanilla PINN remains accurate even when
$\varepsilon$ is reduced to $10^{-4}$, provided that the solution remains
smooth and that no boundary layer is present. Within this baseline test, then,
a small viscosity by itself does not produce a loss of accuracy comparable
with the singular-layer difficulty that we consider later.

\subsection{Smooth Navier--Stokes test problem}
\label{subsec:smooth-nse}

We next test the full nonlinear Navier--Stokes vorticity equation, with no
boundary layer structure. The purpose is to verify that the PINN resolves the
nonlinear convection term in a smooth setting, before the singular layers are
introduced.

We again use
\[
\Omega=[0,2\pi]\times[0,1],
\qquad
0\le t\le1,
\]
with periodicity in $x$ and the homogeneous wall conditions
\[
\omega(x,0,t)=\omega(x,1,t)=0,
\qquad
\psi(x,0,t)=\psi(x,1,t)=0.
\]
As for the Stokes problem above, these wall conditions are imposed exactly
through
\[
\omega_\theta=y(1-y)\widehat\omega_\theta,
\qquad
\psi_\theta=y(1-y)\widehat\psi_\theta.
\]

\subsubsection{Manufactured Navier--Stokes solution}

Let
\[
\alpha=\frac14,
\qquad
\lambda_1=1+\pi^2,
\qquad
\lambda_2=4+\pi^2,
\]
and define
\begin{equation}
\psi(x,y,t)
=e^{-t}\left[
\sin x\sin(\pi y)
+\alpha\sin(2x)\sin(\pi y)
\right].
\label{eq:psi-exp2}
\end{equation}
Then
\begin{equation}
\omega(x,y,t)
=e^{-t}\left[
\lambda_1\sin x\sin(\pi y)
+\alpha\lambda_2\sin(2x)\sin(\pi y)
\right],
\label{eq:omega-exp2}
\end{equation}
which satisfies $-\Delta\psi=\omega$.  The corresponding velocity is
\begin{align}
u(x,y,t)
&=\partial_y\psi
=e^{-t}\pi\cos(\pi y)
\left[\sin x+\alpha\sin(2x)\right],
\label{eq:u-exp2}\\
v(x,y,t)
&=-\partial_x\psi
=-e^{-t}\sin(\pi y)
\left[\cos x+2\alpha\cos(2x)\right].
\label{eq:v-exp2}
\end{align}

The two Fourier modes have different Laplacian eigenvalues, so that $\omega$
differs from every constant multiple of $\psi$, and
consequently the nonlinear convection term
\[
\partial_y\psi\,\partial_x\omega-\partial_x\psi\,\partial_y\omega
\]
does not vanish.  The forcing is manufactured from the complete
Navier--Stokes operator,
\begin{equation}
f
=
\partial_t\omega-
\varepsilon\Delta\omega
+
\partial_y\psi\,\partial_x\omega-
\partial_x\psi\,\partial_y\omega,
\label{eq:f-exp2}
\end{equation}
and it is evaluated by automatic differentiation rather than by formulas
derived by hand.

This example is again deliberately \emph{boundary-layer free}. The spatial
structure of this solution consists of the two fixed smooth
Fourier modes in
\eqref{eq:psi-exp2} only. No spatial scale shrinks with $\varepsilon$, and no
thin viscous layer develops as $\varepsilon\to0$.

\subsubsection{PINN residual and training setup}

We keep the architecture of the network, the hard wall ansatz, the sets of
collocation points, the weights in the loss, and the parameters of the
optimizer the same as for the Stokes problem. The only change in the
differential residual is the addition of the nonlinear convection term,
\begin{equation}
R_\omega
=
\frac{\partial\omega_\theta}{\partial t}
-
\varepsilon\Delta\omega_\theta
+
\frac{\partial\psi_\theta}{\partial y}
\frac{\partial\omega_\theta}{\partial x}
-
\frac{\partial\psi_\theta}{\partial x}
\frac{\partial\omega_\theta}{\partial y}
-f,
\label{eq:Rw-exp2}
\end{equation}
while
\begin{equation}
R_\psi=-\Delta\psi_\theta-\omega_\theta.
\label{eq:Rpsi-exp2}
\end{equation}
The loss and all of the other settings are the settings (i)--(vii) of
\S\ref{subsubsec:settings}.

\subsubsection{Numerical results}

We consider two widely separated values of the viscosity, $\varepsilon=1$ and
$\varepsilon=10^{-4}$, while keeping the exact solution and all of the
training settings fixed, and we collect the global space--time relative $L^2$
errors in Table~\ref{tab:nse-global-rel}.

\begin{table}[h!]
\centering
\begin{tabular}{ccccc}
\toprule
$\varepsilon$ & $E_\omega$ & $E_\psi$ & $E_u$ & $E_v$ \\
\midrule
$1$
& $3.166575\times10^{-4}$
& $2.659901\times10^{-3}$
& $3.112354\times10^{-3}$
& $7.323640\times10^{-3}$ \\
$10^{-4}$
& $5.559928\times10^{-3}$
& $5.124730\times10^{-3}$
& $6.384537\times10^{-3}$
& $1.293362\times10^{-2}$ \\
\bottomrule
\end{tabular}
\caption{Global space--time relative $L^2$ errors for the smooth,
boundary-layer-free Navier--Stokes problem.}
\label{tab:nse-global-rel}
\end{table}

The maximum absolute vorticity error is $1.18\times10^{-2}$ at $\varepsilon=1$
and $7.90\times10^{-2}$ at $\varepsilon=10^{-4}$.

Since the problem is time-dependent, we examine as well the spatial relative
$L^2$ errors at selected times.

The computation at $\varepsilon=1$ gives sub-percent global errors for all of
the four fields, while at $\varepsilon=10^{-4}$ the global errors remain
approximately $0.56\%$, $0.51\%$, $0.64\%$, and $1.29\%$ for $\omega$, $\psi$,
$u$, and $v$, respectively. We observe that the selected-time errors increase
moderately toward $t=1$ in the case of the small viscosity, but that no severe
loss of accuracy occurs.

Taken together with the Stokes computations above, these results establish a
smooth, boundary-layer-free baseline for the two problems. The vanilla PINN
remains effective at every small $\varepsilon$, as long as the exact solution
contains no thin spatial layer. In the experiments that follow we therefore
isolate the
effect of the boundary layers, by introducing exact solutions which possess an
$\varepsilon$-dependent thin spatial scale.

\subsection{Finite-volume residual for the smooth test problems}
\label{subsec:fv-residual}

We revisit the smooth Stokes problem at $\varepsilon=10^{-4}$ (cf.\
Table~\ref{tab:stokes-eps-rel}). In \eqref{eq:Rw-exp1}--\eqref{eq:Rpsi-exp1}
we replace the Laplacian computed by automatic differentiation with a
finite-volume approximation, namely a balance of the fluxes through the faces
of a control volume. That approximation is built from the first derivatives of
the network itself, and it is evaluated on a fixed tensor-product grid rather
than on i.i.d.\ random collocation points. This construction follows the
finite volume neural networks of \cite{GieKim2025} and of
\cite{GieJungKimFVNN,GieJungKimBurgers}, developed for singularly perturbed
problems in one and in two dimensions, and we adapt it to the coupled
vorticity--stream function system.

\subsubsection{Finite-volume grid}

Let
\[
N_x, N_y, N_t \in \mathbb N,
\qquad
\Delta x=\frac{2\pi}{N_x},
\qquad
\Delta y=\frac1{N_y-1},
\]
and define the tensor-product grid
\[
x_i=i\Delta x,\ i=0,\dots,N_x-1,\qquad
y_j=j\Delta y,\ j=0,\dots,N_y-1,
\]
\[
t_k=\frac{k}{N_t-1},\ k=0,\dots,N_t-1.
\]
Since the problem is $2\pi$-periodic in $x$, the grid does not duplicate the
endpoint $x=2\pi$, and the index $i$ is understood modulo $N_x$. The wall rows
$j=0$ and $j=N_y-1$ are retained as stencil neighbors even though
\eqref{eq:hard-bc-exp1} already forces
$\omega_\theta=\psi_\theta=0$ there. For the present computation, $N_x=50$,
$N_y=21$, $N_t=10$, giving $N_xN_yN_t=10{,}500$ collocation points, comparable
to the $N_{\Omega}=10{,}000$ used above.

\subsubsection{Finite-volume Laplacian}

Only first derivatives
$\partial_x\omega_\theta,\partial_y\omega_\theta,\partial_x\psi_\theta,\partial_y\psi_\theta$
are computed by automatic differentiation. Writing $g^x_{i,j,k},g^y_{i,j,k}$
for any pair of such first-derivative fields evaluated at grid node
$(x_i,y_j,t_k)$, define the discrete flux divergence
\begin{equation}
\big(\nabla_h\!\cdot\!(g^x,g^y)\big)_{i,j,k}
=
\frac{g^x_{i+1,j,k}-g^x_{i-1,j,k}}{2\Delta x}
+
\frac{g^y_{i,j+1,k}-g^y_{i,j-1,k}}{2\Delta y},
\qquad j=1,\dots,N_y-2,
\label{eq:fv-div}
\end{equation}
where $i\pm1$ is taken modulo $N_x$. The formula \eqref{eq:fv-div} is the
balance of the fluxes through the four faces of the cell centered at
$(x_i,y_j)$, each face flux being the arithmetic mean of the nodal fluxes on its
two sides, for instance $g^y_{i,j+1/2,k}=(g^y_{i,j,k}+g^y_{i,j+1,k})/2$.
Applying \eqref{eq:fv-div} to
$(g^x,g^y)=(\partial_x\omega_\theta,\partial_y\omega_\theta)$ approximates
$\Delta\omega_\theta$, and applying it to
$(\partial_x\psi_\theta,\partial_y\psi_\theta)$ approximates
$\Delta\psi_\theta$; write $\Delta_h\omega_\theta,\Delta_h\psi_\theta$ for
these two approximations. The finite-volume residuals, restricted to the
interior rows $j=1,\dots,N_y-2$, are
\begin{equation}
R_\omega^h=\partial_t\omega_\theta-\varepsilon\,\Delta_h\omega_\theta,
\qquad
R_\psi^h=-\Delta_h\psi_\theta-\omega_\theta.
\label{eq:fv-residuals-stokes}
\end{equation}
Since the flux balance \eqref{eq:fv-div} is second-order accurate,
$R_\omega^h$ and $R_\psi^h$ carry an $\mathcal{O}(\Delta x^2+\Delta y^2)$
truncation error
even when $(\omega_\theta,\psi_\theta)$ is replaced by the exact solution. For
the present grid we have $\Delta x^2+\Delta y^2\approx1.83\times10^{-2}$.
The exact solution attains $\max|R_\omega^h|\approx4.7\times10^{-5}$ and
$\max|R_\psi^h|\approx4.3\times10^{-2}$, which is consistent with that
estimate. These values are a reference level for the finite-volume residual
rather than a bound. A network which approximates the exact solution cannot be
expected to reduce $R_\omega^h$ and $R_\psi^h$ below the truncation error of
the exact solution, whereas a network which approximates the solution of the
discrete equations instead could.

The periodic loss $L_{\mathrm{per}}$ of \eqref{eq:Lper-exp1} is retained,
since the wraparound in \eqref{eq:fv-div} constrains the residual at the grid
nodes only and does not enforce $\omega_\theta(0,y,t)=\omega_\theta(2\pi,y,t)$
pointwise. The total loss is
\[
L(\theta)=L_\omega^h+L_\psi^h+10L_{\mathrm{per}}+10L_{\mathrm{IC}},
\qquad
L_\omega^h=\operatorname{MSE}(R_\omega^h),
\quad
L_\psi^h=\operatorname{MSE}(R_\psi^h).
\]
All of the other settings are the settings (i)--(vii) of
\S\ref{subsubsec:settings}.

\subsubsection{Numerical results}

Table~\ref{tab:fv-stokes-rel} compares the finite-volume residual against the
automatic-differentiation PINN of \S\ref{subsec:smooth-stokes}, both at
$\varepsilon=10^{-4}$.

\begin{table}[h!]
\centering
\begin{tabular}{lccc}
\toprule
Quantity & PINN ($E$) & FV residual ($E^h$) & ratio $E^h/E$ \\
\midrule
$\omega$ & $1.789485\times10^{-3}$ & $3.282677\times10^{-3}$ & $1.83$ \\
$\psi$ & $1.261701\times10^{-3}$ & $4.854801\times10^{-3}$ & $3.85$ \\
$u$ & $1.727749\times10^{-3}$ & $5.272278\times10^{-3}$ & $3.05$ \\
$v$ & $3.751198\times10^{-3}$ & $1.027956\times10^{-2}$ & $2.74$ \\
\bottomrule
\end{tabular}
\caption{Global space--time relative $L^2$ errors, PINN vs.\ finite-volume
residual, smooth Stokes problem, $\varepsilon=10^{-4}$.}
\label{tab:fv-stokes-rel}
\end{table}

The finite-volume residual is $2$--$4$ times less accurate than the residual
computed by automatic differentiation, and the loss is largest for $\psi$.
This distribution is consistent with the
truncation error identified above. In $R_\omega^h$
the truncation error of $\Delta_h$ is multiplied by $\varepsilon=10^{-4}$ and
is negligible, whereas in $R_\psi^h$ it enters at full size. The error of
$\psi$ therefore grows by the factor $3.9$, and the errors of $u$
and $v$, which are derivatives of $\psi$, by $3.0$ and $2.7$. The error of
$\omega$, which the Stokes equations do not couple to $\psi$, grows by $1.8$
only, through the shared network. The cost of the training, however, drops
substantially, since the first
derivatives alone are required.
One run of $5{,}000$ Adam epochs followed by $500$ L-BFGS iterations takes
$236\,\mathrm s$, against $1026\,\mathrm s$ for the automatic-differentiation
residual under identical optimizer settings and hardware. The speedup is by
the factor $4.3\times$. For this smooth, boundary-layer-free problem, the
finite-volume residual
therefore trades a modest loss of accuracy, concentrated in $\psi$, for a
substantial reduction in the cost of the training.

\subsubsection{Finite-volume residual for the Navier--Stokes problem}
\label{subsubsec:fv-nonlinear}

We repeat the finite-volume construction for the nonlinear Navier--Stokes
system of \S\ref{subsec:smooth-nse}, again at $\varepsilon=10^{-4}$. The grid,
the finite-volume Laplacians
$\Delta_h\omega_\theta,\Delta_h\psi_\theta$, and the retained periodic loss
are exactly as for the Stokes system, and the nonlinear convection term is the
only new element.

The convection term in \eqref{eq:Rw-exp2},
\[
\frac{\partial\psi_\theta}{\partial y}\frac{\partial\omega_\theta}{\partial x}
-\frac{\partial\psi_\theta}{\partial x}\frac{\partial\omega_\theta}{\partial y}
=u_\theta\,\partial_x\omega_\theta+v_\theta\,\partial_y\omega_\theta,
\qquad
u_\theta:=\partial_y\psi_\theta,
\quad
v_\theta:=-\partial_x\psi_\theta,
\]
is a pointwise product and not yet a divergence, so that \eqref{eq:fv-div} is
not directly applicable to it. Since
\[
\partial_xu_\theta+\partial_yv_\theta
=\partial_x\partial_y\psi_\theta-\partial_y\partial_x\psi_\theta=0
\]
identically, for \emph{any} $\psi_\theta$ produced by the network, and not for
the exact solution only, integrating by parts gives
\begin{equation}
u_\theta\,\partial_x\omega_\theta+v_\theta\,\partial_y\omega_\theta
=
\partial_x(u_\theta\omega_\theta)+\partial_y(v_\theta\omega_\theta)
=
\nabla\cdot(u_\theta\omega_\theta,v_\theta\omega_\theta),
\label{eq:conservative-form}
\end{equation}
the substitution introducing no additional error. The right-hand side of
\eqref{eq:conservative-form} is now a genuine flux divergence, and we
therefore approximate it by \eqref{eq:fv-div} applied to
$(g^x,g^y)=(u_\theta\omega_\theta,v_\theta\omega_\theta)$, writing
$\nabla_h\cdot(u_\theta\omega_\theta,v_\theta\omega_\theta)$ for the result. A
single finite-volume divergence operator therefore serves the three physical
fluxes, namely the viscous flux $\nabla\omega_\theta$, the Poisson flux
$\nabla\psi_\theta$, and the convective flux
$(u_\theta\omega_\theta,v_\theta\omega_\theta)$.

With $f$ as in \eqref{eq:f-exp2}, the finite-volume residuals, restricted to
interior rows $j=1,\dots,N_y-2$, are
\begin{equation}
R_\omega^h
=
\partial_t\omega_\theta
-\varepsilon\,\Delta_h\omega_\theta
+
\nabla_h\cdot(u_\theta\omega_\theta,v_\theta\omega_\theta)
-f,
\qquad
R_\psi^h=-\Delta_h\psi_\theta-\omega_\theta.
\label{eq:fv-residuals-nse}
\end{equation}
The periodic loss $L_{\mathrm{per}}$ is again retained unchanged, and the
total loss is
\[
L(\theta)=L_\omega^h+L_\psi^h+10L_{\mathrm{per}}+10L_{\mathrm{IC}}.
\]
All of the settings of the network, of the grid and of the optimizer are
unchanged.

Table~\ref{tab:fv-nse-rel} compares the finite-volume residual against the
automatic-differentiation PINN of \S\ref{subsec:smooth-nse}, both at
$\varepsilon=10^{-4}$.

\begin{table}[h!]
\centering
\begin{tabular}{lccc}
\toprule
Quantity & PINN ($E$) & FV residual ($E^h$) & ratio $E^h/E$ \\
\midrule
$\omega$ & $5.559928\times10^{-3}$ & $4.756854\times10^{-3}$ & $0.86$ \\
$\psi$   & $5.124730\times10^{-3}$ & $5.690362\times10^{-3}$ & $1.11$ \\
$u$      & $6.384537\times10^{-3}$ & $6.681201\times10^{-3}$ & $1.05$ \\
$v$      & $1.293362\times10^{-2}$ & $1.203249\times10^{-2}$ & $0.93$ \\
\bottomrule
\end{tabular}
\caption{Global space--time relative $L^2$ errors, PINN vs.\ finite-volume
residual, smooth Navier--Stokes problem, $\varepsilon=10^{-4}$.}
\label{tab:fv-nse-rel}
\end{table}

Unlike the linear Stokes system above, the finite-volume residual here matches
the PINN of \S\ref{subsec:smooth-nse} to within $\pm15\%$ on every field,
sometimes slightly better and sometimes slightly
worse, with no systematic degradation. The cost of the training again drops
substantially, from $1146\,\mathrm s$ to
$243\,\mathrm s$, which is a speedup by the factor $4.7\times$. Taken together
with the Stokes case, this suggests that whether the finite-volume residual matches,
improves upon, or falls behind the automatic-differentiation residual is
problem-dependent. In the Stokes case the truncation bias of the Poisson
equation by itself
degraded $\psi,u,v$. Here the nonlinear convection term is present as well,
and it is subject to the same $\mathcal{O}(\Delta x^2+\Delta y^2)$ truncation
and is likewise undamped by $\varepsilon$. The Poisson bias is then
comparatively less consequential. In both cases the finite-volume residual
reduces the cost of the training by a factor of several, since the first-order
automatic derivatives alone are required.

\subsection{Error measures: modal decomposition and evaluation grids}
\label{subsec:diagnostics}

In \S\ref{sec:failure}--\S\ref{sec:exp10} we use two diagnostics in addition
to the global relative $L^2$ error, since a global relative $L^2$ error on a
uniform grid is a poor measure for a boundary layer.

\emph{The modal decomposition.}  A relative $L^2$ error divides by the norm of
the exact solution.  Suppose that the layer occupies a set of measure
$\mathcal{O}(\sqrt{\varepsilon})$ and that the rest of the solution is smooth.
An error of size $M$ committed entirely inside the layer then contributes only
$\mathcal{O}(M\varepsilon^{1/4})$ to the global norm. A much smaller error
spread over the interior can contribute more.  The global number
therefore mixes the two questions, how well the layer is resolved and how well
everything else is resolved, and answers neither question.  Since the
singular-layer enrichment of \S\ref{sec:slpinn} acts on the layer alone, the
quantity which has to improve is the first, and a diagnostic that does not
measure it separately is of little use for measuring that enrichment.  The
manufactured solutions of the present article admit an exact separation,
because the $x$-modes $e^{\mathrm{i}kx}$ are orthogonal on $[0,2\pi]$ and the
layer is built into one or two of them.  Writing $P_k$ for the projection onto
the $x$-wavenumber $k$, we define
\begin{equation}
E^{(k)}_\phi
=
\frac{\bigl\|P_k(\phi_\theta-\phi)\bigr\|_{L^2(\Omega\times[0,T])}}
     {\bigl\|P_k\,\phi\bigr\|_{L^2(\Omega\times[0,T])}}
\label{eq:modal-error}
\end{equation}
for $\phi\in\{\omega,\psi,u,v\}$, and we write $E_\phi$ for the corresponding
error of the whole field.  The projection is computed by a discrete Fourier
transform in $x$ on a grid of
its own. This grid has the wall-normal nodes of the grid~B described next,
$n_x=64$ equally spaced points over one period and $n_t=21$ time levels. The
periodic endpoint is excluded, so that the transform sees the period exactly
once.  No retraining is
involved at any point, since \eqref{eq:modal-error} is evaluated from the saved
network which produced the global errors.

\emph{The two evaluation grids.}  Errors are reported on a tensor-product
space--time grid, and we use two of them throughout.  The grid~A is the fixed
uniform mesh with $\Delta y=0.0125$, such as a study
of the problem would ordinarily employ. The grid~B is graded toward both
walls, with the matching nonuniform quadrature weights, so that it resolves a
layer of width $\sqrt{\varepsilon}$ at every viscosity of the sweep. Both
grids have $81$ uniform nodes in $x$, the periodic endpoint included, and
$21$ levels in $t$. In $y$, the grid~A has $81$ uniform nodes. The grid~B
consists of the wall
$y=0$, $720$ nodes in geometric progression from $L/2000$ to
$L=\min(1/2,\,10^{3}\sqrt{\varepsilon})$, a uniform core of $960$ nodes between
$L$ and $1-L$, the mirror image of the geometric nodes at the upper wall, and
the wall $y=1$. The weights of the grid~B are the trapezoidal weights of the
resulting nonuniform grid.
The graded region extends far beyond the layer, since the error of a
network which misses the layer is not confined to the layer. We checked the
resolution of the grid~B on every trained network of
\S\ref{sec:failure}--\S\ref{sec:exp10} against three coarser graded grids, of
$120$, $240$ and $480$ geometric nodes with $L=12\sqrt{\varepsilon}$,
$48\sqrt{\varepsilon}$ and $192\sqrt{\varepsilon}$. The values of $E_\omega$
on the grid~B and on the grid with $480$ geometric nodes agree to within
$0.7\%$ in every case. The coarsest grid, which is graded over the layer only, overstates $E_\omega$
by up to $15\%$ for the vanilla PINN at $\varepsilon\le10^{-8}$ and by a factor
of two for the FVNN of \S\ref{sec:exp10} at $\varepsilon=10^{-10}$. For the
scale-only enrichments of \S\ref{sec:slpinn} at $\varepsilon=10^{-10}$ it
overstates $E_\omega$ by a factor of up to $4.8$. The error of
these approximations extends beyond $12\sqrt{\varepsilon}$, into the region
where that grid is coarse. The enriched approximations of \S\ref{sec:slpinn}
and \S\ref{sec:exp10} are insensitive to the choice among the four grids, to
$0.3\%$.
All of the errors reported in this article are computed on the grid~B
unless
the grid~A is named explicitly, and we also report the ratio of the two.
The smooth solutions of \S\ref{sec:framework} contain no layer, and for
them we retain the values computed on the coarsest of the four grids, since
the four grids agree to within $0.05\%$ on such a solution. A
ratio close to one
indicates that the uniform grid resolves the error correctly, and a large
ratio indicates that the approximation contains structure that the uniform
grid does not resolve.

\section{Failure of the vanilla PINN for a vorticity boundary layer}
\label{sec:failure}

In \S\ref{sec:framework} we verified, on exact solutions which are fixed
Fourier modes and whose length scale does not shrink with $\varepsilon$, that
a small viscosity by itself does not degrade the vanilla PINN. In the present
section we keep the equations, the domain, the boundary conditions and the
settings (i)--(vii) of \S\ref{subsubsec:settings}, and we change the exact
solution only, together with the forcing and the initial data which it
determines. The exact solution now carries a vorticity layer of width
$\sqrt{\varepsilon}$ at the wall $y=0$, and the deterioration that we observe
in this section is therefore attributable to the thin layer and not to the
small parameter.

As in \S\ref{subsec:smooth-stokes} we work on
\[
\Omega=[0,2\pi]\times[0,1],
\qquad
t\in[0,1],
\]
with $x$-periodicity and the wall conditions \eqref{eq:wall-bc-exp1}, and we
sweep
\[
\varepsilon=1,\qquad 10^{-2},\qquad 10^{-4},\qquad 10^{-8}.
\]

\subsection{Location and structure of the vorticity layer}

Throughout the article the wall condition is the slip condition
$\mathbf{u}\cdot\mathbf{n}=0$ together with $\omega=0$ on $\Gamma=\partial\Omega$.
As $\varepsilon\to0$ the viscous solution approaches an Euler solution whose
vorticity does not vanish on the wall in general, and the mismatch is
corrected inside a layer of width $\sqrt{\varepsilon}$.  The singular
structure therefore
lives at the level of the \emph{vorticity},
\begin{equation}
\omega^{\varepsilon}
=
\underbrace{\omega^{0}}_{\text{Euler limit}}
+
\underbrace{\Theta\!\left(x,\frac{y}{\sqrt{\varepsilon}},t\right)}_{\text{vorticity layer}},
\label{eq:bl-structure-exp5}
\end{equation}
while the stream function, recovered from the smoothing Poisson solve
$-\Delta\psi=\omega$, inherits a layer that is weaker by a factor
$\mathcal{O}(\varepsilon)$.  The manufactured solution below reproduces exactly
this structure, and the numerical results of \S\ref{subsec:exp5-results}
confirm it, the loss
of accuracy being concentrated almost entirely in $\omega$.

\subsection{Manufactured Stokes solution with a $\sqrt{\varepsilon}$ vorticity layer}

Set
\begin{equation}
\omega(x,y,t)=e^{-t}\sin(x)\,g(y;\varepsilon),
\label{eq:omega-exp5}
\end{equation}
with
\begin{equation}
g(y;\varepsilon)
=
\underbrace{\bigl[\cos(\pi y)-1+2y\bigr]}_{\textstyle g_{\mathrm{smooth}}}
+
\underbrace{\Bigl[e^{-y/\sqrt{\varepsilon}}-(1-y)-y\,e^{-1/\sqrt{\varepsilon}}\Bigr]}_{\textstyle g_{\mathrm{layer}}}.
\label{eq:g-exp5}
\end{equation}
Both of the brackets are produced by the same device, namely that, given any
profile $p$ on $[0,1]$, the map
\begin{equation}
p(y)\;\longmapsto\;p(y)-(1-y)\,p(0)-y\,p(1)
\label{eq:lift-exp5}
\end{equation}
subtracts the linear interpolant between the two walls, so the result vanishes
at $y=0$ and $y=1$ identically, for every $\varepsilon$.  Applying
\eqref{eq:lift-exp5} to $p(y)=\cos(\pi y)$ gives $g_{\mathrm{smooth}}$, and
applying it to $p(y)=e^{-y/\sqrt{\varepsilon}}$ gives $g_{\mathrm{layer}}$.
Consequently the boundary condition $\omega=0$ on $\Gamma$ holds to machine
precision uniformly in $\varepsilon$, with no leftover of size
$\mathcal{O}(e^{-1/\sqrt{\varepsilon}})$ to be absorbed elsewhere.

The exponential $e^{-y/\sqrt{\varepsilon}}$ is the layer, since it falls from
$1$ to $e^{-1}$ over the distance $\sqrt{\varepsilon}$, and it is the sharp
structure whose resolution we are testing. We retain the smooth part
$g_{\mathrm{smooth}}$ so that the solution is not purely a boundary-layer
profile.

We now state where the layer is located, and in what sense it has the
structure \eqref{eq:bl-structure-exp5}.  Discarding from \eqref{eq:g-exp5} the two
terms that are exponentially small away from $y=0$ leaves the outer profile
\begin{equation}
g_{\mathrm{outer}}(y)=\cos(\pi y)-2+3y,
\qquad
g_{\mathrm{outer}}(0)=-1,
\qquad
g_{\mathrm{outer}}(1)=0,
\label{eq:g-outer-exp5}
\end{equation}
so that \eqref{eq:g-exp5} is exactly an outer solution plus a corrector of
amplitude $-g_{\mathrm{outer}}(0)$ carried by $e^{-y/\sqrt{\varepsilon}}$,
which is the structure that \eqref{eq:bl-structure-exp5} describes. At $y=1$,
however, the outer profile already satisfies the slip condition, so that no
correction is needed there and no second layer forms, and the experiment
therefore contains one layer only, at $y=0$. None of the results depends on
this asymmetry. The evaluation grid and the wall diagnostics
are applied at both walls, and the second wall reports no layer. The
asymmetry is a property of the solution and not of the plotting.

For the stream function, writing
\begin{equation}
\psi(x,y,t)=e^{-t}\sin(x)\,h(y;\varepsilon),
\label{eq:psi-exp5}
\end{equation}
the Poisson equation $-\Delta\psi=\omega$ reduces to the two-point boundary
value problem
\begin{equation}
-h''+h=g,
\qquad
h(0)=h(1)=0,
\label{eq:h-ode-exp5}
\end{equation}
which we solve in closed form, as a particular solution corrected by a
homogeneous combination of $\sinh$. The pair $(\omega,\psi)$ is therefore
exact, and not itself a numerical approximation. The forcing
\[
f=\partial_t\omega-\varepsilon\Delta\omega
\]
is then generated by the automatic differentiation applied to
\eqref{eq:omega-exp5}, rather than by differentiation by hand, which removes a
class of silent algebraic errors from the construction of the manufactured
solution.

\begin{remark}\label{rem:double-precision}
All of the computations in this article are carried out in double precision.
The Laplacian $\Delta e^{-y/\sqrt{\varepsilon}}$ carries the factor
$1/\varepsilon$, so that the automatic differentiation assembles the term
$\varepsilon\Delta\omega$ of the residual
$\partial_t\omega-\varepsilon\Delta\omega-f$ from quantities of the size
$\varepsilon^{-1}$, which is $10^{8}$ at $\varepsilon=10^{-8}$. In single
precision the residual of the exact solution cannot be evaluated to the
accuracy required to verify the manufactured solution, and double precision is
therefore necessary.
\end{remark}

\subsection{Training setup and evaluation grid}
\label{subsec:exp5-setup}

The hard wall ansatz \eqref{eq:hard-bc-exp1} and the settings (i)--(vii) of
\S\ref{subsubsec:settings} are kept unchanged, so that the exact solution, and
with it the forcing and the initial data, is the only difference between this
experiment and the smooth one.

Two consequences of this choice should be noted. First, the hard wall ansatz
removes the \emph{boundary} difficulty but not the
\emph{layer} difficulty. The network satisfies $\omega_\theta=0$ on $\Gamma$
exactly. The network has still to represent a function that varies over the scale
$\sqrt{\varepsilon}$ inside the layer. Second, the interior collocation points
are drawn uniformly on
$\Omega\times[0,1]$. The expected fraction of them lying within one layer
width of a wall is $\mathcal{O}(\sqrt{\varepsilon})$. This amounts to roughly
$10{,}000$, $1{,}000$, $100$ and $1$ points, respectively, for
$\varepsilon=1,10^{-2},10^{-4},10^{-8}$. The layer therefore receives very few
collocation points when $\varepsilon$ is
small, and this is the mechanism by which the vanilla PINN is expected to
fail. The singular layer trial functions of \S\ref{sec:slpinn} are designed to
remove this mechanism.

We report the errors on the wall-graded grid~B of \S\ref{subsec:diagnostics},
with $n_y=1441$ for $\varepsilon\ge10^{-4}$ and $n_y=2400$ for
$\varepsilon=10^{-8}$. At $\varepsilon=10^{-8}$ the first node lies at the
distance $L/2000=5.0\times10^{-5}$ from the wall, and the smallest spacing
between consecutive nodes, which is between the first two geometric nodes, is
$5.3\times10^{-7}$. The grid has to resolve the layer
if the comparison is to be meaningful across
the rows. This requirement is already visible here. The uniform grid~A reproduces the
graded value of $E_\omega$ to three digits while $\sqrt{\varepsilon}\ge\Delta
y$, and it agrees to within $6\%$ at $\varepsilon=10^{-4}$. At
$\varepsilon=10^{-8}$ it understates that value by a factor of $2.3$, the
layer being no longer resolved by the mesh.

\subsection{Numerical results for the Stokes problem}
\label{subsec:exp5-results}

The global space--time relative $L^2$ errors are collected in
Table~\ref{tab:bl-rel}.

\begin{table}[h!]
\centering
\begin{tabular}{ccccc}
\toprule
$\varepsilon$ & $E_\omega$ & $E_\psi$ & $E_u$ & $E_v$ \\
\midrule
\multicolumn{5}{l}{\emph{Stokes}}\\
    $1$ & $1.235218\times10^{-3}$ & $2.313406\times10^{-2}$ & $2.121734\times10^{-2}$ & $3.324478\times10^{-2}$ \\
    $10^{-2}$ & $5.232934\times10^{-3}$ & $5.345931\times10^{-3}$ & $6.415900\times10^{-3}$ & $1.407269\times10^{-2}$ \\
    $10^{-4}$ & $1.918475\times10^{-2}$ & $1.500317\times10^{-2}$ & $1.886733\times10^{-2}$ & $3.321186\times10^{-2}$ \\
    $10^{-8}$ & $1.153404\times10^{-1}$ & $4.442234\times10^{-2}$ & $4.896123\times10^{-2}$ & $7.783941\times10^{-2}$ \\
\addlinespace
\multicolumn{5}{l}{\emph{Navier--Stokes}}\\
    $1$ & $4.743132\times10^{-4}$ & $4.891837\times10^{-3}$ & $6.084222\times10^{-3}$ & $5.611120\times10^{-3}$ \\
    $10^{-2}$ & $3.648580\times10^{-3}$ & $3.606493\times10^{-3}$ & $5.115432\times10^{-3}$ & $6.474931\times10^{-3}$ \\
    $10^{-4}$ & $1.888660\times10^{-2}$ & $8.060322\times10^{-3}$ & $1.240837\times10^{-2}$ & $1.272138\times10^{-2}$ \\
    $10^{-8}$ & $4.132712\times10^{-2}$ & $1.178201\times10^{-2}$ & $1.724690\times10^{-2}$ & $1.463375\times10^{-2}$ \\
\bottomrule
\end{tabular}
\caption{Global space--time relative $L^2$ errors for the two problems with a
$\sqrt{\varepsilon}$ vorticity boundary layer, vanilla PINN, on the
layer-resolved grid~B.}
\label{tab:bl-rel}
\end{table}

The maximum absolute errors are given in Table~\ref{tab:bl-max}.

\begin{table}[h!]
\centering
\begin{tabular}{ccccc}
\toprule
$\varepsilon$ & $\omega$ & $\psi$ & $u$ & $v$ \\
\midrule
\multicolumn{5}{l}{\emph{Stokes}}\\
    $1$ & $7.680325\times10^{-4}$ & $3.680082\times10^{-4}$ & $1.672139\times10^{-3}$ & $5.248376\times10^{-4}$ \\
    $10^{-2}$ & $3.567381\times10^{-3}$ & $6.049811\times10^{-4}$ & $3.676477\times10^{-3}$ & $1.438712\times10^{-3}$ \\
    $10^{-4}$ & $9.454122\times10^{-2}$ & $2.153838\times10^{-3}$ & $1.382157\times10^{-2}$ & $7.922700\times10^{-3}$ \\
    $10^{-8}$ & $9.443155\times10^{-1}$ & $5.206569\times10^{-3}$ & $3.711390\times10^{-2}$ & $1.139275\times10^{-2}$ \\
\addlinespace
\multicolumn{5}{l}{\emph{Navier--Stokes}}\\
    $1$ & $5.959758\times10^{-3}$ & $2.503604\times10^{-3}$ & $1.352765\times10^{-2}$ & $8.663708\times10^{-3}$ \\
    $10^{-2}$ & $1.736307\times10^{-2}$ & $2.074353\times10^{-3}$ & $1.521717\times10^{-2}$ & $9.708051\times10^{-3}$ \\
    $10^{-4}$ & $1.929486\times10^{-1}$ & $4.143435\times10^{-3}$ & $6.938878\times10^{-2}$ & $1.573175\times10^{-2}$ \\
    $10^{-8}$ & $9.596944\times10^{-1}$ & $6.383427\times10^{-3}$ & $1.540494\times10^{-1}$ & $2.241684\times10^{-2}$ \\
\bottomrule
\end{tabular}
\caption{Maximum absolute errors for the two problems with a
$\sqrt{\varepsilon}$ vorticity boundary layer, vanilla PINN, on the
layer-resolved grid~B.}
\label{tab:bl-max}
\end{table}

We observe three features in these results.

\emph{The error in the vorticity grows rapidly as $\varepsilon$ decreases.}
Across the eight
decades from $\varepsilon=1$ to $\varepsilon=10^{-8}$, $E_\omega$ grows by a
factor of $93$ and the maximum vorticity error by a factor of $1230$, from
$7.7\times10^{-4}$ to $9.4\times10^{-1}$.  At $\varepsilon=10^{-8}$ the
pointwise error is of the same order as the
solution itself, and the vanilla PINN has therefore not approximated the layer
at all. In the smooth sweep of Table~\ref{tab:stokes-eps-rel}, $E_\omega$ varied by
less than a factor of four between $\varepsilon=1$ and $\varepsilon=10^{-4}$. The network, the sampling, the
schedule of the optimizer and the loss coincide in the two computations, and
the exact solutions differ in the layer, together with the forcing and the
initial data which the layer entails. The growth of the error is therefore
attributable to the layer.

\emph{The error in the stream function and in the velocity stays small.} Over
the same eight decades $E_\psi$, $E_u$ and $E_v$ change by less than a
factor of $2.5$. They do not even vary monotonically, since $E_\psi$ is
smaller at $\varepsilon=10^{-2}$ than at $\varepsilon=1$. The reason is the structure
\eqref{eq:bl-structure-exp5}. The stream function is obtained from the
vorticity by solving the Poisson equation, and the Poisson equation smooths
out the layer, so that the layer in $\psi$ is weaker than the layer in
$\omega$ by the factor $\mathcal{O}(\varepsilon)$. An error of the vorticity
inside a set of width $\sqrt{\varepsilon}$ is therefore largely averaged away
before it reaches $\psi$ and $\mathbf{u}$. A diagnostic which looks at the
velocity only would miss the failure almost entirely.

\emph{The failure is present from the initial time.} At $\varepsilon=10^{-8}$
the spatial relative error of the vorticity is already $1.0\times10^{-1}$ at
$t=0$, where the vorticity is pinned by the
initial-condition loss, and grows only to $1.6\times10^{-1}$ by $t=1$.  The
network is not losing accuracy through the integration in time, but is
unable to represent the profile of the layer at any time, including the one
time level at which it is supervised directly.

\begin{figure}[h!]
\centering
\includegraphics[width=0.75\textwidth]{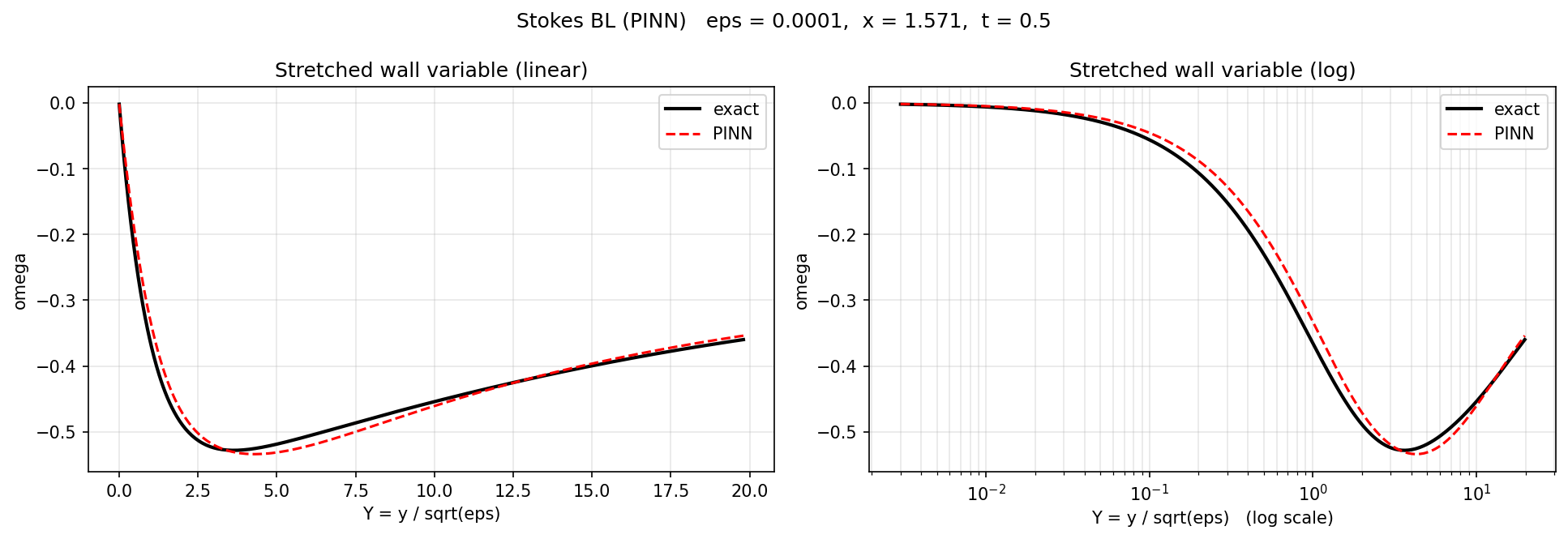}\\[1ex]
\includegraphics[width=0.75\textwidth]{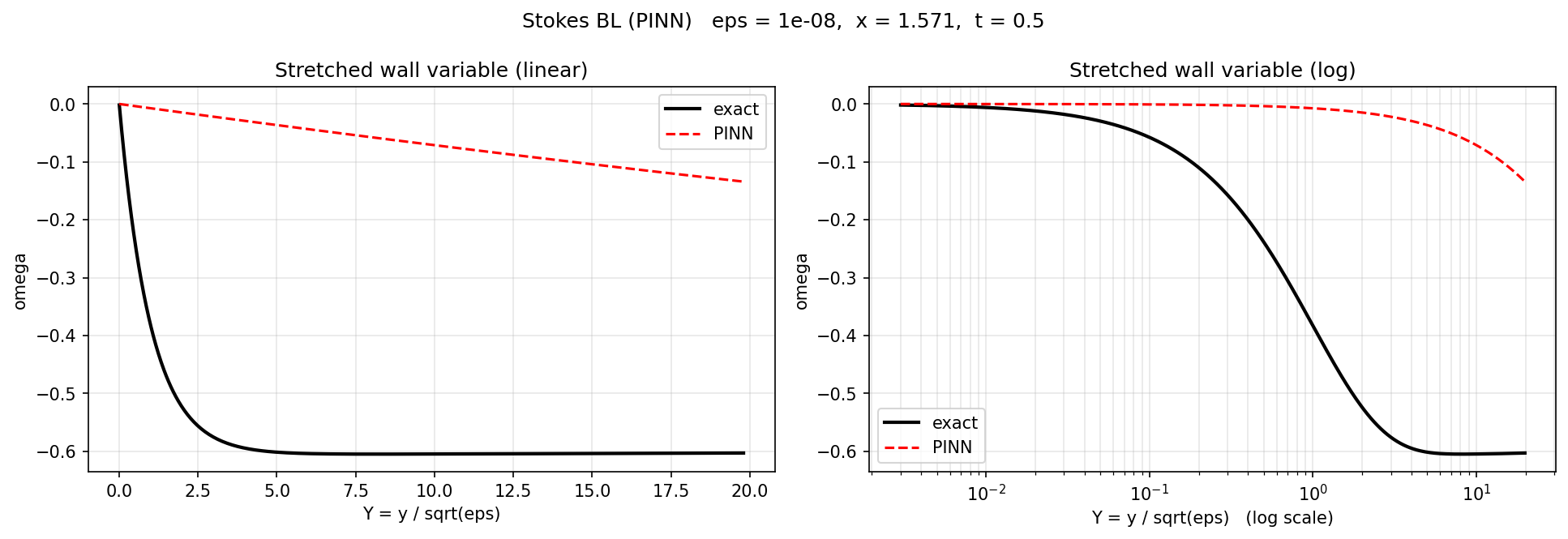}
\caption{Wall-normal vorticity profile at $x=\pi/2$, $t=0.5$, plotted against
the stretched variable $Y=y/\sqrt{\varepsilon}$ on linear (left) and
logarithmic (right) axes.  Top: $\varepsilon=10^{-4}$.  Bottom:
$\varepsilon=10^{-8}$, where the vanilla PINN misses the layer profile
entirely.}
\label{fig:bl-stokes-profile}
\end{figure}

Figure~\ref{fig:bl-stokes-profile} shows the mechanism.  Plotted against the
stretched wall variable $Y=y/\sqrt{\varepsilon}$, the exact vorticity profile
is the same curve at every $\varepsilon$ by construction.  At
$\varepsilon=10^{-4}$ the network still tracks it, with a visible but modest
phase and amplitude error near the minimum.  At $\varepsilon=10^{-8}$ the
prediction has collapsed onto an almost linear
ramp, which satisfies $\omega_\theta=0$ at the wall, as the hard ansatz
guarantees, and then rises through the entire layer without reproducing any
part of the exponential descent. The error is therefore not a small
perturbation of the exact profile, but a
qualitatively different shape.

\begin{figure}[h!]
\centering
\includegraphics[width=0.75\textwidth]{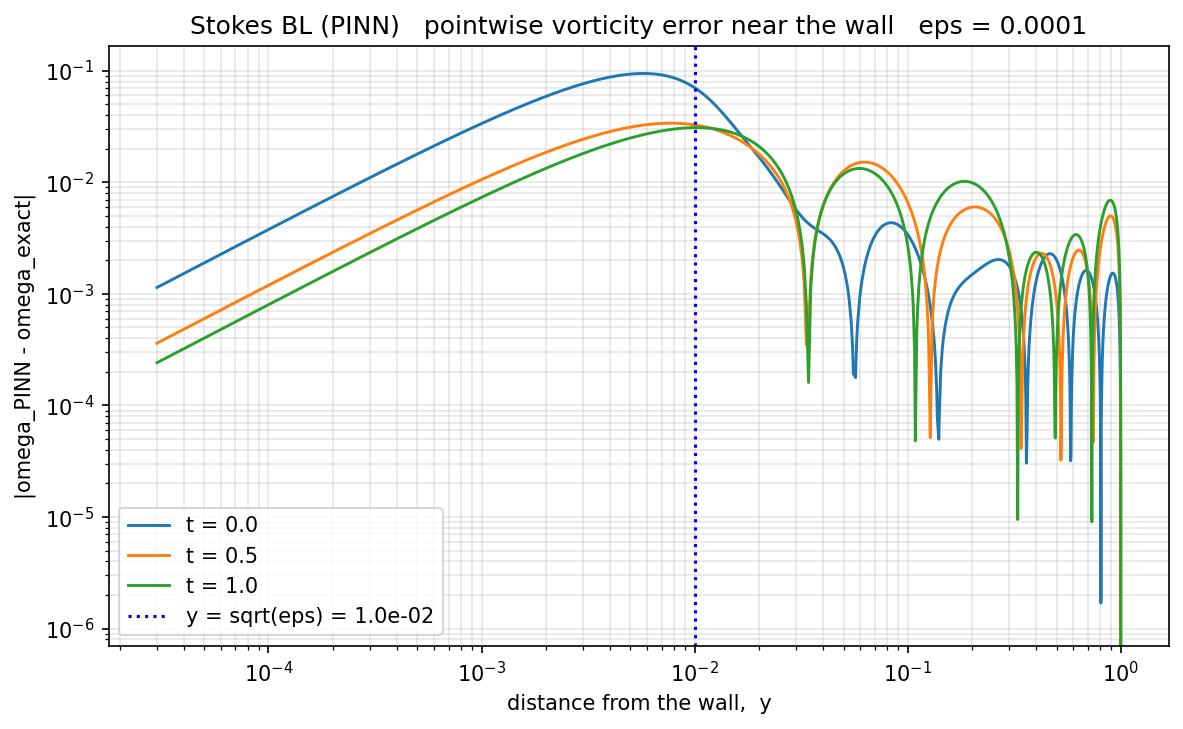}\\[1ex]
\includegraphics[width=0.75\textwidth]{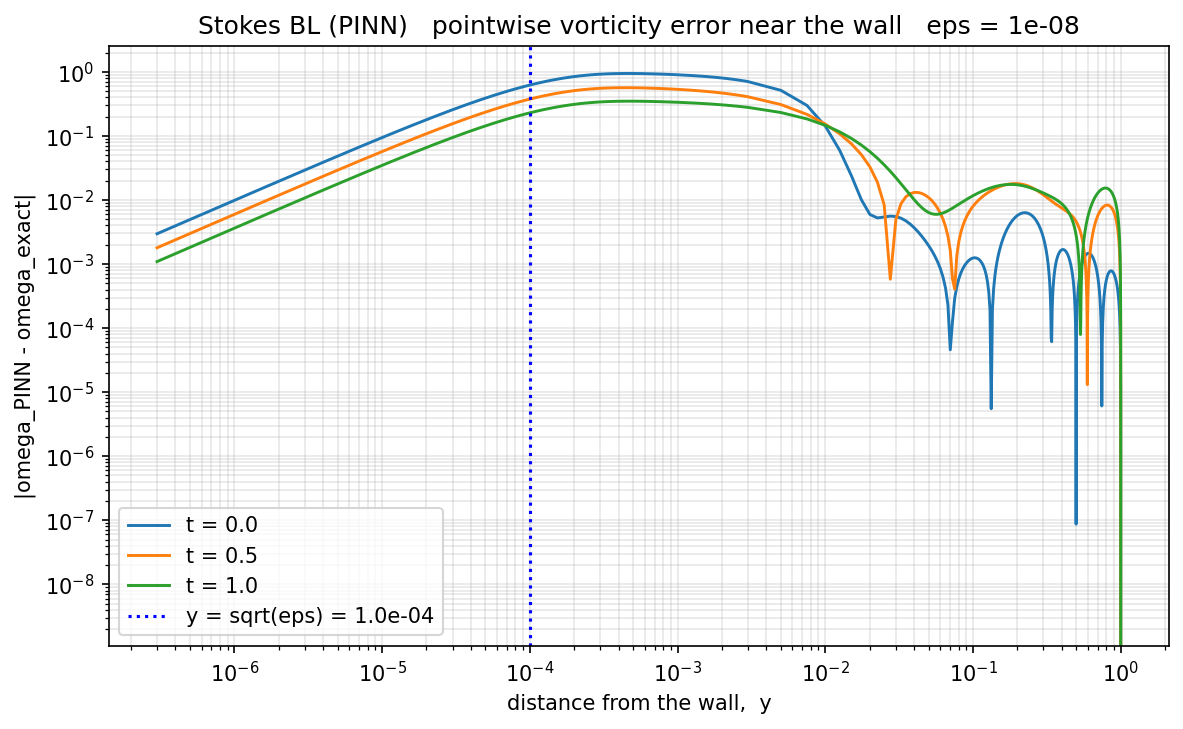}
\caption{Pointwise vorticity error $|\omega_\theta-\omega|$ as a function of
distance from the wall at $x=\pi/2$, for $t=0,0.5,1$.  Top:
$\varepsilon=10^{-4}$.  Bottom: $\varepsilon=10^{-8}$.  The dotted vertical
line marks $y=\sqrt{\varepsilon}$.  In both cases the error is concentrated
within a distance of order $\sqrt{\varepsilon}$ of the wall.}
\label{fig:bl-stokes-error}
\end{figure}

Figure~\ref{fig:bl-stokes-error} localizes the error. The pointwise vorticity
error rises monotonically from the wall and peaks at a
distance of the order $\sqrt{\varepsilon}$, which is marked by the dotted
line. The error then falls by one to two orders of magnitude in the interior, where
it oscillates at the level of the smooth experiments. The error is therefore
held inside a set of measure
$\mathcal{O}(\sqrt{\varepsilon})$, rather than distributed over the domain.
For the same reason, the global relative error in Table~\ref{tab:bl-rel} grows
more slowly than the maximum
error in Table~\ref{tab:bl-max}: the $L^2$ norm dilutes the contribution of
the layer by a factor $\varepsilon^{1/4}$.

\subsubsection{Modal errors}
\label{subsec:exp5-modal}

The two tables above report the error of the whole field, which is a poor
measure for a boundary-layer problem, and we therefore decompose the error by
the $x$-wavenumber, in the sense of \eqref{eq:modal-error} of
\S\ref{subsec:diagnostics}. In the present experiment the exact solution
consists of the single mode $\sin(x)$, so $P_1\omega=\omega$ and
$\|P_1\omega\|=\|\omega\|$. The \emph{error}, however, is under no such
constraint, since the
network is free to place error in the modes which the solution does not
occupy. Consequently $E^{(1)}_\omega\le E_\omega$ always, and the ratio
measures how much of the error sits in the mode which carries the layer.
Writing
\[
f_1=\frac{\|P_1(\omega_\theta-\omega)\|^2}{\|\omega_\theta-\omega\|^2}
=\left(\frac{E^{(1)}_\omega}{E_\omega}\right)^{\!2},
\]
Table~\ref{tab:bl-stokes-modal} reports both.

\begin{table}[h!]
\centering
\begin{tabular}{ccccc}
\toprule
$\varepsilon$ & $E_\omega$ (global) & $E^{(1)}_\omega$ & $E^{(1)}_\omega/E_\omega$ & $f_1$ \\
\midrule
    $1$ & $1.235218\times10^{-3}$ & $6.174407\times10^{-4}$ & $0.500$ & $25.0\%$ \\
    $10^{-2}$ & $5.232934\times10^{-3}$ & $2.027953\times10^{-3}$ & $0.388$ & $15.0\%$ \\
    $10^{-4}$ & $1.918475\times10^{-2}$ & $1.417098\times10^{-2}$ & $0.739$ & $54.6\%$ \\
    $10^{-8}$ & $1.153404\times10^{-1}$ & $1.124565\times10^{-1}$ & $0.975$ & $95.1\%$ \\
\bottomrule
\end{tabular}
\caption{Vorticity error split by $x$-wavenumber for the Stokes problem with a
$\sqrt{\varepsilon}$ vorticity layer.  $f_1$ is the fraction of the total error
energy carried by the mode that holds the layer.}
\label{tab:bl-stokes-modal}
\end{table}

Two things follow.

First, \emph{the error concentrates in the layer mode as $\varepsilon$
decreases}.
At $\varepsilon=1$ only a quarter of the error energy lies in $k=1$; the rest is
spurious content in modes the exact solution does not contain, which is the
ordinary approximation error of an under-trained network.  By
$\varepsilon=10^{-8}$ that fraction is $95\%$. The failure at a small
$\varepsilon$ is therefore not only larger but also of
a different type. Almost all of it is now the inability of the network to
represent the single mode that varies on the scale $\sqrt{\varepsilon}$.

Second, \emph{the global number understates the degradation}.  Across the sweep
$E^{(1)}_\omega$ grows by a factor of $182$, against $93$ for $E_\omega$.
The two disagree because $\|\omega\|$ itself depends on $\varepsilon$ in this
family. The norm $\|\omega\|$ rises from $0.184$ at $\varepsilon=1$ to $0.621$ at
$\varepsilon=10^{-8}$ as the layer sharpens. Part of the apparent stability of
the relative error is therefore a growing denominator, and not a stable
numerator.
The maximum absolute error of Table~\ref{tab:bl-max}, which involves no
denominator at all, grows by $1230$.

The other fields behave as \eqref{eq:bl-structure-exp5} predicts.  At
$\varepsilon=10^{-8}$,
\[
E^{(1)}_\psi=3.48\times10^{-2},
\qquad
E^{(1)}_u=3.63\times10^{-2},
\qquad
E^{(1)}_v=4.86\times10^{-2},
\]
all of which are smaller than $E^{(1)}_\omega=1.12\times10^{-1}$, by factors
between $2.3$ and $3.2$. The Poisson solve damps the error of the layer before
it reaches $\psi$ and $\mathbf{u}$, as the estimate of the order
$\mathcal{O}(\varepsilon)$ says that it should.

Taken together with the smooth baseline of \S\ref{subsec:smooth-stokes}, these
results isolate the difficulty.  A small viscosity with a smooth solution
costs nothing, whereas a small viscosity with a $\sqrt{\varepsilon}$ layer
costs two orders of magnitude in the vorticity and produces a wall profile
which is qualitatively wrong.  The remedy suggested by
\eqref{eq:bl-structure-exp5} is to stop asking the network to discover the
scale $\sqrt{\varepsilon}$ from uniformly sampled data, and to build the known
structure of the layer into the trial space instead,
\begin{equation}
\omega_\theta
=
\omega_{\theta,\mathrm{reg}}
+
\chi\!\left(\frac{y}{\sqrt{\varepsilon}}\right)\omega_{\theta,\mathrm{layer}},
\label{eq:slpinn-ansatz-exp5}
\end{equation}
leaving to the network a remainder which is regular uniformly in
$\varepsilon$.  Before the cut-off $\chi$ and the layer component
$\omega_{\theta,\mathrm{layer}}$ are constructed, we have to determine whether
the convection term introduces a second difficulty, independent of the layer,
which a corrector built from the linear analysis would not address.
Section~\ref{sec:exp6} repeats the experiment on the Navier--Stokes problem
with the same layer.

\subsection{Navier--Stokes problem with a $\sqrt{\varepsilon}$ vorticity layer}
\label{sec:exp6}

In \S\ref{subsec:exp5-results} the vanilla PINN failed to resolve the
$\sqrt{\varepsilon}$ vorticity layer for the Stokes problem. In the present
subsection we add the convection term and ask whether the nonlinearity adds a
difficulty of its own, or whether the layer remains the only difficulty. The
answer decides how the singular layer enrichment has to be designed. If
the convection is a second and independent difficulty, the corrector has to be
derived for the nonlinear problem, whereas if the layer dominates, a corrector
built from the linear analysis can be carried over.

We construct the experiment so that this question is the only one which it is
able to answer. The domain, the boundary conditions, the hard wall ansatz
\eqref{eq:hard-bc-exp1} and the settings (i)--(vii) of
\S\ref{subsubsec:settings} are taken unchanged from the Stokes problem of
\S\ref{subsec:exp5-setup}.
The first $x$-mode of the exact solution is \emph{identical} to the one
employed there.
We add the convection term
$\partial_y\psi\,\partial_x\omega-\partial_x\psi\,\partial_y\omega$ in the
residual, together with one smooth second mode in the exact solution, whose
purpose is explained in \S\ref{subsec:exp6-design}.

\subsubsection{Design of the manufactured solution}
\label{subsec:exp6-design}

Let $\lambda_2=4+\pi^2$ and $\alpha=1/4$, and set
\begin{align}
\psi(x,y,t)
&=e^{-t}\Bigl[\sin(x)\,h(y;\varepsilon)+\alpha\sin(2x)\sin(\pi y)\Bigr],
\label{eq:psi-exp6}\\[0.5ex]
\omega(x,y,t)
&=e^{-t}\Bigl[\sin(x)\,g(y;\varepsilon)+\alpha\lambda_2\sin(2x)\sin(\pi y)\Bigr],
\label{eq:omega-exp6}
\end{align}
with $g$ and $h$ exactly as in \eqref{eq:g-exp5} and \eqref{eq:h-ode-exp5}.
The Poisson coupling closes mode by mode, since the first mode satisfies
$-h''+h=g$ by construction. For the second we have $-\Delta[\sin(2x)\sin(\pi
y)]=\lambda_2\sin(2x)\sin(\pi y)$, so that scaling its vorticity by
$\lambda_2$ makes $-\Delta\psi=\omega$ an identity rather than an
approximation. Both of the modes vanish at $y=0$ and $y=1$, the
functions $g$ and $h$ by the linear-interpolant construction
\eqref{eq:lift-exp5} and $\sin(\pi y)$ outright, so that $\omega=\psi=0$ on
$\Gamma$ holds exactly for every $\varepsilon$. The forcing
\[
f=\partial_t\omega+\partial_y\psi\,\partial_x\omega-\partial_x\psi\,\partial_y\omega-\varepsilon\Delta\omega
\]
is generated by automatic differentiation of
\eqref{eq:psi-exp6}--\eqref{eq:omega-exp6},
and all computations are carried out in double precision, for the reason given
in Remark~\ref{rem:double-precision}.

Three of the choices made in \eqref{eq:psi-exp6}--\eqref{eq:omega-exp6}
require justification, since each choice is needed for a particular
interpretation of the results.

\emph{Why the first mode is reused verbatim.} The comparison of interest is
one between two runs which differ in one respect only. The functions $g$ and
$h$ are the same as in the Stokes problem. The layer has
therefore the same width, the same amplitude, the same wall mismatch, and the
same damping of the order $\mathcal{O}(\varepsilon)$ through the Poisson
solve. Any difference
between the errors of the two experiments is therefore attributable to the
convection term and to the presence of the second mode, and not to a layer of
a different shape.

\emph{Why a second mode is necessary.}  With a single mode
$\psi=e^{-t}\sin(x)h$, $\omega=e^{-t}\sin(x)g$, the Jacobian does not vanish:
\begin{equation}
\partial_y\psi\,\partial_x\omega-\partial_x\psi\,\partial_y\omega
=\tfrac12 e^{-2t}\sin(2x)\,\bigl(h'g-hg'\bigr),
\label{eq:one-mode-jacobian}
\end{equation}
and $h$ differs from every multiple of $g$, since $h$ solves $-h''+h=g$. The quantity
\eqref{eq:one-mode-jacobian} is small, however, for, measured on the exact
solution, it reaches only about $13\%$ of the viscous term inside the layer,
at every $\varepsilon$ that we test. Such a problem would be a Stokes boundary
layer carrying a perturbation, and a
null result would be inconclusive. The second mode raises the convective term
to the same order as the
viscous one (\S\ref{subsec:exp6-budget}) while remaining smooth, so that it
adds a nonlinear interaction without adding a second thin structure. We note
as well that $-\Delta[\sin(2x)\sin(\pi y)]$ is proportional to
$\sin(2x)\sin(\pi y)$. The second mode therefore contributes nothing to
\eqref{eq:one-mode-jacobian} by itself. All of the added nonlinearity comes
from the cross terms between the two modes, that is, from the smooth mode
transporting the layer.

\emph{Why the second mode makes the global vorticity error insensitive to the
layer.} Driving the convection requires a second mode with an $\mathcal{O}(1)$
\emph{velocity},
and hence with an $\mathcal{O}(1)$ stream function, but its vorticity is then
$\lambda_2=13.87$ times larger. This ratio follows from the geometry of the
modes rather than from a
particular choice of the constants, since $\lambda_2$ is the smallest
eigenvalue available to any mode whose $x$-wavenumber differs from the first.
The consequence is that the
mode which carries the layer holds only
\[
0.41\%,\quad 2.85\%,\quad 4.29\%,\quad 4.51\%
\]
of the energy of the vorticity at $\varepsilon=1,10^{-2},10^{-4},10^{-8}$, so
that a global relative $L^2$ error in $\omega$ is overwhelmingly a measurement
of the smooth mode. The modal diagnostic \eqref{eq:modal-error} of
\S\ref{subsec:exp5-modal} was
introduced precisely for this situation, and we report the modal errors in
\S\ref{subsec:exp6-modal} alongside the global tables.

\subsubsection{Size of each term inside the layer}
\label{subsec:exp6-budget}

We evaluate each term of $\partial_t\omega+J(\psi,\omega)-\varepsilon\Delta\omega=f$,
where
$J(\psi,\omega)=\partial_y\psi\,\partial_x\omega-\partial_x\psi\,\partial_y\omega$,
directly from the exact
solution, as a root-mean-square over the slab $0<y<3\sqrt{\varepsilon}$, by
Monte-Carlo sampling with $4000$ points, and the ratio of the convective to
the viscous term is

\[
\frac{\|J(\psi,\omega)\|}{\|\varepsilon\Delta\omega\|}
= 0.003,\qquad 0.45,\qquad 0.78,\qquad 0.96
\]

at $\varepsilon=1$, $10^{-2}$, $10^{-4}$ and $10^{-8}$ respectively.

The ratio approaches unity, so that on the $\sqrt{\varepsilon}$ scale the
convective and the viscous terms become comparable in magnitude, which is the
property that the second mode was introduced to produce. The scaling
behind it is elementary. With $\delta=\sqrt{\varepsilon}$ we have
$\partial_y\omega=\mathcal{O}(\delta^{-1})$ and
$\partial_y^2\omega=\mathcal{O}(\delta^{-2})$, so that
$\varepsilon\Delta\omega=\mathcal{O}(1)$, while near the wall
$\psi=\mathcal{O}(y)$ gives $\partial_x\psi=\mathcal{O}(\delta)$, so that
$\partial_x\psi\,\partial_y\omega=\mathcal{O}(1)$ as well, and
$\partial_y\psi=\mathcal{O}(1)$ and
$\partial_x\omega=\mathcal{O}(1)$ trivially. Neither of the terms degenerates
or blows
up as $\varepsilon\to0$.

Two qualifications should be attached to this measurement, both of which
matter for the way in which the result is described. First, a comparable
root-mean-square size is a weaker statement than a mutual cancellation in the
equation, since $\|\partial_t\omega\|$ and $\|f\|$ are themselves $\mathcal{O}(1)$ on
the same slab and are, at small $\varepsilon$, the larger terms. That is to be
expected of a
manufactured solution, in which $f$ absorbs the part of the remaining terms
that does not cancel, but it means that the statement concerns the relative
magnitude and not a
self-balancing layer. Second, the structure that we have here differs from a
classical Prandtl
layer. A Prandtl layer is the no-slip velocity layer governed by the Prandtl
equations \cite{GieHamoudaJungTemam2018}. Under the slip conditions used
throughout the present article, what
is present is a \emph{vorticity} layer on the scale $\sqrt{\varepsilon}$.
Inside it the convective and the viscous terms are of the same order.

\subsubsection{Global errors}
\label{subsec:exp6-results}

Tables~\ref{tab:bl-rel} and~\ref{tab:bl-max} report the global
space--time relative $L^2$ errors and the maximum absolute errors on the
layer-resolved grid of \S\ref{subsec:exp5-results}.

A direct comparison of the Stokes and the Navier--Stokes columns of
Table~\ref{tab:bl-rel} would suggest that the nonlinear problem is
\emph{easier} than the linear one at every viscosity, but that comparison is
misleading. The exact solution has here a much larger norm than the Stokes one. At
$\varepsilon=10^{-8}$ the norm $\|\omega\|$ is $4.71$ times larger, and
$\|\psi\|$, $\|u\|$ and $\|v\|$ are larger by the factors $4.38$, $4.32$ and
$8.58$. The two relative errors therefore have different denominators, and
they are not comparable quantities.
Multiplying each relative error by its own norm removes the ambiguity, and we
report the result in \S\ref{subsec:exp6-compare}. The maximum absolute errors
of Table~\ref{tab:bl-max}, which carry no denominator, already give the
correct comparison, being $9.6\times10^{-1}$ at $\varepsilon=10^{-8}$ against
$9.4\times10^{-1}$ for the Stokes problem, in a solution whose vorticity is
nearly five times larger.

Within the Navier--Stokes family alone the degradation is unambiguous. The
maximum vorticity error grows by a factor of $161$ across the sweep, and at
$\varepsilon=10^{-8}$ it is of the same order as the amplitude of the layer
itself. The selected-time errors show the same pattern as in the Stokes case. At
$\varepsilon=10^{-8}$ the relative vorticity error is already
$2.81\times10^{-2}$ at $t=0$, where the vorticity is pinned directly by the
initial-condition loss, and it reaches only $6.40\times10^{-2}$ at $t=1$. The
network is not losing accuracy by integrating in time, but does not
represent the layer at any time.

\subsubsection{Modal errors}
\label{subsec:exp6-modal}

Table~\ref{tab:bl-nse-modal} applies \eqref{eq:modal-error}.

\begin{table}[h!]
\centering
\footnotesize
\setlength{\tabcolsep}{3pt}
\begin{tabular}{ccccccc}
\toprule
$\varepsilon$ & $E^{(1)}_\omega$ & $E^{(2)}_\omega$ & $E^{(1)}_\psi$ & $E^{(1)}_u$ & $E^{(1)}_v$ & $k=1$ share \\
\midrule
    $1$ & $2.914202\times10^{-3}$ & $1.990922\times10^{-4}$ & $6.378920\times10^{-2}$ & $5.972659\times10^{-2}$ & $7.576769\times10^{-2}$ & $0.41\%$ \\
    $10^{-2}$ & $1.007221\times10^{-2}$ & $1.073898\times10^{-3}$ & $1.020755\times10^{-2}$ & $1.339151\times10^{-2}$ & $1.616367\times10^{-2}$ & $2.85\%$ \\
    $10^{-4}$ & $7.994863\times10^{-2}$ & $3.766452\times10^{-3}$ & $1.948841\times10^{-2}$ & $3.041523\times10^{-2}$ & $2.455049\times10^{-2}$ & $4.29\%$ \\
    $10^{-8}$ & $1.876691\times10^{-1}$ & $4.458653\times10^{-3}$ & $2.858840\times10^{-2}$ & $3.981480\times10^{-2}$ & $2.764125\times10^{-2}$ & $4.51\%$ \\
\bottomrule
\end{tabular}
\caption{Error by $x$-wavenumber for the Navier--Stokes problem with a
$\sqrt{\varepsilon}$ vorticity boundary layer.  Mode $k=1$ carries the layer;
mode $k=2$ is smooth.  The last column is the share of the exact vorticity energy
held by $k=1$.}
\label{tab:bl-nse-modal}
\end{table}

Table~\ref{tab:bl-nse-modal} is the central table of the experiment, and it
shows what the global tables cannot show. At $\varepsilon=10^{-8}$ the same
network is trained by the same loss on the
same collocation points. The network commits an error of $19\%$ on the mode
that carries the layer and an error of $0.45\%$ on the mode that does not. The
two differ
by the factor $42$. Across the entire sweep $E^{(2)}_\omega$ grows by a factor
of $22$ only, and never leaves the range of the smooth baselines of
\S\ref{sec:framework},
while $E^{(1)}_\omega$ grows by $64$
and ends a factor of $42$ larger. The two modes are approximated by one
network in one optimization, so that the
smooth mode is a control that requires no separate run. The degradation of the
first mode is therefore specific to the thin scale that
this mode contains. The cause is neither the small parameter nor the
nonlinearity, since both act
on the two modes alike.

The global relative error understates the failure in the layer by the factor
$E^{(1)}_\omega/E_\omega=4.54$ at $\varepsilon=10^{-8}$, as we anticipated in
\S\ref{subsec:exp6-design}. As in the Stokes case, the error concentrates in
the layer mode as the layer
becomes thinner, the fraction $f_1$ rising from $15.6\%$ at $\varepsilon=1$ to
$93.0\%$ at $\varepsilon=10^{-8}$,
against $25.0\%$ and $95.1\%$ for the Stokes problem.

The velocity components require a separate comment. The shares of the velocity
components in $k=1$ at $\varepsilon=10^{-8}$ are $5.35\%$ for $u$ and only
$1.36\%$ for $v$. The share for $v$ is smaller because $v=-\partial_x\psi$ multiplies
each mode by its own $x$-wavenumber, and so amplifies the smooth mode by a
further factor of two. The global errors of the velocity are therefore even
less sensitive to the
layer than the global errors of the vorticity, and they should not be used as
evidence for or against the resolution of the layer.

\subsection{Comparison of the Stokes and Navier--Stokes results}
\label{subsec:exp6-compare}

Table~\ref{tab:bl-compare} places the two boundary-layer experiments side by
side, in the two quantities which are legitimately comparable, namely the
absolute space--time $L^2$ error of the vorticity and the modal relative error
of the layer mode.

\begin{table}[h!]
\centering
\begin{tabular}{ccccccc}
\toprule
& \multicolumn{3}{c}{$\|\omega_\theta-\omega\|_{L^2}$} & \multicolumn{3}{c}{$E^{(1)}_\omega$} \\
\cmidrule(lr){2-4}\cmidrule(lr){5-7}
$\varepsilon$ & Stokes & NSE & ratio & Stokes & NSE & ratio \\
\midrule
    $1$ & $2.272554\times10^{-4}$ & $1.358115\times10^{-3}$ & $5.98$ & $6.174407\times10^{-4}$ & $2.914202\times10^{-3}$ & $4.72$ \\
    $10^{-2}$ & $2.559907\times10^{-3}$ & $1.057717\times10^{-2}$ & $4.13$ & $2.027953\times10^{-3}$ & $1.007221\times10^{-2}$ & $4.97$ \\
    $10^{-4}$ & $1.161185\times10^{-2}$ & $5.516422\times10^{-2}$ & $4.75$ & $1.417098\times10^{-2}$ & $7.994863\times10^{-2}$ & $5.64$ \\
    $10^{-8}$ & $7.163436\times10^{-2}$ & $1.208459\times10^{-1}$ & $1.69$ & $1.124565\times10^{-1}$ & $1.876691\times10^{-1}$ & $1.67$ \\
\bottomrule
\end{tabular}
\caption{The Stokes against the Navier--Stokes layer, vorticity,
layer-resolved grid.  The absolute error is the relative error of
Table~\ref{tab:bl-rel} multiplied
by the norm of the corresponding exact solution.  The two measures are
computed independently and agree on the trend of the ratio.}
\label{tab:bl-compare}
\end{table}

The two columns of ratios were obtained by different routes. The first removes
the discrepancy of the norms, and the second removes the modal dilution. The
two ratios differ by at most $21\%$ for $\varepsilon\ge10^{-4}$ and by $1\%$
at $\varepsilon=10^{-8}$, and they lead to the same two observations.

\emph{The Navier--Stokes benchmark costs accuracy, and the cost is independent
of the layer.} At $\varepsilon=1$ there is no layer at all, and the
Navier--Stokes run is already worse by the factor $5.98$. That factor reflects
a harder problem, with two modes in place of one, a nonlinear residual, and a
forcing that no longer vanishes, and it has nothing to do with the thin
structures.

\emph{The gap closes as the layer sharpens.} By $\varepsilon=10^{-8}$ the two
problems differ by the factor $1.67$ in the layer mode and by $1.69$ in
absolute terms, and by $1.02$ only in the maximum vorticity error
(Table~\ref{tab:bl-max}).
The cost of the Navier--Stokes benchmark, which does not depend on
$\varepsilon$, is therefore overtaken by the difficulty of the layer, whose
normal derivative grows like $1/\sqrt{\varepsilon}$ and which is common to the
two problems. Equivalently, $E^{(1)}_\omega$ grows by $182$ for the Stokes problem
and by $64$ for the Navier--Stokes problem, the smaller factor reflecting the
worse starting point only, and not a gentler degradation.

The consequence for the present article is the following. The dominant
difficulty at small $\varepsilon$ is the layer and not the
convection. An enrichment derived from the linear boundary layer analysis can
therefore be expected to transfer to the nonlinear system, and the two
problems can share one corrector.

\subsubsection{Wall-normal profiles}

The wall-normal profile, measured against the stretched variable
$Y=y/\sqrt{\varepsilon}$ as in Figure~\ref{fig:bl-stokes-profile}, shows the
same behavior. At $\varepsilon=10^{-4}$ the network follows the exponential
descent, but it
undershoots the minimum by about $12\%$ and lags it in $Y$. The Navier--Stokes
network is already worse than the Stokes network at the
same viscosity. At $\varepsilon=10^{-8}$
the prediction has collapsed to a nearly linear ramp, which reproduces no part
of the descent. The hard ansatz still enforces $\omega_\theta=0$ at the wall,
and the network then rises through the entire layer without resolving it.

The pointwise error behaves as in the Stokes case. At $\varepsilon=10^{-8}$ it
rises from the wall, peaks at $\mathcal{O}(1)$ immediately outside
$y=\sqrt{\varepsilon}$, and falls by roughly two orders of magnitude in the
interior. The error is therefore held inside a set of measure
$\mathcal{O}(\sqrt{\varepsilon})$.  At $\varepsilon=10^{-4}$ the contrast
between the layer and the interior is
weaker. This weaker contrast is consistent with the modal split, since $f_1$
is only $77\%$
there against $93\%$ at $\varepsilon=10^{-8}$. A larger share of the error
still lies outside the layer mode.

The two problems of this section together isolate the difficulty, across
four viscosities and two equations. A smooth solution at a small
$\varepsilon$ costs nothing (\S\ref{sec:framework}), while a
$\sqrt{\varepsilon}$ vorticity layer costs two orders of magnitude in the
layer mode, for the linear and for the nonlinear system alike. The
Navier--Stokes benchmark adds a factor of about five for
$\varepsilon\ge10^{-4}$, which the difficulty of the layer overtakes by
$\varepsilon=10^{-8}$. The remedy indicated by \eqref{eq:slpinn-ansatz-exp5}
is therefore the same in
the two cases: the known structure of the layer should be built into the trial
function, rather than left for the network to discover from uniformly sampled
data. The modal error $E^{(1)}$ is the quantity which has to improve. In
\S\ref{sec:exp7} we construct the exact solutions on which such an
enrichment is tested.

\section{Exact solutions consistent with the vanishing viscosity limit}
\label{sec:exp7}

The problem of this article is the vanishing viscosity limit of
\eqref{eq:vorticity-NS}--\eqref{eq:stream-poisson} under the slip condition
\eqref{eq:slip}. In that problem the forcing $\mathbf{F}$ and the initial data
are smooth and independent of $\varepsilon$, and the initial vorticity vanishes
on $\Gamma$ \cite[(1.7), (2.8)]{GieJung2013}. The viscous solution then
converges to the solution $(\omega^0,\psi^0)$ of the limit problem
\eqref{eq:limit-problem}, and the only singular structure of the limit is the
vorticity layer created by the mismatch $\omega^0|_\Gamma\neq0$. Following the
boundary layer analysis of \cite{GieJung2013}, that layer is the corrector
$\Theta$ of \S\ref{sec:corrector}: by \cite[Theorem 2.3]{GieJung2013},
$\psi^\varepsilon-(\psi^0+\theta)$, where $-\Delta\theta=\Theta$, converges to
zero in $L^\infty(0,T;H^2(\Omega))$ at the rate $\varepsilon^{3/4}$. The
corrector is therefore available as analytic information on the solution
before any computation, and the enrichment of \S\ref{sec:slpinn} embeds the
corrector in the trial function for this reason.

The exact solutions of \S\ref{sec:failure} are simple examples on which the
vanilla PINN fails, and they serve that purpose, since the failure does not
depend on the origin of the layer. They are not solutions of the vanishing
viscosity problem above. In \S\ref{sec:failure} the layer is prescribed as the
exponential $e^{-y/\sqrt{\varepsilon}}$ and the forcing is computed from the
solution. The forcing therefore carries the layer: for the Stokes problem,
\[
f-f^0=-(2-\varepsilon)\,e^{-t}\sin(x)\,e^{-y/\sqrt{\varepsilon}}
+\mathcal{O}(\varepsilon),
\]
where $f^0=\partial_t\omega^0$ is the forcing of the limit solution
$\omega^0=e^{-t}\sin(x)\,g_{\mathrm{outer}}(y)$ in \eqref{eq:g-outer-exp5}.
At the wall $f-f^0$ equals $(-2+\varepsilon\pi^2)\,e^{-t}\sin(x)$, which does
not vanish as $\varepsilon\to0$. The data
of \S\ref{sec:failure} thus depend on $\varepsilon$ inside the layer, the layer
is created by the forcing and not by the boundary condition, and the expansion
$\omega^\varepsilon\simeq\omega^0+\Theta$ of \cite{GieJung2013} does not apply.
Indeed, the exponential does not solve the corrector equation
\eqref{eq:corrector-eq}: for the Stokes problem its residual equals
$-2\Theta$.

In the present section we therefore construct a family of exact solutions
consistent with the vanishing viscosity limit, in the following sense. The
forcing of these solutions converges, as $\varepsilon\to0$, to a limit forcing
independent of $\varepsilon$, and their layer is the corrector of
\S\ref{sec:corrector} up to the order $\mathcal{O}(\varepsilon)$. One point of
the construction departs from \cite{GieJung2013}. The limit
solution that we choose has a wall vorticity $\omega^0|_\Gamma$ which does not
vanish at $t=0$, so that the compatibility condition \cite[(2.8)]{GieJung2013}
does not hold and the corrector cannot start from zero. We therefore give up
the initial condition $\Theta|_{t=0}=0$ of \eqref{eq:corrector-eq}, and we use
the corrector of \cite{GieJung2013} started at an earlier time $-T_0<0$, so
that the initial vorticity $\omega^0(\cdot,0)+\Theta(\cdot,0)$ satisfies the
boundary condition $\omega=0$ on $\Gamma$. The initial data then converge to
$\omega^0(\cdot,0)$ in $L^2(\Omega)$ at the
rate $\varepsilon^{1/4}$, but they depend on $\varepsilon$ through the
corrector, and we do not invoke Theorem 2.3 of \cite{GieJung2013}, which is
stated for data independent of $\varepsilon$. The family provides, by
construction, the expansion
$\omega^\varepsilon=\omega^0+\Theta+\mathcal{O}(\varepsilon)$ with the
corrector of \eqref{eq:corrector-eq}, in which the remainder is the tangential
derivative $-\partial_x^2\Psi^{\rm lay}$ of \S\ref{subsec:exp7-design} and is
of the order $\mathcal{O}(\varepsilon)$ uniformly in $\Omega\times(0,T)$. This
corrector is the analytic information that the enrichment of
\S\ref{sec:slpinn} uses.

\subsection{Construction of the exact solutions}
\label{subsec:exp7-design}

We take the limit stream function in the form
\begin{equation}
\psi^0(x,y,t) = U_0\,y(1-y) + A_0\sin(k\zeta)\,q(y),
\qquad q(y)=y^2(1-y)^3,
\qquad \zeta = x - c\,t,
\label{eq:psi0-exp7}
\end{equation}
where $U_0>0$ is the slip speed of the shear flow $U_0\,y(1-y)$ at the walls,
$A_0$ and $k$ are the amplitude and the wavenumber of the perturbation, and
$c=U_0$ for the Navier--Stokes problem and $c=0$ for the Stokes problem.
Throughout the article we take $U_0=1$, $A_0=1/2$ and $k=1$.
The polynomial $q$ is chosen with care, since each of its five conditions
serves a specific purpose and the construction requires all of the five,
namely
\begin{align*}
q(0)=q(1)=0
&\quad\Longrightarrow\quad \psi^0=0 \text{ on both walls;}\\
q'(0)=q'(1)=0
&\quad\Longrightarrow\quad \partial_y\psi^0 = \pm U_0 \text{ on the walls, \emph{constant in} } x;\\
q(y)=\mathcal{O}(y^2) \text{ at } y=0
&\quad\Longrightarrow\quad \partial_x\psi^0\,\partial_y\Theta = \mathcal{O}(\sqrt{\varepsilon});\\
q''(1)=0
&\quad\Longrightarrow\quad \omega^0(x,1,t)=2U_0, \text{ independent of } x;\\
q''(0)=2\neq0
&\quad\Longrightarrow\quad \omega^0(x,0,t) \text{ does depend on } x.
\end{align*}
The second is what makes \eqref{eq:corrector-eq} solvable in closed form,
since with $U$ constant the equation is the heat equation in the frame
$\zeta$. The third matters because of the term
$\partial_x\psi^0\,\partial_y\Theta$, which the asymptotic
derivation evaluates at the wall and therefore discards. Inside the layer this
term is in general $\mathcal{O}(1)$, since $\partial_x\psi^0$
vanishes at the wall while $\partial_y\Theta=\mathcal{O}(\varepsilon^{-1/2})$, and it
becomes negligible only because $q$ is quadratic there. The fourth condition
avoids a second travelling frame at the opposite wall. Writing
\[
\omega^0=-\Delta\psi^0,
\qquad
\omega^0(x,0,t) = W(\zeta) := 2U_0 - 2A_0\sin(k\zeta),
\qquad
\omega^0(x,1,t) = 2U_0,
\]
the solution of \eqref{eq:corrector-eq} is, with $s=t+T_0$ for a time shift
$T_0>0$ and a cut-off $\sigma(y)=S((2\delta-y)/\delta)$ equal to $1$ within
$\delta=1/5$ of a wall and $0$ beyond $2\delta$,
\begin{equation}
\Theta = -\,\sigma(y)\,W(\zeta)\operatorname{erfc}\!\Big(\frac{y}{2\sqrt{\varepsilon s}}\Big)
         -\,\sigma(1-y)\,2U_0\operatorname{erfc}\!\Big(\frac{1-y}{2\sqrt{\varepsilon s}}\Big).
\label{eq:Theta-exp7}
\end{equation}
Here $S(u)=u^5(126-420u+540u^2-315u^3+70u^4)$ for $0\le u\le1$, extended by
$0$ for $u<0$ and by $1$ for $u>1$, is a $C^4$ step from $0$ to $1$, so that
the forcing, which takes four derivatives of $\sigma$, remains smooth.
The wall vorticity $W(\zeta)$ of the limit solution does not vanish at $t=0$,
so that the corrector cannot start from zero, as it does in \cite{GieJung2013}
under the condition \cite[(2.8)]{GieJung2013}, without violating $\omega=0$ on
$\Gamma$ at $t=0$. The shift $T_0>0$ therefore replaces the initial condition
$\Theta|_{t=0}=0$ of \eqref{eq:corrector-eq} by the profile
\eqref{eq:Theta-exp7} at $t=0$, which has the thickness $2\sqrt{\varepsilon
T_0}$. In other words, \eqref{eq:Theta-exp7} is the corrector of
\cite{GieJung2013} started at the time $-T_0$. The initial vorticity
$\omega^0(\cdot,0)+\Theta(\cdot,0)$
therefore vanishes on $\Gamma$, as the boundary condition \eqref{eq:slip}
requires. At $T_0=0$ the initial vorticity would be $\omega^0(\cdot,0)$, which
does not vanish on $\Gamma$, and such non-compatible initial data require a
separate analysis \cite{Gie2014}. Throughout the article we take $T_0=1$.

To obtain the stream function without solving a two-point boundary value problem,
we use the second repeated integral of the error function, which is, with the
stretched variable $z=y/(2\sqrt{\varepsilon s})$,
\begin{equation}
i^2\!\operatorname{erfc}(z)
=\tfrac14\Big[(1+2z^2)\operatorname{erfc}(z)-\tfrac{2z}{\sqrt{\pi}}e^{-z^2}\Big],
\qquad
\partial_y^2\, i^2\!\operatorname{erfc}\!\Big(\frac{y}{2\sqrt{\varepsilon s}}\Big)
=\frac{\operatorname{erfc}(z)}{4\varepsilon s},
\label{eq:i2erfc}
\end{equation}
which is elementary.  Setting
\[
\Psi^{\rm lay}_{\rm raw}
 = \sigma(y)\,4\varepsilon s\,W(\zeta)\, i^2\!\operatorname{erfc}(z),
\]
together with its twin at the far wall, gives $-\partial_y^2\Psi^{\rm
lay}_{\rm raw}=\Theta$ in closed form where $\sigma\equiv1$. In the transition
region $\delta<y<2\delta$ the derivatives of $\sigma$ contribute the terms
$-2\sigma'\partial_y(\cdot)-\sigma''(\cdot)$, which carry the factor
$\operatorname{erfc}(z)$ with $z\ge\delta/(2\sqrt{\varepsilon s})$ and are
therefore below $10^{-22}$ at $\varepsilon=10^{-4}$ and smaller at the smaller
viscosities. Subtracting the linear interpolant, namely setting
\begin{equation}
\Psi^{\rm lay}
 = \Psi^{\rm lay}_{\rm raw}
 -(1-y)\,\Psi^{\rm lay}_{\rm raw}\big|_{y=0}
 -y\,\Psi^{\rm lay}_{\rm raw}\big|_{y=1},
\label{eq:psi-lay-exp7}
\end{equation}
leaves $\Psi^{\rm lay}_{yy}$ untouched, so that the corrector is unchanged,
while making $\psi=\psi^0+\Psi^{\rm lay}$ vanish on both of the walls exactly.
Since $\Psi^{\rm lay}$ then vanishes identically \emph{along} each wall, its
tangential derivatives vanish there as well, and
\[
\omega(x,0,t) = \omega^0(x,0,t)-\Psi^{\rm lay}_{yy}(x,0,t) = W-W = 0
\]
holds exactly as well.  Finally we define
\begin{equation}
\omega := -\Delta\psi
\label{eq:omega-exp7}
\end{equation}
by automatic differentiation, which makes the Poisson coupling exact to machine
precision, and the forcing as in \S\ref{sec:failure}.

\begin{remark}
The Stokes and the Navier--Stokes correctors differ through $c$ in
\eqref{eq:psi0-exp7} only: without the convection the wall data remain fixed,
whereas with the
convection they travel at the slip speed. Both of them are
the \emph{same} profile $\operatorname{erfc}(y/2\sqrt{\varepsilon s})$
multiplied by an amplitude, and the argument of the amplitude alone differs.
This coincidence follows from the constant slip speed of the construction. For
a wall slip velocity which
varies in $x$, the Navier--Stokes corrector is a Duhamel integral and is no
longer a single profile times an amplitude, and the experiments of
\S\ref{sec:slpinn} say nothing about that case.
\end{remark}

\subsection{Validity and verification}
\label{subsec:exp7-valid}

The construction is the leading-order layer only while the layer sits well
inside the cut-off. Table~\ref{tab:blc-valid} measures the two quantities
which say so. The first is the residual of \eqref{eq:corrector-eq} evaluated
on the layer $\omega-\omega^0$ of the exact solution, relative to
$\|\omega-\omega^0\|$. The closed-form profile \eqref{eq:Theta-exp7} satisfies
\eqref{eq:corrector-eq} exactly where $\sigma\equiv1$, and the residual of the
order $\mathcal{O}(\varepsilon)$ comes from the tangential derivative
$-\partial_x^2\Psi^{\rm lay}$, which the definition \eqref{eq:omega-exp7}
includes in the layer. The second is the distance between the forcing of the exact
solution and the $\varepsilon$-independent limit forcing
$f^0=\partial_t\omega^0+J(\psi^0,\omega^0)$. Both are root-mean-squares over
$0<y<3\sqrt{\varepsilon}$.

\begin{table}[h!]
\centering
\begin{tabular}{ccc}
\toprule
$\varepsilon$ & residual of \eqref{eq:corrector-eq} $/\,\|\Theta\|$ & $\|f-f^0\|/\|\partial_t\omega\|$ \\
\midrule
$1$        & $1.7\times10^{3}$    & $2.1\times10^{3}$ \\
$10^{-2}$  & $2.5$                & $7.4$ \\
$10^{-4}$  & $6.4\times10^{-5}$   & $3.0\times10^{-2}$ \\
$10^{-6}$  & $6.5\times10^{-7}$   & $2.5\times10^{-3}$ \\
$10^{-8}$  & $6.5\times10^{-9}$   & $2.5\times10^{-4}$ \\
$10^{-10}$ & $6.5\times10^{-11}$  & $2.5\times10^{-5}$ \\
\bottomrule
\end{tabular}
\caption{Consistency of the exact solutions of \S\ref{subsec:exp7-design}.
The residual is
$\mathcal{O}(\varepsilon)$ and the departure of the forcing from its limit is
$\mathcal{O}(\sqrt{\varepsilon})$.}
\label{tab:blc-valid}
\end{table}

At $\varepsilon=10^{-2}$ the half-width $2\sqrt{\varepsilon(1+T_0)}$ of the
layer reaches $1.4\delta$ and the construction breaks down, and at
$\varepsilon=1$ it exceeds the channel. No choice of the parameters removes
this limitation. At $\varepsilon=1$ there is no boundary layer,
and the construction of \S\ref{subsec:exp7-design} does not apply. The control
at large $\varepsilon$ comes from \S\ref{sec:framework}
and from the row $\varepsilon=1$ of \S\ref{sec:failure}. We therefore sweep
\[
\varepsilon = 10^{-4},\quad 10^{-6},\quad 10^{-8},\quad 10^{-10},
\]
and the boundary conditions hold exactly at every one of these values, as
shown in \S\ref{subsec:exp7-design}.

All of the other settings are the settings (i)--(vii) of
\S\ref{subsubsec:settings}, with the evaluation grids of
\S\ref{subsec:diagnostics}. The runs below are therefore comparable with one
another, and, with the caveats of
\S\ref{subsec:exp7-compare}, with those of \S\ref{sec:failure}.

\subsection{Numerical results for the vanilla PINN}
\label{subsec:exp7-results}

\begin{table}[h!]
\centering
\begin{tabular}{ccccc}
\toprule
$\varepsilon$ & $E_\omega$ & $E_\psi$ & $E_u$ & $E_v$ \\
\midrule
\multicolumn{5}{l}{\emph{Stokes}}\\
    $10^{-4}$ & $1.469593\times10^{-2}$ & $6.220036\times10^{-3}$ & $7.557591\times10^{-3}$ & $1.006841\times10^{-1}$ \\
    $10^{-6}$ & $6.627423\times10^{-2}$ & $8.108030\times10^{-3}$ & $1.271656\times10^{-2}$ & $2.072855\times10^{-1}$ \\
    $10^{-8}$ & $8.438753\times10^{-2}$ & $9.684711\times10^{-3}$ & $1.443022\times10^{-2}$ & $3.825112\times10^{-1}$ \\
    $10^{-10}$ & $8.957216\times10^{-2}$ & $8.273946\times10^{-3}$ & $1.521910\times10^{-2}$ & $2.790531\times10^{-1}$ \\
\midrule
\multicolumn{5}{l}{\emph{Navier--Stokes}}\\
    $10^{-4}$ & $1.712628\times10^{-2}$ & $5.394385\times10^{-3}$ & $7.165500\times10^{-3}$ & $1.488572\times10^{-1}$ \\
    $10^{-6}$ & $6.936433\times10^{-2}$ & $9.308087\times10^{-3}$ & $1.500936\times10^{-2}$ & $2.749230\times10^{-1}$ \\
    $10^{-8}$ & $8.499346\times10^{-2}$ & $1.586798\times10^{-2}$ & $2.024412\times10^{-2}$ & $3.638793\times10^{-1}$ \\
    $10^{-10}$ & $9.154314\times10^{-2}$ & $1.170270\times10^{-2}$ & $1.627791\times10^{-2}$ & $2.864537\times10^{-1}$ \\
\bottomrule
\end{tabular}
\caption{Global space--time relative $L^2$ errors on the problem of
\S\ref{sec:exp7}, vanilla PINN, layer-resolved grid.  $E_v$ is a poor
diagnostic here; see \S\ref{subsec:exp7-compare}.}
\label{tab:blc-rel}
\end{table}

\begin{table}[h!]
\centering
\begin{tabular}{ccccc}
\toprule
& \multicolumn{2}{c}{$\max|\omega_\theta-\omega|$} & \multicolumn{2}{c}{$\max|u_\theta-u|$}\\
\cmidrule(lr){2-3}\cmidrule(lr){4-5}
$\varepsilon$ & Stokes & NSE & Stokes & NSE \\
\midrule
    $10^{-4}$ & $2.210392\times10^{-1}$ & $2.329010\times10^{-1}$ & $1.781129\times10^{-2}$ & $1.489678\times10^{-2}$ \\
    $10^{-6}$ & $2.193626\times10^{0}$ & $2.225671\times10^{0}$ & $2.814963\times10^{-2}$ & $3.530383\times10^{-2}$ \\
    $10^{-8}$ & $2.885349\times10^{0}$ & $2.886244\times10^{0}$ & $2.872682\times10^{-2}$ & $4.586702\times10^{-2}$ \\
    $10^{-10}$ & $2.987186\times10^{0}$ & $2.987898\times10^{0}$ & $3.200865\times10^{-2}$ & $2.824445\times10^{-2}$ \\
\bottomrule
\end{tabular}
\caption{Maximum absolute errors on the problem of \S\ref{sec:exp7}.  The layer
amplitude is $W\in[1,3]$, so the vorticity entries at $\varepsilon\le10^{-6}$ are
the whole layer.}
\label{tab:blc-max}
\end{table}

\emph{The layer is missed entirely.} The amplitude of the corrector is
$W(\zeta)=2-\sin\zeta\in[1,3]$, and by $\varepsilon=10^{-6}$ the maximum
vorticity error is $2.19$, while by $\varepsilon=10^{-10}$ it is $2.99$, which
is the full amplitude of
the layer. The relative error then saturates, at about $9\%$ for both
equations. The selected-time errors are uniform in $t$, since at
$\varepsilon=10^{-10}$ the Navier--Stokes run gives $8.97\times10^{-2}$ at
$t=0$ and $9.46\times10^{-2}$ at $t=1$. The network therefore does not
approximate the layer at any time, rather than losing it as $t$ increases.

\emph{The uniform grid does not detect the failure.} On the fixed mesh of
grid~A the same runs report $E_\omega\approx2.8\times10^{-2}$ at every
viscosity below $10^{-6}$. This value is a factor of $2.5$ to $3.3$ below the
value obtained on the layer-resolved grid (Table~\ref{tab:blc-gridAB}). A
study of
this problem on a uniform grid would conclude that the error has stopped
growing.

\begin{table}[h!]
\centering
\begin{tabular}{ccccccc}
\toprule
& \multicolumn{3}{c}{Stokes} & \multicolumn{3}{c}{Navier--Stokes}\\
\cmidrule(lr){2-4}\cmidrule(lr){5-7}
$\varepsilon$ & grid A & grid B & B/A & grid A & grid B & B/A \\
\midrule
    $10^{-4}$ & $1.442\times10^{-2}$ & $1.470\times10^{-2}$ & $1.02$ & $1.694\times10^{-2}$ & $1.713\times10^{-2}$ & $1.01$ \\
    $10^{-6}$ & $2.201\times10^{-2}$ & $6.627\times10^{-2}$ & $3.01$ & $2.777\times10^{-2}$ & $6.936\times10^{-2}$ & $2.50$ \\
    $10^{-8}$ & $2.545\times10^{-2}$ & $8.439\times10^{-2}$ & $3.32$ & $2.720\times10^{-2}$ & $8.499\times10^{-2}$ & $3.12$ \\
    $10^{-10}$ & $2.797\times10^{-2}$ & $8.957\times10^{-2}$ & $3.20$ & $2.804\times10^{-2}$ & $9.154\times10^{-2}$ & $3.26$ \\
\bottomrule
\end{tabular}
\caption{$E_\omega$ on the fixed uniform grid against the layer-resolved grid,
problem of \S\ref{sec:exp7}.}
\label{tab:blc-gridAB}
\end{table}

\subsection{Modal errors and the limits of the comparison}
\label{subsec:exp7-compare}

In \S\ref{sec:exp6} the modal decomposition \eqref{eq:modal-error} separated
the layer mode from the smooth modes. On the present problem it separates
different quantities, since the amplitude $W(\zeta)=2U_0-2A_0\sin(k\zeta)$ of
the layer occupies the wavenumbers $k=0$ and $k=1$. Both of the reported modes
therefore contain a
part of the layer. The mode $k=0$ carries the $x$-independent part of the
layer, together with the constant outer vorticity $2U_0$, and holds $98.9\%$
of the energy of the vorticity. The mode $k=1$ carries the travelling part and
holds $1.1\%$. We report both modal errors in Table~\ref{tab:blc-modal}.

\begin{table}[h!]
\centering
\begin{tabular}{cccccc}
\toprule
& \multicolumn{2}{c}{$E^{(0)}_\omega$} & \multicolumn{2}{c}{$E^{(1)}_\omega$} & \\
\cmidrule(lr){2-3}\cmidrule(lr){4-5}
$\varepsilon$ & Stokes & NSE & Stokes & NSE & ratio \\
\midrule
    $10^{-4}$ & $1.009217\times10^{-2}$ & $1.013389\times10^{-2}$ & $8.297612\times10^{-2}$ & $1.146337\times10^{-1}$ & $1.38$ \\
    $10^{-6}$ & $6.043399\times10^{-2}$ & $6.203246\times10^{-2}$ & $2.543356\times10^{-1}$ & $2.893598\times10^{-1}$ & $1.14$ \\
    $10^{-8}$ & $7.852624\times10^{-2}$ & $7.845655\times10^{-2}$ & $2.920628\times10^{-1}$ & $3.063173\times10^{-1}$ & $1.05$ \\
    $10^{-10}$ & $8.384025\times10^{-2}$ & $8.484443\times10^{-2}$ & $3.013240\times10^{-1}$ & $3.209438\times10^{-1}$ & $1.07$ \\
\bottomrule
\end{tabular}
\caption{Vorticity error by $x$-wavenumber on the problem of \S\ref{sec:exp7}.
The last column is $E^{(1)}_{\rm NSE}/E^{(1)}_{\rm Stokes}$.}
\label{tab:blc-modal}
\end{table}

Two readings of this table are possible, and only one of them is supported by
the data.

The table shows that \emph{the travelling component is the harder one}, since
$E^{(1)}_\omega$ exceeds $E^{(0)}_\omega$ by a factor of at least three at
every viscosity, and by a factor of eight to eleven at $\varepsilon=10^{-4}$,
in the two equations alike. The table shows as well that the
two equations
fail indistinguishably at small $\varepsilon$, the steady components
agreeing to within $2\%$ and the travelling ones to within $7\%$ at
$\varepsilon\le10^{-8}$.

Table~\ref{tab:blc-modal} does \emph{not} support the conclusion that the two
problems are equally difficult, nor the conclusion that an enrichment designed
for the linear
problem will transfer to the nonlinear one. The agreement is measured in a
saturated regime. Table~\ref{tab:blc-max} shows the pointwise error at the
amplitude of the
layer. A method that produces no layer gives the same error for either
equation. The agreement therefore reflects the complete failure, and
not a property of the equations. The trend supports this interpretation, since
the ratio moves from $1.38$ at
$\varepsilon=10^{-4}$, the one viscosity at which the network still resolves
part of the layer, toward $1.07$ as the resolution is lost. The single
informative comparison is therefore the row $\varepsilon=10^{-4}$,
where the nonlinear problem is $38\%$ worse in the travelling mode. All of the
runs here use one fixed random seed, so the size of that figure is
not established
either. Whether the enrichment transfers between the two systems has to be measured
directly, in the regime in which the layer is resolved, and that is the
subject of \S\ref{sec:slpinn}.

\begin{remark}
The error $E_v$ in Table~\ref{tab:blc-rel} runs from $10\%$ to $38\%$ and does
not indicate a failure in the velocity. In \eqref{eq:psi0-exp7} the same
polynomial $q$ sets two quantities. The polynomial $q$ sets the wall vorticity
through $q''(0)=2$, and the $x$-dependent velocity in the
interior through $\max q=0.0346$. The ratio of the two is $58$, so that
$\|v\|=\|\partial_x\psi\|$ is about $58$ times smaller than $\|u\|$. An
absolute error
of $7\times10^{-3}$, which is the order of magnitude of
Table~\ref{tab:blc-max}, then appears as a large relative error. The
modal errors are the meaningful statement, for at $\varepsilon=10^{-10}$ we
have $E^{(1)}_v = 2.51\times10^{-1}$ against $E^{(1)}_u=1.38\times10^{-1}$ for
the Stokes problem, and $2.39\times10^{-1}$ against $2.39\times10^{-1}$ for
the Navier--Stokes problem.
\end{remark}

\subsection{Wall-normal profiles}

\begin{figure}[h!]
\centering
\includegraphics[width=0.75\textwidth]{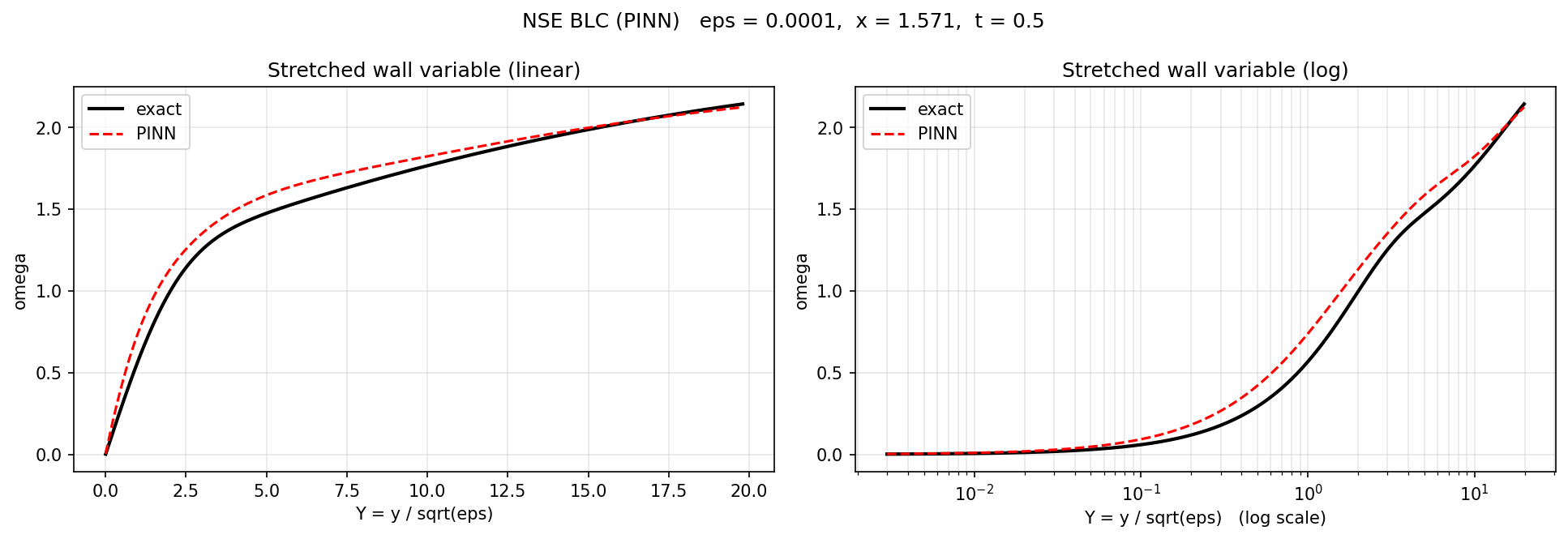}\\[1ex]
\includegraphics[width=0.75\textwidth]{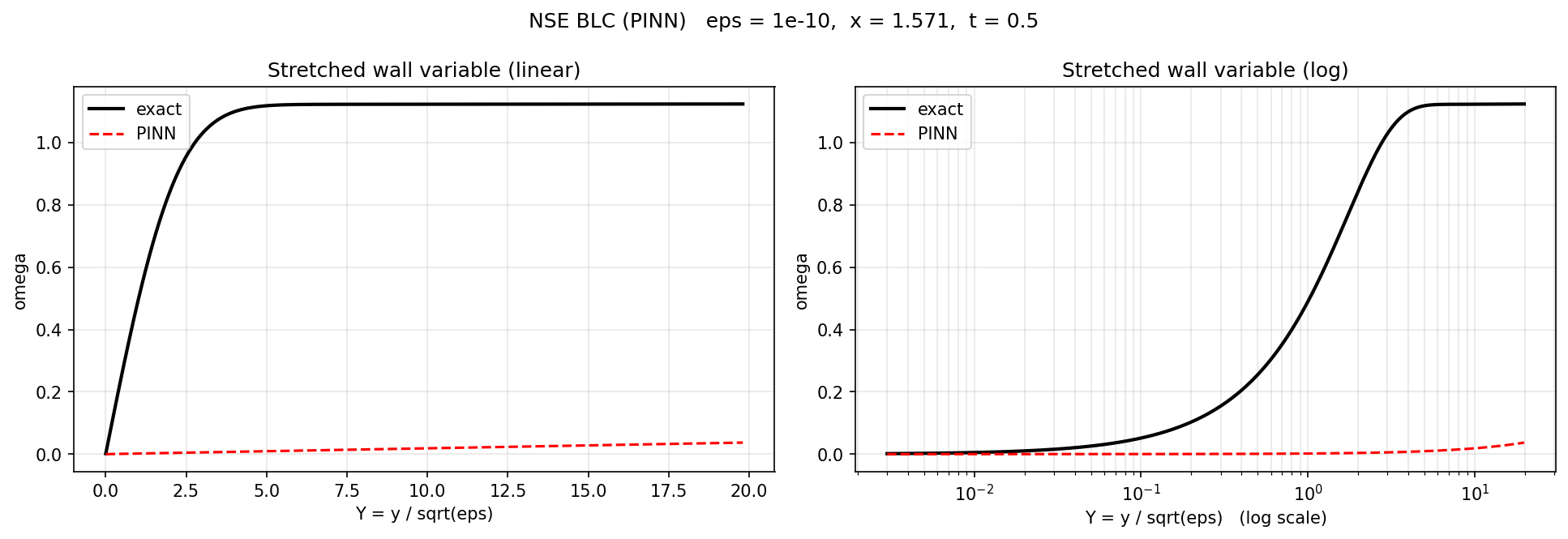}
\caption{Navier--Stokes problem of \S\ref{sec:exp7}: wall-normal vorticity
profile at $x=\pi/2$, $t=0.5$, against $Y=y/\sqrt{\varepsilon}$.  Top:
$\varepsilon=10^{-4}$, where the network still tracks part of the
$\operatorname{erfc}$ descent.  Bottom: $\varepsilon=10^{-10}$, where it does
not.}
\label{fig:blc-profile}
\end{figure}

Figure~\ref{fig:blc-profile} shows the wall-normal profile, the same
diagnostic as in \S\ref{sec:failure}, now on the exact solution of
\S\ref{subsec:exp7-design}.

The present section supplies what \S\ref{sec:failure} could not supply. The
layer of the new solution is the corrector of the boundary layer analysis,
to the order $\mathcal{O}(\varepsilon)$, and the forcing of the new solution
approaches the limit forcing at the rate $\mathcal{O}(\sqrt{\varepsilon})$.
Both boundary conditions are exact at every viscosity. On that problem the
vanilla PINN fails completely for $\varepsilon\le10^{-6}$, in the linear and in
the nonlinear system alike, and the diagnostic computed on the uniform grid
understates that failure by a factor of about three. These errors are the
numbers which the enrichment of \S\ref{sec:slpinn} has to
improve upon. The quantity to watch is $E^{(1)}_\omega$, which exceeds
$E^{(0)}_\omega$ by a
factor of $3.6$ to $3.9$ at $\varepsilon\le10^{-8}$ and by a larger factor at the
two larger viscosities. Two forms of the enrichment are separated there. The
first takes
the scale
$\sqrt{\varepsilon}$ alone from the analysis and leaves the profile to a
sub-network of the stretched variable, whereas the second inserts the analytic
profile $\operatorname{erfc}$ of \eqref{eq:Theta-exp7} with a learned
amplitude. The second is legitimate on the present problem only, where that
profile is
the true layer. Even there the comparison requires care, since the trial space
then contains the exact solution up to the order $\mathcal{O}(\varepsilon)$
and only the amplitude remains to be found.

\section{Analysis-informed singular-layer PINNs}
\label{sec:slpinn}

In \S\ref{sec:exp7} we constructed exact solutions consistent with the
vanishing viscosity limit, and the vanilla PINN missed the layer on these
solutions once $\varepsilon\le10^{-6}$, by a margin which the uniform evaluation grid
understates by a factor of about three. In the present section we embed the
boundary layer analysis in the trial function, instead of using the analysis
only to design the test problem.  The analysis supplies two distinct pieces of
information, and the question is which of the two the enrichment requires. The
first is the
\emph{scale} $\sqrt{\varepsilon}$ on which the layer varies, whereas the
second is the \emph{corrector} itself, a specific profile with a specific
amplitude, tied to the outer solution through
$\Theta|_\Gamma=-\omega^0|_\Gamma$. The methods of the first kind in
\S\ref{sec:intro}
\cite{ZhangSuYuan2025,HuangQiuWangWang2024,PangXiaoYang2026} use the first of
the two only. We therefore build both of the enrichments, following the
methodology of our
singular layer PINNs
\cite{GieHongJung2024,GieHongJungMunkhjin2024,GieHongJungPolygonal,GieHongJungLee2026,ChangGieHongJungBurgers,ChangGieHongJung2024,MoonGieHongJungPipe}.
The two enrichments share the regular part, the scale $\sqrt{\varepsilon}$ and
the exact boundary conditions, and we compare them on the same problem.
The scale-only enrichment comes in two forms, with one layer network shared by the two walls and with one layer network per
wall. The second form separates the coupling of the walls, which the
corrector-informed enrichment does not have, from the profile and from the
amplitude law.

\subsection{Two singular-layer trial functions: scale-only and corrector-informed}
\label{subsec:exp8-ansatz}

Both of the trial functions of this section are built from a regular part
$(a,b)$, which approximates the outer solution, and a layer part, which carries
the structure of \eqref{eq:corrector-eq}.  The regular part is written in the
split form
\begin{equation}
a(x,y,t)=(1-y)\,P(x,t)+y\,Q(x,t)+y(1-y)\,R(x,y,t),
\qquad
b(x,y,t)=y(1-y)\,V(x,y,t),
\label{eq:reg-split}
\end{equation}
where $R$ and $V$ are the two outputs of the trunk network of
\eqref{eq:raw-network-exp1}, and $P$ and $Q$ are the two outputs of a second and
smaller network of the two variables $(x,t)$.  The reason for
\eqref{eq:reg-split} is that the trial functions \eqref{eq:ansatz-B} and
\eqref{eq:ansatz-A} need the traces of the regular part on the two walls,
\begin{equation}
a_0(x,t):=a(x,0,t)=P(x,t),
\qquad
a_1(x,t):=a(x,1,t)=Q(x,t),
\qquad
b_0=b_1=0,
\label{eq:wall-traces}
\end{equation}
and \eqref{eq:reg-split} makes those traces \emph{outputs} of the network rather
than three separate evaluations of one network, at $y=0$, at $y=1$ and at the
collocation point.  Since the residual is built from second derivatives, the
three evaluations would require three second-derivative graphs per training
step, which we measured at $3.3$ times the cost of a vanilla step.  The factor
$y(1-y)$ in front of $R$ plays a different role from the hard
boundary condition of \eqref{eq:hard-bc-exp1}.  The factor multiplies the
interior remainder only. The behavior at the wall is carried
by $P$ and $Q$ through the layer profile. No part of the network is therefore
asked to produce a slope of the size $\varepsilon^{-1/2}$.

With $s=t+T_0$ and the same cut-off $\sigma$ as in \eqref{eq:Theta-exp7}, we define
\begin{equation}
\begin{aligned}
E_0 &= \sigma(y)\operatorname{erfc}\!\Big(\tfrac{y}{2\sqrt{\varepsilon s}}\Big), &
I_0 &= \sigma(y)\,4\varepsilon s\; i^2\!\operatorname{erfc}\!\Big(\tfrac{y}{2\sqrt{\varepsilon s}}\Big),\\
E_1 &= \sigma(1-y)\operatorname{erfc}\!\Big(\tfrac{1-y}{2\sqrt{\varepsilon s}}\Big), &
I_1 &= \sigma(1-y)\,4\varepsilon s\; i^2\!\operatorname{erfc}\!\Big(\tfrac{1-y}{2\sqrt{\varepsilon s}}\Big).
\end{aligned}
\label{eq:layer-factors}
\end{equation}

The factors $E_0$ and $E_1$ are the profile of \eqref{eq:Theta-exp7} at the two
walls, and the factors $I_0$ and $I_1$ are their second primitives in $y$, in the
sense of the identity \eqref{eq:i2erfc}.

\textbf{Variant B, corrector-informed (the enrichment of this article).}  We
define the layer part of the vorticity
as
\begin{equation}
\Theta_\theta := -\,a_0E_0-a_1E_1,
\label{eq:Theta-net}
\end{equation}
which is \eqref{eq:Theta-exp7} with the amplitudes $W(\zeta)$ and $2U_0$ of the
exact corrector replaced by the wall traces $a_0$ and $a_1$ of the network, and
we set
\begin{equation}
\omega_\theta = a + \Theta_\theta,
\qquad
\psi_\theta = b + a_0I_0 + a_1I_1 - (1-y)\big(b_0+a_0\varepsilon s\big)
                                  - y\big(b_1+a_1\varepsilon s\big).
\label{eq:ansatz-B}
\end{equation}
Since $i^2\!\operatorname{erfc}(0)=1/4$, the factors $I_0$ and $I_1$ satisfy
$I_0|_{y=0}=I_1|_{y=1}=\varepsilon s$, and the subtracted linear interpolant in
\eqref{eq:ansatz-B} is exactly their traces on the walls.  With the split
\eqref{eq:reg-split} and the traces \eqref{eq:wall-traces}, the second equation
of \eqref{eq:ansatz-B} collapses to the form that we implement,
\begin{equation}
\psi_\theta = y(1-y)\,V + P\,I_0 + Q\,I_1
              -\varepsilon s\big[(1-y)P+yQ\big].
\label{eq:ansatz-B-psi}
\end{equation}
We state three properties of \eqref{eq:ansatz-B} separately.

\emph{(i) The boundary conditions are exact and require no boundary loss.} At
$y=0$ we
have $E_0=1$ while $E_1=0$, because the cut-off $\sigma(1-y)$ of the far wall
vanishes there, and hence $\omega_\theta(x,0,t)=a_0-a_0=0$ identically, for
every value of every parameter. The same holds at $y=1$, and $\psi_\theta$
vanishes at both of the walls by the subtracted linear interpolant. The factor
$y(1-y)$ employed up to this point is not needed
and is not present.

\emph{(ii) The amplitude of the layer is the wall trace of the regular
network.} The amplitude is $a_0$, the trace of the regular network itself on
the wall, and this is precisely the defining property of the corrector,
$\Theta|_\Gamma=-\omega^0|_\Gamma$, transcribed into the ansatz. The limit
solution $\omega^0$ does not enter the trial function. The amplitude law is
imposed through the trace of the network itself, so that the enrichment
requires no a priori knowledge of the limit solution, and the network learns
the outer solution and the amplitude of the layer together. The consequence in
practice is the following. The amplitude is determined by
the residual in the interior, where there are ten thousand collocation points,
rather than by the residual
inside the layer, where there is almost no information. At
$\varepsilon=10^{-8}$ about $3$ of the $10{,}000$ uniform points fall within
$3\sqrt{\varepsilon}$ of a wall, and at $\varepsilon=10^{-10}$ about $0.3$ of
them do, so that most of the runs at that viscosity have none at all.

\emph{(iii) The enrichment of the stream function is necessary.} The layer
perturbs $\psi$ itself by $\mathcal{O}(\varepsilon)$ only, but it
perturbs $\partial_y^2\psi$ at the full amplitude, and the Poisson residual is
built from the second derivatives. Pairing $I$ with $E$ through the identity
in \eqref{eq:i2erfc} makes $-\Delta\psi_\theta=\omega_\theta$ consistent at
the leading order by construction, with the \emph{same} coefficient $a_0$ on
the two sides.

A refinement term $y\,E_0\,c_L(x,t)+(1-y)\,E_1\,c_R(x,t)$, where $c_L$ and
$c_R$ are two further outputs of the wall network, carries the next order of
the hierarchy. The refinement vanishes at the walls, and it therefore
preserves the property (i). For the constant slip speed of the problem of \S\ref{sec:exp7}, the corrector
equation
is the heat equation with wall data independent of $t$ in the travelling
frame, and \eqref{eq:Theta-exp7} solves the leading-order corrector equation
exactly. We therefore set the refinement to zero in the present experiments.

\textbf{Variant A, scale-only (for comparison).}  In the variant~A the
analysis supplies the
scale $\sqrt{\varepsilon}$ only.  Let
\begin{equation}
Y_0=\frac{y}{\sqrt{\varepsilon}},
\qquad
Y_1=\frac{1-y}{\sqrt{\varepsilon}},
\label{eq:stretched-A}
\end{equation}
let $\chi(Y)=S((24-Y)/12)$ be a cut-off in the stretched variable, with the
step $S$ of \S\ref{subsec:exp7-design}, equal to $1$ for $Y\le12$ and to $0$
for $Y\ge24$. Let $\Lambda=(\Lambda_\omega,\Lambda_\psi)$ be a
second network of the \emph{stretched} variable, of the same size as the wall
network of the variant~B. We set
\begin{equation}
G=(G_\omega,G_\psi)
 =(a,b)+\chi(Y_0)\,\Lambda(x,Y_0,t)+\chi(Y_1)\,\Lambda(x,Y_1,t),
\label{eq:G-A}
\end{equation}
and impose the boundary conditions by subtracting the linear interpolant of the
two wall traces,
\begin{equation}
\omega_\theta = G_\omega - (1-y)\,G_\omega|_{y=0} - y\,G_\omega|_{y=1},
\qquad
\psi_\theta = G_\psi - (1-y)\,G_\psi|_{y=0} - y\,G_\psi|_{y=1}.
\label{eq:ansatz-A}
\end{equation}
The scale alone enters \eqref{eq:ansatz-A}, since the profile in $Y$ and the
amplitude are both produced by $\Lambda$ and are free.  The property (i) of
the variant~B survives, since the subtracted interpolant
equals $G_\omega$ and $G_\psi$ at each of the walls, whereas the variant~A
gives up the property (ii). Two further differences from the variant~B follow
from \eqref{eq:G-A}--\eqref{eq:ansatz-A}. The traces $P$ and $Q$ cancel in
\eqref{eq:ansatz-A}, since the subtracted interpolant contains them, so that
the wall network plays no role in the variant~A and the amplitude of the layer
is produced by $\Lambda$ alone. The same network $\Lambda$ serves the two
walls, so that the layers at the two
walls share one profile in $x$ and $t$. The exact solution, however, has the
amplitude $W(\zeta)$ at the lower wall and $2U_0$ at the upper wall. The
variant~A therefore differs from the variant~B
in the profile of the layer, in the amplitude law and in the coupling of the
two walls, and the comparison between the two measures the three together.

\textbf{Variant A$'$, scale-only with independent walls (for
comparison).} In the variant~A$'$ we replace the shared network $\Lambda$ of
\eqref{eq:G-A} by two networks of the same size, $\Lambda_0$ for the lower
wall and $\Lambda_1$ for the upper wall,
\begin{equation}
G=(a,b)+\chi(Y_0)\,\Lambda_0(x,Y_0,t)+\chi(Y_1)\,\Lambda_1(x,Y_1,t),
\label{eq:G-A2}
\end{equation}
and we impose the boundary conditions by \eqref{eq:ansatz-A} as before. Each
wall now has its own profile and its own amplitude, both free, and the
variant~A$'$ differs from the variant~B in the profile and in the amplitude law
only. The comparison between the variants~A and~A$'$ measures the coupling of
the walls, and the comparison between the variants~A$'$ and~B measures the
profile and the amplitude law together.

The variant~B contains the exact solution of \S\ref{subsec:exp7-design} up to
the order $\mathcal{O}(\varepsilon)$. Substituting the outer solution
$(a,b)=(\omega^0,\psi^0)$ into \eqref{eq:ansatz-B} gives $a_0=W$ and
$a_1=2U_0$, and reproduces the exact solution except for the term
$-\partial_x^2\Psi^{\rm lay}$ of \S\ref{sec:exp7}, so that the amplitude alone
remains to be found. We state here the circularity of the variant~B on the
present problem, namely that the profile inserted into the trial function
\emph{is} the true layer. No such statement holds for the variant~A, whose
trial space does not contain
the exact solution, since the two walls share one profile. The
variant~A$'$ approximates the two layers independently, and it is therefore
free of the structural restriction of the variant~A. Unlike the variant~B,
however, the variant~A$'$ has to learn the profiles and the amplitudes of the
two layers from the training.

\begin{remark}
The subtraction in \eqref{eq:ansatz-A} is linear in $y$. A function linear in
$y$ has no second derivative in $y$, and its contribution to $-\Delta\psi$ is
the smooth term
$(1-y)\,\partial_x^2G_\psi|_{y=0}+y\,\partial_x^2G_\psi|_{y=1}$, which the
automatic differentiation includes and which carries no large factor. A
subtraction of the form $\sigma(y)\,G_\psi|_{y=0}+\sigma(1-y)\,G_\psi|_{y=1}$
with the cut-off $\sigma$ would be equally exact at the walls, but its second
derivative would inject the term $\sigma''(y)\,G_\psi|_{y=0}$, of the size
$\delta^{-2}=25$, into the Poisson residual. The same device makes the exact
solution of \S\ref{sec:exp7} satisfy its boundary conditions, for the same
reason.
\end{remark}

\subsection{Training setup}
\label{subsec:exp8-setup}

The loss is given by \eqref{eq:loss-exp1}, with the residuals
\eqref{eq:Rw-exp1}--\eqref{eq:Rpsi-exp1} for the Stokes problem and
\eqref{eq:Rw-exp2} for the Navier--Stokes problem, evaluated on the trial
functions \eqref{eq:ansatz-B} or \eqref{eq:ansatz-A} in place of the hard ansatz
\eqref{eq:hard-bc-exp1}.  No term is added to the loss for the enrichment, and no
wall term is needed, since the boundary conditions hold identically.  All of
the settings other than the trial function are the settings (i)--(vii) of
\S\ref{subsubsec:settings}, and in particular the trunk for $(R,V)$ is the
$4\times50$ tanh network of (i). The new elements are the auxiliary networks,
namely the wall network $(P,Q)$ of the variant~B and the stretched network
$\Lambda$ of the variant~A, or the two stretched networks $\Lambda_0$ and
$\Lambda_1$ of the variant~A$'$. All of them are $3\times20$. A small
network suffices, since the structure of the layer is essentially
one-dimensional in the stretched variable. A small network also keeps the
memory low, since the memory is dominated by the second-derivative graph of a
sub-network, which is evaluated several times per forward pass. The enrichment
therefore adds few parameters. There are $7{,}952$ parameters for the vanilla
network, $8{,}894$ for the variant~B ($+11.8\%$),
$9{,}856$ for the variant~A ($+23.9\%$) and $10{,}818$ for the
variant~A$'$ ($+36.0\%$). The enrichment adds more to the time
per step, since each forward pass evaluates the layer factors and, for the
variants~A and~A$'$, the sub-networks. Under identical conditions one training step costs about twice the vanilla
step for the variant~B and about $2.8$ times for the variant~A, and the
variant~A$'$ costs about $4\%$ more than the variant~A;
Table~\ref{tab:fv-cost} gives the measured values.

\subsection{Numerical results for the Stokes problem} \label{subsec:exp8-results}

We run the four trial spaces, the unenriched one and the three
enrichments, on
the Stokes problem of \S\ref{subsec:exp7-design} at the
four viscosities of the sweep, and we collect the global errors in
Tables~\ref{tab:sl-rel} and~\ref{tab:slpinn-max}. Four observations follow.

\begin{table}[h!]
\centering
\begin{tabular}{ccccc}
\toprule
$\varepsilon$ & $E_\omega$ & $E_\psi$ & $E_u$ & $E_v$ \\
\midrule
\multicolumn{5}{l}{\emph{vanilla PINN}}\\
    $10^{-4}$ & $1.469593\times10^{-2}$ & $6.220036\times10^{-3}$ & $7.557591\times10^{-3}$ & $1.006841\times10^{-1}$ \\
    $10^{-6}$ & $6.627423\times10^{-2}$ & $8.108030\times10^{-3}$ & $1.271656\times10^{-2}$ & $2.072855\times10^{-1}$ \\
    $10^{-8}$ & $8.438753\times10^{-2}$ & $9.684711\times10^{-3}$ & $1.443022\times10^{-2}$ & $3.825112\times10^{-1}$ \\
    $10^{-10}$ & $8.957216\times10^{-2}$ & $8.273946\times10^{-3}$ & $1.521910\times10^{-2}$ & $2.790531\times10^{-1}$ \\
\midrule
\multicolumn{5}{l}{\emph{sl-PINN, corrector variant}}\\
    $10^{-4}$ & $3.870849\times10^{-3}$ & $1.630285\times10^{-3}$ & $2.020239\times10^{-3}$ & $6.184603\times10^{-2}$ \\
    $10^{-6}$ & $3.798262\times10^{-3}$ & $1.100614\times10^{-3}$ & $1.630483\times10^{-3}$ & $4.285034\times10^{-2}$ \\
    $10^{-8}$ & $3.947600\times10^{-3}$ & $1.471562\times10^{-3}$ & $1.807135\times10^{-3}$ & $4.710111\times10^{-2}$ \\
    $10^{-10}$ & $4.063590\times10^{-3}$ & $1.457937\times10^{-3}$ & $1.797969\times10^{-3}$ & $4.874462\times10^{-2}$ \\
\midrule
\multicolumn{5}{l}{\emph{sl-PINN, scale variant}}\\
    $10^{-4}$ & $1.029169\times10^{-2}$ & $2.208899\times10^{-3}$ & $3.792926\times10^{-3}$ & $9.299087\times10^{-2}$ \\
    $10^{-6}$ & $2.526786\times10^{-2}$ & $3.729917\times10^{-3}$ & $5.519888\times10^{-3}$ & $1.254102\times10^{-1}$ \\
    $10^{-8}$ & $2.839520\times10^{-2}$ & $4.400670\times10^{-3}$ & $5.753863\times10^{-3}$ & $1.360560\times10^{-1}$ \\
    $10^{-10}$ & $3.067910\times10^{-2}$ & $3.923541\times10^{-3}$ & $5.203948\times10^{-3}$ & $1.439406\times10^{-1}$ \\
\midrule
\multicolumn{5}{l}{\emph{sl-PINN, scale variant with independent walls}}\\
    $10^{-4}$ & $4.444267\times10^{-3}$ & $1.660162\times10^{-3}$ & $2.292495\times10^{-3}$ & $8.176482\times10^{-2}$ \\
    $10^{-6}$ & $3.678220\times10^{-3}$ & $1.367116\times10^{-3}$ & $1.831835\times10^{-3}$ & $8.374606\times10^{-2}$ \\
    $10^{-8}$ & $1.024140\times10^{-2}$ & $8.373839\times10^{-4}$ & $1.153442\times10^{-3}$ & $4.807372\times10^{-2}$ \\
    $10^{-10}$ & $1.389843\times10^{-2}$ & $6.562982\times10^{-4}$ & $9.102740\times10^{-4}$ & $3.553013\times10^{-2}$ \\
\bottomrule
\end{tabular}
\caption{Global space--time relative $L^2$ errors on the Stokes problem of
\S\ref{sec:exp7}, layer-resolved grid~B.  The vanilla rows repeat
Table~\ref{tab:blc-rel} for comparison.}
\label{tab:sl-rel}
\end{table}

\begin{table}[h!]
\centering
\scriptsize
\setlength{\tabcolsep}{2.5pt}
\begin{tabular}{ccccccccc}
\toprule
& \multicolumn{4}{c}{Stokes} & \multicolumn{4}{c}{Navier--Stokes}\\
\cmidrule(lr){2-5}\cmidrule(lr){6-9}
$\varepsilon$ & vanilla & corrector & scale & scale$'$ & vanilla & corrector & scale & scale$'$ \\
\midrule
    $10^{-4}$ & $2.21\times10^{-1}$ & $3.34\times10^{-2}$ & $1.57\times10^{-1}$ & $4.59\times10^{-2}$ & $2.33\times10^{-1}$ & $8.68\times10^{-2}$ & $1.18\times10^{-1}$ & $1.03\times10^{-1}$ \\
    $10^{-6}$ & $2.19\times10^{0}$ & $5.70\times10^{-2}$ & $6.34\times10^{-1}$ & $1.98\times10^{-1}$ & $2.23\times10^{0}$ & $1.12\times10^{-1}$ & $6.49\times10^{-1}$ & $2.12\times10^{-1}$ \\
    $10^{-8}$ & $2.89\times10^{0}$ & $6.39\times10^{-2}$ & $1.65\times10^{0}$ & $1.69\times10^{0}$ & $2.89\times10^{0}$ & $1.17\times10^{-1}$ & $1.86\times10^{0}$ & $2.21\times10^{0}$ \\
    $10^{-10}$ & $2.99\times10^{0}$ & $6.41\times10^{-2}$ & $4.29\times10^{0}$ & $2.81\times10^{0}$ & $2.99\times10^{0}$ & $1.24\times10^{-1}$ & $3.07\times10^{0}$ & $2.94\times10^{0}$ \\
\bottomrule
\end{tabular}
\caption{Maximum absolute vorticity errors, $\max|\omega_\theta-\omega|$.  The
amplitude of the corrector is $W(\zeta)=2-\sin\zeta\in[1,3]$, so that the
vanilla entry at $\varepsilon=10^{-10}$ is the full amplitude of the layer and the
Stokes entry of the scale variant exceeds it. The column scale$'$ is the
scale variant with independent wall networks, the variant~A$'$ of
\S\ref{subsec:exp8-ansatz}.}
\label{tab:slpinn-max}
\end{table}

\emph{The error of the corrector variant is uniform in $\varepsilon$.} Across
six decades of
the viscosity $E_\omega$ takes the values $3.871\times10^{-3}$,
$3.798\times10^{-3}$, $3.948\times10^{-3}$ and $4.064\times10^{-3}$, which is
the same number to two figures. This uniformity is the signature which the
analysis predicts. If the ansatz carries the layer, the network approximates
the smooth
outer solution only, whose difficulty does not depend on $\varepsilon$. Over
the same range the vanilla error grows by a factor of six, from
$1.470\times10^{-2}$ to $8.957\times10^{-2}$, and the ratio between the two
methods widens from $3.8$ to $22.0$.

\emph{The maximum error is bounded by the enrichment.} Table~\ref{tab:slpinn-max}
is the more direct statement. The amplitude of the corrector is
$W(\zeta)=2-\sin\zeta\in[1,3]$. At
$\varepsilon=10^{-10}$ the maximum vanilla vorticity error reaches $2.99$,
which is the full amplitude of the layer. The corrector variant stays at
$6.4\times10^{-2}$, which is two percent of the amplitude. The enrichment
therefore removes the layer from the error, rather than reducing the error by
a factor.

\emph{The scale variant deteriorates as $\varepsilon\to0$.} At
$\varepsilon=10^{-4}$ it is $1.4$ times better than the vanilla network, and at
$\varepsilon=10^{-10}$ it is still $2.9$ times better. Its error, however, grows
by a factor of three over the sweep, from $1.029\times10^{-2}$ to
$3.068\times10^{-2}$, while the error of the corrector variant does not grow.
At $\varepsilon=10^{-10}$ it is $7.5$ times the error of the corrector
variant. The maximum vorticity error of the scale
variant there is $4.29$,
which is larger than the largest amplitude of the layer present in the
problem. This inequality is important for the following reason. A method that
fails to produce the layer has an error equal to the amplitude
of the layer and no larger, whereas an error that exceeds that amplitude means that a
spurious layer has been produced. The scale alone is therefore insufficient.
With the profile and the amplitude left free, the error of the
enrichment grows as $\varepsilon\to0$, as the error of the unenriched network
does, and the enrichment produces a layer which the solution does not
contain.

\emph{Independent wall networks remove the coupling and not the growth.}
The variant~A$'$ reduces the error of the variant~A at every viscosity, by the factors $2.3$, $6.9$, $2.8$ and $2.2$. At $\varepsilon=10^{-4}$ and
$10^{-6}$ its error equals that of the corrector variant, $4.444\times10^{-3}$
and $3.678\times10^{-3}$ against $3.871\times10^{-3}$ and $3.798\times10^{-3}$.
At $\varepsilon=10^{-8}$ and $10^{-10}$ it is $2.6$ and $3.4$ times the error
of the corrector variant, and its growth over the sweep, a factor of $3.1$, is
the growth of the variant~A. Its maximum vorticity error reaches $1.69$ at
$\varepsilon=10^{-8}$ and $2.81$ at $\varepsilon=10^{-10}$, which is the
amplitude of the layer, against $6.4\times10^{-2}$ for the corrector variant.
The coupling of the walls therefore accounts for the difference between the two
scale-only variants, whereas the free profile and the free amplitude account
for the loss of uniformity, which the independent networks do not remove.

\subsubsection{Errors on the uniform grid}
\label{subsec:exp8-grid}

\begin{table}[h!]
\centering
\small
\setlength{\tabcolsep}{3pt}
\begin{tabular}{ccccccccc}
\toprule
& \multicolumn{2}{c}{vanilla} & \multicolumn{2}{c}{corrector} & \multicolumn{2}{c}{scale} & \multicolumn{2}{c}{scale$'$}\\
\cmidrule(lr){2-3}\cmidrule(lr){4-5}\cmidrule(lr){6-7}\cmidrule(lr){8-9}
$\varepsilon$ & grid A & B/A & grid A & B/A & grid A & B/A & grid A & B/A \\
\midrule
    $10^{-4}$ & $1.4418\times10^{-2}$ & $1.02$ & $3.8710\times10^{-3}$ & $1.00$ & $1.0285\times10^{-2}$ & $1.00$ & $4.4239\times10^{-3}$ & $1.00$ \\
    $10^{-6}$ & $2.2013\times10^{-2}$ & $3.01$ & $3.6549\times10^{-3}$ & $1.04$ & $1.9628\times10^{-2}$ & $1.29$ & $2.8277\times10^{-3}$ & $1.30$ \\
    $10^{-8}$ & $2.5453\times10^{-2}$ & $3.32$ & $3.7549\times10^{-3}$ & $1.05$ & $2.1130\times10^{-2}$ & $1.34$ & $1.9161\times10^{-3}$ & $5.34$ \\
    $10^{-10}$ & $2.7966\times10^{-2}$ & $3.20$ & $3.8642\times10^{-3}$ & $1.05$ & $1.3333\times10^{-2}$ & $2.30$ & $1.6444\times10^{-3}$ & $8.45$ \\
\bottomrule
\end{tabular}
\caption{$E_\omega$ on the fixed uniform grid~A, and the ratio to the
layer-resolved value.  A ratio near one means that the uniform grid resolves
the error.}
\label{tab:sl-gridAB}
\end{table}

The two grids of \S\ref{subsec:diagnostics} separate the four trial spaces,
as Table~\ref{tab:sl-gridAB} shows. For the corrector variant the grids~A
and~B agree to within $5\%$ at every viscosity, so that no error is hidden
between the uniform nodes, since the prediction has no structure there that
the uniform grid does not resolve.
For the vanilla network the ratio is $3.2$ to $3.3$, as we established in \S\ref{sec:exp7},
and for the scale variant it reaches $2.3$ at $\varepsilon=10^{-10}$.
For the variant~A$'$ the ratio reaches $5.3$ at $\varepsilon=10^{-8}$ and
$8.5$ at $\varepsilon=10^{-10}$, so that most of the error of that variant is
invisible on the uniform grid. On the grid~A the variant~A$'$ reports the
smallest error of the four trial spaces at $\varepsilon\le10^{-8}$,
$1.9\times10^{-3}$ and $1.6\times10^{-3}$, while on the grid~B its error is
$2.6$ and $3.4$ times that of the corrector variant. This
hidden error is the spurious layer of Table~\ref{tab:slpinn-max}. The free
sub-network places structure where neither the residual points during the
training nor the uniform nodes during the evaluation observe it.

\subsubsection{Modal errors}
\label{subsec:exp8-modal}

\begin{table}[h!]
\centering
\scriptsize
\setlength{\tabcolsep}{1.6pt}
\begin{tabular}{ccccccccc}
\toprule
& \multicolumn{2}{c}{vanilla} & \multicolumn{2}{c}{corrector} & \multicolumn{2}{c}{scale} & \multicolumn{2}{c}{scale$'$}\\
\cmidrule(lr){2-3}\cmidrule(lr){4-5}\cmidrule(lr){6-7}\cmidrule(lr){8-9}
$\varepsilon$ & $E^{(0)}_\omega$ & $E^{(1)}_\omega$ & $E^{(0)}_\omega$ & $E^{(1)}_\omega$ & $E^{(0)}_\omega$ & $E^{(1)}_\omega$ & $E^{(0)}_\omega$ & $E^{(1)}_\omega$ \\
\midrule
\multicolumn{9}{l}{\emph{Stokes}}\\
    $10^{-4}$ & $1.009\times10^{-2}$ & $8.298\times10^{-2}$ & $6.628\times10^{-4}$ & $3.012\times10^{-2}$ & $2.987\times10^{-3}$ & $6.836\times10^{-2}$ & $1.362\times10^{-3}$ & $2.743\times10^{-2}$ \\
    $10^{-6}$ & $6.043\times10^{-2}$ & $2.543\times10^{-1}$ & $7.883\times10^{-4}$ & $2.216\times10^{-2}$ & $4.989\times10^{-3}$ & $2.304\times10^{-1}$ & $1.886\times10^{-3}$ & $1.599\times10^{-2}$ \\
    $10^{-8}$ & $7.853\times10^{-2}$ & $2.921\times10^{-1}$ & $5.887\times10^{-4}$ & $2.126\times10^{-2}$ & $6.742\times10^{-3}$ & $2.516\times10^{-1}$ & $6.877\times10^{-3}$ & $6.886\times10^{-2}$ \\
    $10^{-10}$ & $8.384\times10^{-2}$ & $3.013\times10^{-1}$ & $7.016\times10^{-4}$ & $2.209\times10^{-2}$ & $2.144\times10^{-2}$ & $1.980\times10^{-1}$ & $1.332\times10^{-2}$ & $2.304\times10^{-2}$ \\
\addlinespace
\multicolumn{9}{l}{\emph{Navier--Stokes}}\\
    $10^{-4}$ & $1.013\times10^{-2}$ & $1.146\times10^{-1}$ & $9.478\times10^{-4}$ & $3.353\times10^{-2}$ & $3.873\times10^{-3}$ & $6.470\times10^{-2}$ & $1.728\times10^{-3}$ & $3.873\times10^{-2}$ \\
    $10^{-6}$ & $6.203\times10^{-2}$ & $2.894\times10^{-1}$ & $7.967\times10^{-4}$ & $3.205\times10^{-2}$ & $3.998\times10^{-3}$ & $2.289\times10^{-1}$ & $2.354\times10^{-3}$ & $3.528\times10^{-2}$ \\
    $10^{-8}$ & $7.846\times10^{-2}$ & $3.063\times10^{-1}$ & $8.630\times10^{-4}$ & $3.187\times10^{-2}$ & $6.554\times10^{-3}$ & $2.534\times10^{-1}$ & $7.029\times10^{-3}$ & $5.990\times10^{-2}$ \\
    $10^{-10}$ & $8.484\times10^{-2}$ & $3.209\times10^{-1}$ & $8.388\times10^{-4}$ & $3.249\times10^{-2}$ & $1.360\times10^{-2}$ & $2.138\times10^{-1}$ & $1.257\times10^{-2}$ & $3.319\times10^{-2}$ \\
\bottomrule
\end{tabular}
\caption{Vorticity error by $x$-wavenumber, for the two systems.  On the Stokes
problem the mode $0$ carries $99.07\%$ of the exact energy and the mode $1$ the
remaining $0.93\%$, and on the Navier--Stokes problem $98.9\%$ and $1.1\%$
respectively; the layer is present in both of the modes.}
\label{tab:slpinn-modal}
\end{table}

Table~\ref{tab:slpinn-modal} resolves the error by the $x$-wavenumber, for the
present problem and for the nonlinear one of \S\ref{subsec:slpinn-nse} alike.
The mode $0$ carries $99.07\%$ of the energy of the exact vorticity, and the
mode $1$ the remaining $0.93\%$. The layer is present in both of the modes.
The present
problem therefore does not share the modal dilution of \S\ref{sec:failure}.
The corrector variant improves both of the modes, and improves the
dominant one most, since $E^{(0)}_\omega$ falls by a factor of $133$ at
$\varepsilon=10^{-8}$ and stays near $7\times10^{-4}$ throughout, while
$E^{(1)}_\omega$ remains at about $2.2\times10^{-2}$. The error of the layer
mode is thirty times that of the mode $0$, and it is uniform in
$\varepsilon$. Since the trial space contains the exact solution up to the order
$\mathcal{O}(\varepsilon)$ (\S\ref{subsec:exp8-ansatz}), we attribute this
uniform error to the optimizer; we have not tested this attribution by
changing the network or the training.

The modal errors of the two scale-only variants give the same conclusion
as their global ones, with one addition. For the variant~A the layer mode
stays at $E^{(1)}_\omega=0.20$ to $0.25$ for $\varepsilon\le10^{-6}$, against
$0.25$ to $0.30$ for the vanilla network, so that the scale alone brings
little improvement in the mode that carries the layer. For the variant~A$'$
the layer mode is as accurate as that of the corrector variant at
$\varepsilon=10^{-4}$, $10^{-6}$ and $10^{-10}$, with $E^{(1)}_\omega$ equal to
$2.7\times10^{-2}$, $1.6\times10^{-2}$ and $2.3\times10^{-2}$ against
$3.0\times10^{-2}$, $2.2\times10^{-2}$ and $2.2\times10^{-2}$, and three times
larger at $\varepsilon=10^{-8}$. The excess of the variant~A$'$ over the
corrector variant at $\varepsilon=10^{-10}$ lies in the mode $0$, where
$E^{(0)}_\omega=1.33\times10^{-2}$ against $7.0\times10^{-4}$, and in the
maximum error of $2.81$ of Table~\ref{tab:slpinn-max}. At this viscosity the
free amplitude misrepresents the $x$-independent part of the layer and not its
travelling part.

\subsubsection{Effect of layer-refined collocation points}
\label{subsec:exp8-sampling}

A possible cause of the failure of the variant~A is the sampling rather than
the enrichment, since the amplitude of the variant~A is free and almost no
collocation point lies inside the layer. We test this possibility directly, by
drawing half of the interior points inside
$0<y<24\sqrt{\varepsilon}$, or inside its mirror image at the far wall, in
place of uniformly, at $\varepsilon=10^{-8}$, with all of the other
settings unchanged.

The refined sampling does not improve the variant~A. The maximum pointwise
error improves by a factor of $1.3$, while every other measure deteriorates,
$E_\omega$ from $2.840\times10^{-2}$ to $3.486\times10^{-2}$, the layer mode from $2.516\times10^{-1}$ to $3.154\times10^{-1}$, and the mode
$0$ from $6.742\times10^{-3}$ to $8.022\times10^{-3}$. Moving half of the collocation
points into a region of measure
$\mathcal{O}(\sqrt{\varepsilon})$ leaves too few points for the outer problem, where the mode $0$ is determined. The gap to the corrector variant, a
factor of $8.8$ in $E_\omega$, does not close. The difficulty therefore does not lie in a lack of data alone. A profile and an amplitude which are not tied to the outer solution
are not determined by the analysis. Under the present sampling and
training, the residual does not constrain them enough to recover the correct
layers.

\subsubsection{Wall-normal profiles} \label{subsec:exp8-profiles}

\begin{figure}[h!]
\centering
\includegraphics[width=\textwidth]{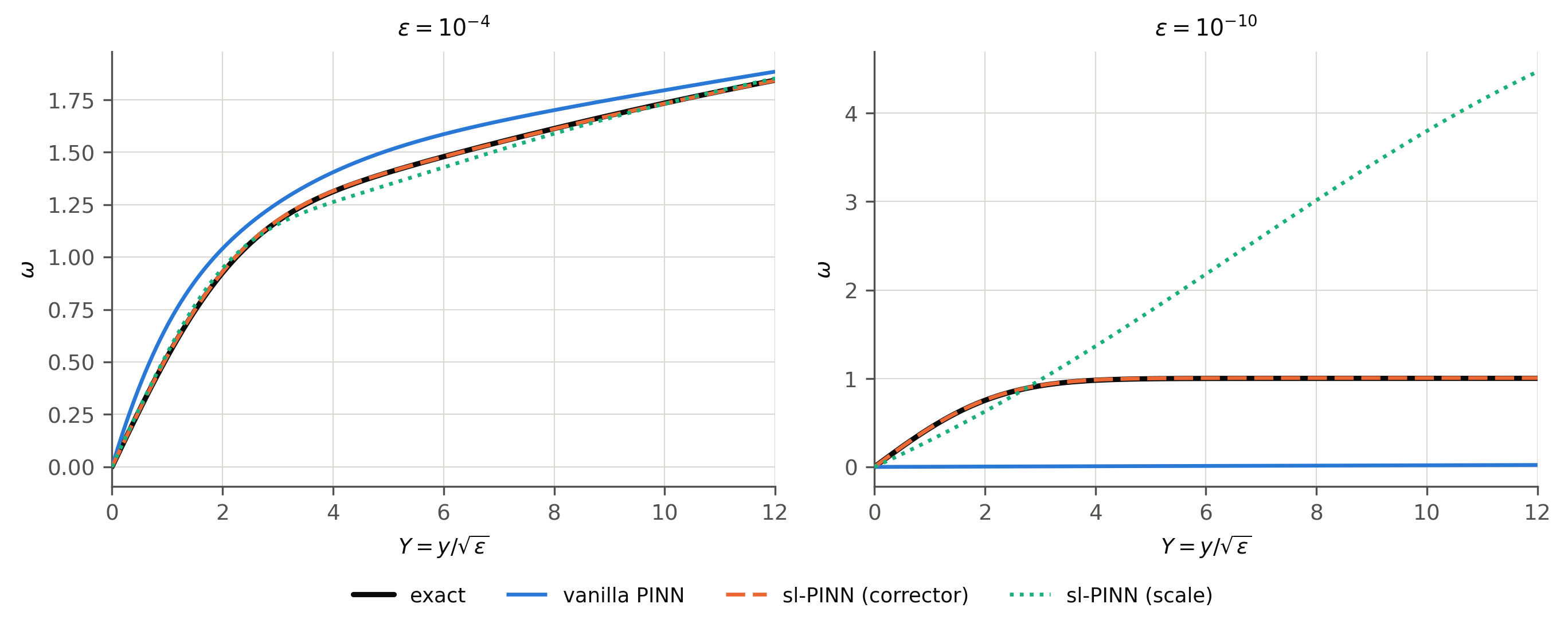}
\caption{Wall-normal vorticity profile at $x=\pi/2$, $t=0.5$, against the
stretched variable $Y=y/\sqrt{\varepsilon}$, so that the exact layer is the same
curve in both panels.  Left: $\varepsilon=10^{-4}$.  Right:
$\varepsilon=10^{-10}$, where the vanilla prediction is flat at zero across the
entire layer and the scale variant overshoots the exact profile, reaching
$4.5$ where the solution has levelled at $1$.  The corrector variant lies
on the exact curve at both viscosities.}
\label{fig:sl-profile}
\end{figure}

\begin{figure}[h!]
\centering
\includegraphics[width=\textwidth]{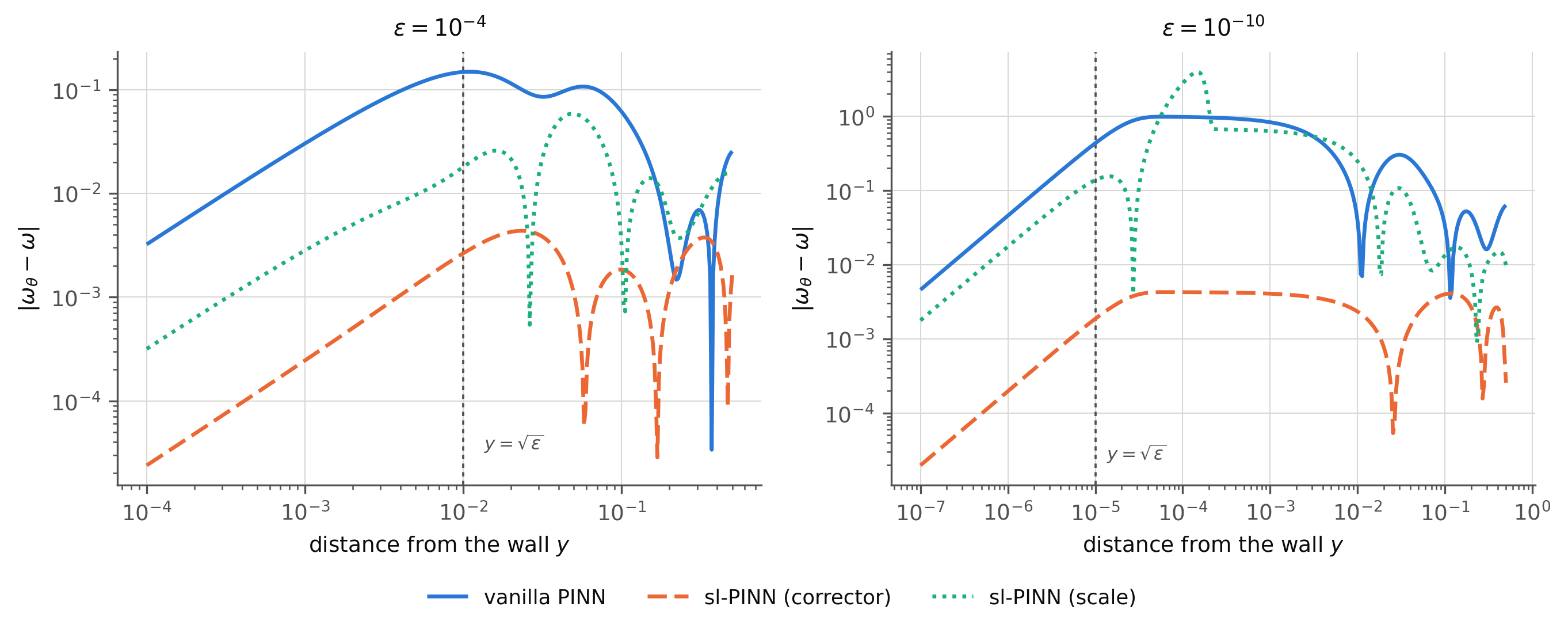}
\caption{Pointwise vorticity error against distance from the wall at $x=\pi/2$,
$t=0.5$.  Left: $\varepsilon=10^{-4}$.  Right: $\varepsilon=10^{-10}$.  The dotted
line marks $y=\sqrt{\varepsilon}$.  The corrector variant is below the other two
at every distance and at both viscosities; the scale variant's error peaks
\emph{inside} the layer, at about $2\sqrt{\varepsilon}$.}
\label{fig:sl-wallerr}
\end{figure}

\begin{figure}[h!]
\centering
\includegraphics[width=\textwidth]{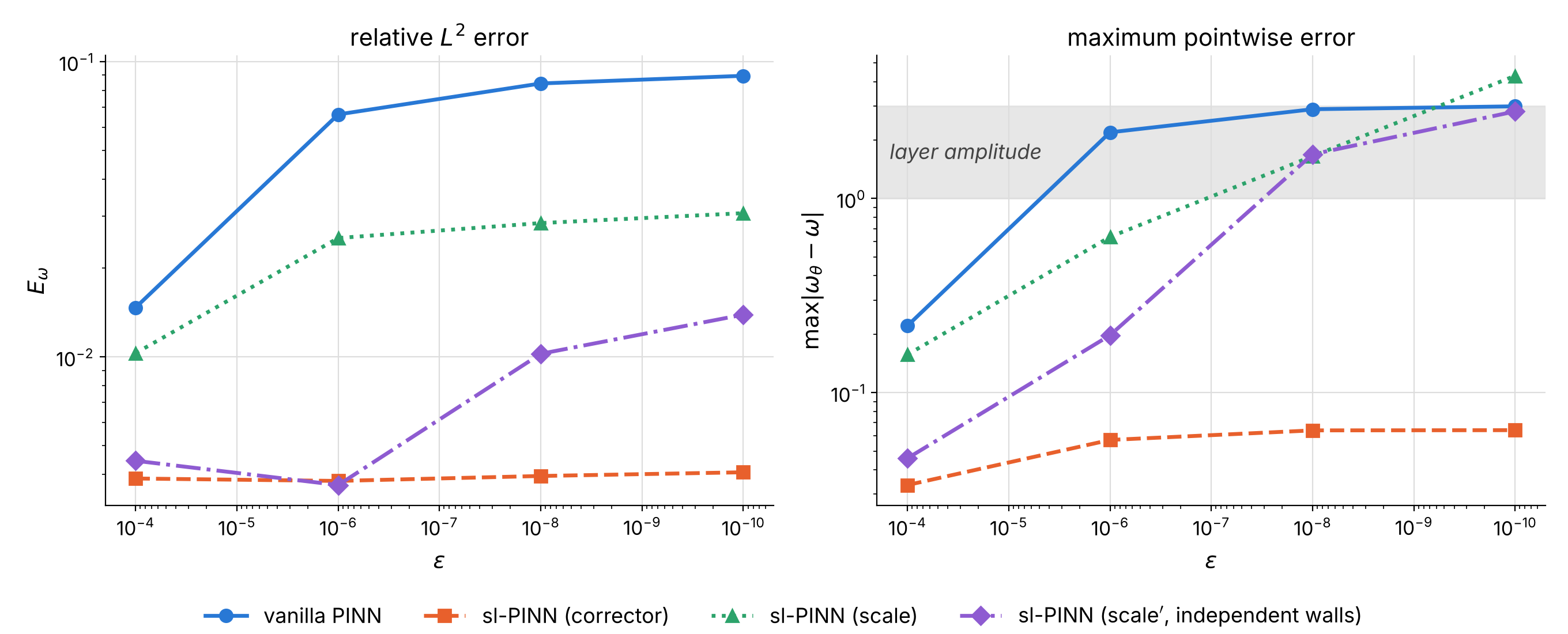}
\caption{Error against viscosity for the four trial spaces.  Left: global
relative $L^2$ vorticity error on the layer-resolved grid.  Right: maximum
pointwise vorticity error, with the band of the layer amplitude $W\in[1,3]$ shaded.
An error inside the band means the layer is absent from the prediction; an error
above it means a spurious layer has been produced.}
\label{fig:sl-sweep}
\end{figure}

Figures~\ref{fig:sl-profile}--\ref{fig:sl-sweep} show the same results in the
wall-normal direction.
Figure~\ref{fig:sl-profile} plots the wall-normal profile of the vorticity
against the stretched variable $Y=y/\sqrt{\varepsilon}$. On that variable the
exact layer is the same curve at every viscosity. At $\varepsilon=10^{-10}$
the vanilla prediction is flat at zero across the
entire layer, while the scale variant overshoots the exact profile and reaches
$4.5$ where the solution has levelled at $1$. The corrector
variant lies on the exact curve. Figure~\ref{fig:sl-wallerr} localizes the
pointwise error. For the corrector variant it is below the other two at every
distance from the
wall, whereas for the scale variant it peaks \emph{inside} the layer, at about
$2\sqrt{\varepsilon}$, where the spurious layer is located.
Figure~\ref{fig:sl-sweep} collects the whole sweep, with the band of the layer
amplitude shaded. An error inside the band means that the layer is absent from
the prediction. An error above the band means that a spurious layer has been
produced.

All three enrichments use the scale $\sqrt{\varepsilon}$, and all
impose the
boundary conditions exactly and structurally. The variant~A differs from
the variant~B in the profile of
the layer, in the amplitude law and in the
coupling of the two walls (\S\ref{subsec:exp8-ansatz}), and the
variant~A$'$ in the first two only. The coupling of the walls accounts for a
factor of $2.2$ to $6.9$ in $E_\omega$. The profile together with the
amplitude law accounts for a further factor of $2.6$ and $3.4$ at
$\varepsilon=10^{-8}$ and $10^{-10}$, and for the growth of the error as
$\varepsilon\to0$, which only the corrector-informed enrichment removes. The full
corrector-informed enrichment,
namely the analytic profile together with the amplitude law, therefore
outperforms both scale-only enrichments at $\varepsilon\le10^{-8}$. The scale alone, with a
free profile and a free amplitude, does
not suffice under the present architecture and training conditions. The
present experiments do not separate the contribution of the
profile from that of the amplitude law. All of the runs of this section use a
single seed; the study
of \S\ref{subsec:exp9-seed} measures the resulting uncertainty on this same
problem, and the separation between the variants~A and~B is larger than that
scatter.

Against the variant~A$'$, which removes the coupling of the walls, the
corrector-informed enrichment has four advantages in the present experiments.
First, its error is uniform in $\varepsilon$, with a drift of $7\%$ over six
decades, whereas the error of the variant~A$'$ grows by a factor of $3.1$.
Second, its maximum error stays at $2\%$ of the amplitude of the layer,
whereas that of the variant~A$'$ reaches $94\%$ at $\varepsilon=10^{-10}$.
Third, the uniform grid resolves its error to within $5\%$, whereas it
understates the error of the variant~A$'$ by a factor of up to $8.5$. Fourth,
it uses $8{,}894$ parameters against $10{,}818$, a training step costs about
$30\%$ less, and its layer part has a Laplacian in closed form, which the
finite-volume residual of \S\ref{sec:exp10} requires. We carry the same
construction, in the remainder of the section, to the Navier--Stokes system.

\subsection{Extension to the Navier--Stokes problem}
\label{subsec:slpinn-nse}

In \S\ref{subsec:exp8-results} we tested the corrector variant on the Stokes
problem, whose exact layer the variant embeds (\S\ref{subsec:exp8-ansatz}).
The uniformity in $\varepsilon$ therefore demonstrates that the mechanism
works, and does not yet demonstrate that the mechanism transfers to the
nonlinear system. In the present subsection we test the same trial function on
the Navier--Stokes problem. There the corrector obeys
$\partial_t\Theta+U_0\partial_x\Theta=\varepsilon\partial_y^2\Theta$ and the
wall data travel at the speed $U_0$, so that the embedded profile is a
prediction rather than a transcription.

In the construction of \S\ref{subsec:exp7-design} the
Navier--Stokes and the Stokes problems differ in exactly one place, namely in
the speed at which the wall data travel. The corrector obeys
\eqref{eq:corrector-eq}. For the flat channel with a
constant slip speed, this reads
$\partial_t\Theta+U_0\partial_x\Theta=\varepsilon\partial_y^2\Theta$
with the convection, and $\partial_t\Theta=\varepsilon\partial_y^2\Theta$
without it. The
wall data are therefore carried on $\zeta=x-U_0t$ in the first case, and stand
still in the second. In
the moving frame the profile is the same $\operatorname{erfc}$.

The ansatz \eqref{eq:ansatz-B} is therefore \emph{unchanged}, and the present
subsection tests whether the unchanged ansatz carries over to the nonlinear
system. The layer factors \eqref{eq:layer-factors} contain no
dependence on $x$ at all, and the structure of the layer in $x$ enters through
$a_0$ only, which is the trace of the regular network on the wall. The trial
function does not encode the fact that $a_0$ has now to travel at
the speed $U_0$. If the enrichment works, then the residual in the interior,
which now
carries the Jacobian, drives the wall trace to the correct travelling profile.

\subsubsection{Size of each term inside the layer}
\label{subsec:exp9-budget}

The nonlinear problem of \S\ref{sec:exp6} showed that such a test can reduce
to a repetition of the linear one: if the convection term is small inside the
layer, the two problems are repeated rather than compared.
Before the training, we measured the root-mean-square size of each term of
$\partial_t\omega+J(\psi,\omega)-\varepsilon\Delta\omega=f$ on a slab of thickness
$3\sqrt{\varepsilon}$ at the wall and, for contrast, on the interior, and we
obtained at $\varepsilon=10^{-8}$
\[
\begin{array}{lcccc}
 & \|\partial_t\omega\| & \|J\| & \|\varepsilon\Delta\omega\| & \|f\|\\[2pt]
\text{layer}    & 5.149\times10^{-1} & 4.333\times10^{-1} & 2.864\times10^{-1} & 1.355\times10^{-4}\\
\text{interior} & 2.071\times10^{-1} & 1.444\times10^{-1} & 4.838\times10^{-3} & 1.535\times10^{-1}
\end{array}
\]
Inside the layer the convection term is $1.5$ times the diffusion term, and it
is of the same order as the derivative in time. The forcing is four orders of
magnitude below all three. The three terms therefore cancel among themselves,
as they do in a boundary layer, and the nonlinearity is active inside the
layer.

The two quantities of Table~\ref{tab:blc-valid} take the same values on the
travelling problem. The training is identical to that of the Stokes runs of
\S\ref{subsec:exp8-setup} in every respect. One step costs $1.15$ times the
Stokes step, and the difference is the cost of the Jacobian.

\subsubsection{Numerical results}
\label{subsec:exp9-results}

\begin{table}[h!]
\centering
\begin{tabular}{ccccc}
\toprule
$\varepsilon$ & $E_\omega$ & $E_\psi$ & $E_u$ & $E_v$ \\
\midrule
\multicolumn{5}{l}{\emph{vanilla PINN}}\\
    $10^{-4}$ & $1.712628\times10^{-2}$ & $5.394385\times10^{-3}$ & $7.165500\times10^{-3}$ & $1.488572\times10^{-1}$ \\
    $10^{-6}$ & $6.936433\times10^{-2}$ & $9.308087\times10^{-3}$ & $1.500936\times10^{-2}$ & $2.749230\times10^{-1}$ \\
    $10^{-8}$ & $8.499346\times10^{-2}$ & $1.586798\times10^{-2}$ & $2.024412\times10^{-2}$ & $3.638793\times10^{-1}$ \\
    $10^{-10}$ & $9.154314\times10^{-2}$ & $1.170270\times10^{-2}$ & $1.627791\times10^{-2}$ & $2.864537\times10^{-1}$ \\
\midrule
\multicolumn{5}{l}{\emph{sl-PINN, corrector variant}}\\
    $10^{-4}$ & $5.038883\times10^{-3}$ & $1.694858\times10^{-3}$ & $2.289625\times10^{-3}$ & $6.210852\times10^{-2}$ \\
    $10^{-6}$ & $5.020663\times10^{-3}$ & $2.007713\times10^{-3}$ & $2.579070\times10^{-3}$ & $8.132246\times10^{-2}$ \\
    $10^{-8}$ & $5.161699\times10^{-3}$ & $2.161418\times10^{-3}$ & $2.804562\times10^{-3}$ & $8.637025\times10^{-2}$ \\
    $10^{-10}$ & $5.223168\times10^{-3}$ & $2.060115\times10^{-3}$ & $2.680246\times10^{-3}$ & $7.420707\times10^{-2}$ \\
\midrule
\multicolumn{5}{l}{\emph{sl-PINN, scale variant}}\\
    $10^{-4}$ & $9.603099\times10^{-3}$ & $4.207412\times10^{-3}$ & $5.233716\times10^{-3}$ & $1.536940\times10^{-1}$ \\
    $10^{-6}$ & $2.490693\times10^{-2}$ & $7.375447\times10^{-3}$ & $8.613681\times10^{-3}$ & $2.210055\times10^{-1}$ \\
    $10^{-8}$ & $2.853990\times10^{-2}$ & $5.069612\times10^{-3}$ & $6.795993\times10^{-3}$ & $1.649619\times10^{-1}$ \\
    $10^{-10}$ & $2.685775\times10^{-2}$ & $4.891086\times10^{-3}$ & $6.304541\times10^{-3}$ & $1.397489\times10^{-1}$ \\
\midrule
\multicolumn{5}{l}{\emph{sl-PINN, scale variant with independent walls}}\\
    $10^{-4}$ & $5.462548\times10^{-3}$ & $1.600996\times10^{-3}$ & $2.423003\times10^{-3}$ & $7.330789\times10^{-2}$ \\
    $10^{-6}$ & $5.243557\times10^{-3}$ & $1.561612\times10^{-3}$ & $2.238766\times10^{-3}$ & $7.992163\times10^{-2}$ \\
    $10^{-8}$ & $9.781289\times10^{-3}$ & $1.788090\times10^{-3}$ & $2.163545\times10^{-3}$ & $8.180310\times10^{-2}$ \\
    $10^{-10}$ & $1.349275\times10^{-2}$ & $1.046493\times10^{-3}$ & $1.362718\times10^{-3}$ & $6.065119\times10^{-2}$ \\
\bottomrule
\end{tabular}
\caption{Global space--time relative $L^2$ errors on the Navier--Stokes
problem of \S\ref{sec:exp7}, layer-resolved grid~B.}
\label{tab:nse-rel}
\end{table}

\emph{The error of the corrector variant is uniform in $\varepsilon$ on the
Navier--Stokes problem as well.} Across six decades
$E_\omega$ takes the values $5.039\times10^{-3}$, $5.021\times10^{-3}$,
$5.162\times10^{-3}$ and $5.223\times10^{-3}$, a total drift of $4.0\%$,
against $7.0\%$ for the same variant on the Stokes problem. The maximum
vorticity error runs from $2.9\%$ to $4.1\%$ of the largest
amplitude of the layer, and is likewise uniform in $\varepsilon$. The vanilla
maximum reaches
$99.6\%$ of that amplitude by $\varepsilon=10^{-10}$, which is the full
amplitude of the layer. The grids~A and~B agree to within $5\%$ for the
corrector variant at every
viscosity, the ratio being $1.00$, $1.03$, $1.04$ and $1.05$. Nothing is
hidden between the uniform nodes. For the vanilla network the same ratio runs
$1.01$, $2.50$, $3.12$ and $3.26$.

\emph{The scale variants degrade as on the Stokes problem.} The
error of the variant~A grows from
$9.603\times10^{-3}$ to $2.686\times10^{-2}$, and its maximum error reaches
$3.07$, which exceeds the largest amplitude of the layer. The error of the
variant~A$'$ grows from $5.463\times10^{-3}$ to $1.349\times10^{-2}$, with a
maximum error of $2.94$ at $\varepsilon=10^{-10}$, and it equals the error of
the corrector variant at $\varepsilon=10^{-4}$ and $10^{-6}$ only, as on the
Stokes problem. The grid~B to grid~A ratio of the variant~A$'$ grows through
$1.00$, $1.19$ and $3.12$ to $6.11$, which is worse than the
$3.26$ of the vanilla network. A study on a uniform grid would therefore
understate the failure of this variant by more than it understates that of the
unenriched one. The qualitative conclusion of \S\ref{subsec:exp8-sampling},
that the scale alone, with a free profile and amplitude, deteriorates as
$\varepsilon\to0$, is thus reproduced on the nonlinear system. The wall-normal
profiles repeat those of Figure~\ref{fig:sl-profile}, with the wall data now
travelling at the speed $U_0$.

\subsection{Comparison of the Stokes and Navier--Stokes results}
\label{subsec:exp9-transfer}

\begin{table}[h!]
\centering
\scriptsize
\setlength{\tabcolsep}{2pt}
\begin{tabular}{cccccccccc}
\toprule
& \multicolumn{4}{c}{$E_\omega$, Stokes} & \multicolumn{4}{c}{$E_\omega$, Navier--Stokes} & \\
\cmidrule(lr){2-5}\cmidrule(lr){6-9}
$\varepsilon$ & vanilla & corr. & scale & scale$'$ & vanilla & corr. & scale & scale$'$ & ratio \\
\midrule
    $10^{-4}$ & $1.47\times10^{-2}$ & $3.87\times10^{-3}$ & $1.03\times10^{-2}$ & $4.44\times10^{-3}$ & $1.71\times10^{-2}$ & $5.04\times10^{-3}$ & $9.60\times10^{-3}$ & $5.46\times10^{-3}$ & $1.30$ \\
    $10^{-6}$ & $6.63\times10^{-2}$ & $3.80\times10^{-3}$ & $2.53\times10^{-2}$ & $3.68\times10^{-3}$ & $6.94\times10^{-2}$ & $5.02\times10^{-3}$ & $2.49\times10^{-2}$ & $5.24\times10^{-3}$ & $1.32$ \\
    $10^{-8}$ & $8.44\times10^{-2}$ & $3.95\times10^{-3}$ & $2.84\times10^{-2}$ & $1.02\times10^{-2}$ & $8.50\times10^{-2}$ & $5.16\times10^{-3}$ & $2.85\times10^{-2}$ & $9.78\times10^{-3}$ & $1.31$ \\
    $10^{-10}$ & $8.96\times10^{-2}$ & $4.06\times10^{-3}$ & $3.07\times10^{-2}$ & $1.39\times10^{-2}$ & $9.15\times10^{-2}$ & $5.22\times10^{-3}$ & $2.69\times10^{-2}$ & $1.35\times10^{-2}$ & $1.29$ \\
\bottomrule
\end{tabular}
\caption{The four trial spaces on the Stokes and on the Navier--Stokes
problem. The last column is the Navier--Stokes error of the corrector variant
divided by its Stokes error.}
\label{tab:nse-transfer}
\end{table}

\begin{figure}[h!]
\centering
\includegraphics[width=\textwidth]{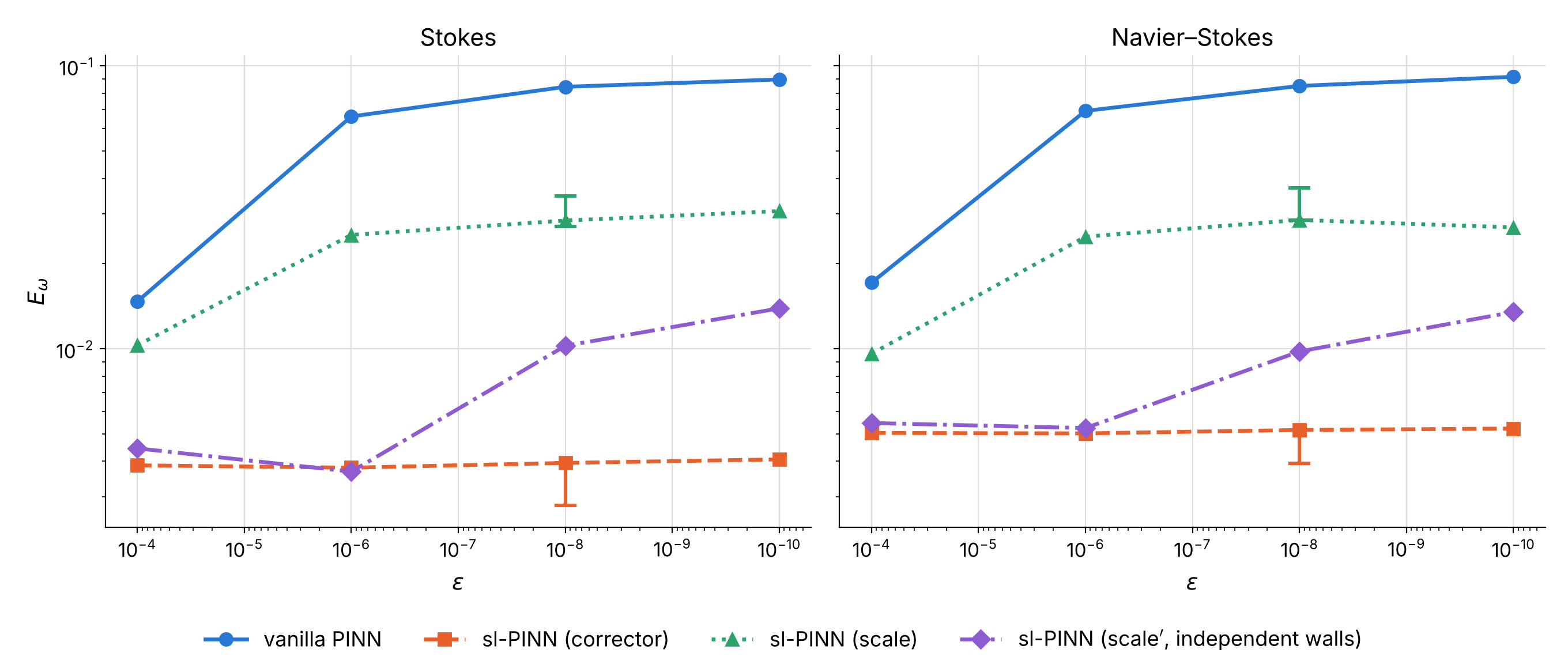}
\caption{The four trial spaces on the Stokes (left) and on the Navier--Stokes
(right) problem, on one $E_\omega$ scale.  The bars at $\varepsilon=10^{-8}$
are the full range over three random seeds (\S\ref{subsec:exp9-seed}),
for the corrector variant and for the scale variant.}
\label{fig:nse-transfer}
\end{figure}

Table~\ref{tab:nse-transfer} and Figure~\ref{fig:nse-transfer} contain the
result of the present section. The two panels of the figure show the same
behavior, and the last column of
the table quantifies the agreement. The Navier--Stokes error of the corrector
variant is
$1.30$, $1.32$, $1.31$ and $1.29$ times its Stokes error at the four
viscosities, all at the same random seed. \emph{The cost of the nonlinearity is a
multiplicative factor that does not depend on $\varepsilon$.} The value of
that factor is less sharply determined than these four numbers suggest. Over
the three seeds of \S\ref{subsec:exp9-seed} at $\varepsilon=10^{-8}$ the
ratio takes the values $1.31$, $1.63$ and $1.33$, with mean $1.42$ and a full
range equal to $22\%$ of the mean. We therefore quote the factor as about
$1.4$.
The $\varepsilon$-independence, on the other hand, is a paired comparison at a
fixed seed, in which the collocation points and the initialization are the
same for the two equations. The scatter of the ratio across the seeds does not
enter this comparison, although we have not repeated the sweep in
$\varepsilon$ at the other seeds.

This constancy is the main point of the comparison. A corrector taken from the
linear analysis and inserted into a nonlinear
solver could have failed in two ways. The error could have grown like
$1/\varepsilon$ as the layer became thinner and the neglected nonlinear terms
inside it grew relative to the retained ones. The error could also have jumped
once the Jacobian was included. We observe that neither failure occurs, and
that the cost of the convection
is a fixed multiplicative factor which does not increase as $\varepsilon$
decreases.

The vanilla column serves as a control.
The ratio of the Navier--Stokes to the Stokes error of the vanilla network
falls from $1.165$ to
$1.02$, so that the two equations become \emph{indistinguishable} as
$\varepsilon\to0$. The vanilla network fails completely on both equations, as
noted in \S\ref{subsec:exp7-compare}. The ratio of the corrector variant does
not converge to one, since neither method is saturated and the
enriched solver still resolves the layer at $\varepsilon=10^{-10}$. The factor
of about $1.4$ is therefore measured on a method which resolves the layer at
every viscosity of the sweep.

\subsubsection{Modal errors}
\label{subsec:exp9-modal}

In the lower block of Table~\ref{tab:slpinn-modal} the corrector variant holds
$E^{(0)}_\omega$ near $8.6\times10^{-4}$ and $E^{(1)}_\omega$ near
$3.2\times10^{-2}$ at every viscosity. Both are uniform in $\varepsilon$, and
the ratio between them is about $38$,
which repeats the structure seen on the Stokes problem. The error that
survives the enrichment is concentrated in the mode which carries the layer.
We attribute this surviving error to the optimizer, as on the Stokes problem,
rather than to a failure to build the layer.

\subsection{Dependence on the random seed}
\label{subsec:exp9-seed}

All of the results reported so far in the present article come from one seed.
Within a fixed seed the comparison across $\varepsilon$ is a paired
comparison. The collocation points and the initialization are the same draw,
so that a change in the error is attributable to $\varepsilon$ alone. The
absolute level of any one number is, however, not determined by a single seed.
We
ran two further seeds at $\varepsilon=10^{-8}$, for the variants~A and~B,
for both of the equations, and for the sl-FVNN of \S\ref{sec:exp10}, and
we collect the results in Table~\ref{tab:seeds}.

\begin{table}[h!]
\centering
\footnotesize
\setlength{\tabcolsep}{3pt}
\begin{tabular}{llcccc}
\toprule
method & seed & $E_\omega$ & $\max|\omega_\theta-\omega|$ & $E^{(1)}_\omega$ & $E^{(0)}_\omega$ \\
\midrule
\multicolumn{6}{l}{\emph{Stokes}}\\
    sl-PINN, corrector & 1234 & $3.948\times10^{-3}$ & $6.390\times10^{-2}$ & $2.126\times10^{-2}$ & $5.887\times10^{-4}$ \\
     & 7 & $2.787\times10^{-3}$ & $3.820\times10^{-2}$ & $1.788\times10^{-2}$ & $6.300\times10^{-4}$ \\
     & 20260916 & $2.963\times10^{-3}$ & $4.101\times10^{-2}$ & $1.494\times10^{-2}$ & $6.891\times10^{-4}$ \\
\addlinespace[2pt]
    sl-PINN, scale & 1234 & $2.840\times10^{-2}$ & $1.653\times10^{0}$ & $2.516\times10^{-1}$ & $6.742\times10^{-3}$ \\
     & 7 & $3.465\times10^{-2}$ & $1.052\times10^{0}$ & $3.095\times10^{-1}$ & $8.684\times10^{-3}$ \\
     & 20260916 & $2.709\times10^{-2}$ & $1.046\times10^{0}$ & $2.385\times10^{-1}$ & $6.195\times10^{-3}$ \\
\addlinespace[2pt]
    sl-FVNN, corrector & 1234 & $2.216\times10^{-3}$ & $3.980\times10^{-2}$ & $1.255\times10^{-2}$ & $4.800\times10^{-4}$ \\
     & 7 & $4.537\times10^{-3}$ & $7.628\times10^{-2}$ & $3.228\times10^{-2}$ & $6.647\times10^{-4}$ \\
     & 20260916 & $3.771\times10^{-3}$ & $6.051\times10^{-2}$ & $2.482\times10^{-2}$ & $1.000\times10^{-3}$ \\
\midrule
\multicolumn{6}{l}{\emph{Navier--Stokes}}\\
    sl-PINN, corrector & 1234 & $5.162\times10^{-3}$ & $1.174\times10^{-1}$ & $3.187\times10^{-2}$ & $8.630\times10^{-4}$ \\
     & 7 & $4.531\times10^{-3}$ & $1.101\times10^{-1}$ & $2.194\times10^{-2}$ & $1.126\times10^{-3}$ \\
     & 20260916 & $3.934\times10^{-3}$ & $1.350\times10^{-1}$ & $1.798\times10^{-2}$ & $9.474\times10^{-4}$ \\
\addlinespace[2pt]
    sl-PINN, scale & 1234 & $2.854\times10^{-2}$ & $1.856\times10^{0}$ & $2.534\times10^{-1}$ & $6.554\times10^{-3}$ \\
     & 7 & $3.705\times10^{-2}$ & $8.875\times10^{-1}$ & $3.371\times10^{-1}$ & $6.496\times10^{-3}$ \\
     & 20260916 & $3.184\times10^{-2}$ & $1.619\times10^{0}$ & $2.822\times10^{-1}$ & $8.362\times10^{-3}$ \\
\addlinespace[2pt]
    sl-FVNN, corrector & 1234 & $4.887\times10^{-3}$ & $1.690\times10^{-1}$ & $3.394\times10^{-2}$ & $8.039\times10^{-4}$ \\
     & 7 & $5.628\times10^{-3}$ & $1.550\times10^{-1}$ & $4.179\times10^{-2}$ & $9.764\times10^{-4}$ \\
     & 20260916 & $6.263\times10^{-3}$ & $1.699\times10^{-1}$ & $4.710\times10^{-2}$ & $1.374\times10^{-3}$ \\
\bottomrule
\end{tabular}
\caption{Errors at $\varepsilon=10^{-8}$ for three random seeds, which set the collocation points and the initialization. The seed $1234$ is the seed of all of the other tables. The sl-FVNN is the method of \S\ref{sec:exp10}.}
\label{tab:seeds}
\end{table}

We measure the spread over the three seeds as the full range as a percentage
of the mean. The spread is $25$ to $36\%$ in $E_\omega$ for the sl-PINN variants,
$66\%$ for the sl-FVNN on the Stokes problem, and as much as $67\%$ in the maximum
error. Computed seed by seed, the ratio of the Navier--Stokes to the Stokes
error of the corrector variant is $1.31$, $1.63$ and $1.33$, with mean $1.42$.
Three numbers do not estimate a distribution, and we give these values as an
order of magnitude and not as a confidence interval. Three conclusions follow.

First, \emph{the conclusions of the present section lie well outside that
spread}. When the means are compared, the gap between the corrector and the scale
variants at $\varepsilon=10^{-8}$ is a factor of $9.3$ on the Stokes problem
and $7.2$ on the Navier--Stokes problem. Comparing the \emph{worst} of the
corrector seeds against the \emph{best} of the scale seeds, it is still $6.9$
and $5.5$. The scatter that we observe does not reorder the two variants.

Second, \emph{the variation in $\varepsilon$ is smaller than that spread}. The
total drift of the corrector variant across six
decades is $4.0\%$ for the Navier--Stokes problem and $7.0\%$ for the Stokes
problem, while the spread across the seeds at a single viscosity is $27\%$ and
$36\%$. The dependence of $E_\omega$ on $\varepsilon$ therefore lies below the
resolution of an experiment with a single draw. The vanilla error, by
contrast, moves by a factor of six over the same range.

Third, \emph{the sl-FVNN and the sl-PINN are of the same accuracy within
the spread}. Over the three seeds the sl-FVNN gives $2.22\times10^{-3}$,
$4.54\times10^{-3}$ and $3.77\times10^{-3}$ on the Stokes problem and
$4.89\times10^{-3}$, $5.63\times10^{-3}$ and $6.26\times10^{-3}$ on the
Navier--Stokes problem. Its mean is $1.09$ times the mean of the sl-PINN on
the Stokes problem and $1.23$ times on the Navier--Stokes problem, and the two
ranges overlap on the Stokes problem. At the seed $1234$ the sl-FVNN
is more accurate than the sl-PINN on the Stokes problem by a factor of $1.8$
(Table~\ref{tab:fv-rel}). This factor is a property of that draw, and we quote
the two methods as equally accurate to within a factor of $1.5$.

Accordingly, the factors quoted throughout the present section are accurate to
about $\pm30\%$, whereas the qualitative statements, uniform against growing,
bounded against
saturated, and a constant penalty for the nonlinearity, lie
far outside that uncertainty.

\subsection{Limits of the conclusion}
\label{subsec:exp9-summary}

The corrector-informed enrichment keeps the vorticity error uniform in
$\varepsilon$ on the Navier--Stokes system, at a cost of about $1.4$ which
does not depend on $\varepsilon$ (\S\ref{subsec:exp9-transfer}). We note the
limitation of this conclusion. The slip speed $U_0$ of the present
construction is constant, so that the corrector equation
\eqref{eq:corrector-eq} is solvable in closed form and the Navier--Stokes
corrector is the Stokes profile in the travelling frame $\zeta=x-U_0t$. The
experiments of this section therefore test the corrector of the linear
analysis, namely the profile together with the amplitude law, on the nonlinear
system, and the uniformity in $\varepsilon$ and the constant cost are
statements about that corrector.
For a slip speed which varies along the wall, the corrector is a Duhamel
integral rather than a single $\operatorname{erfc}$, and the analytic profile
is a leading-order approximation with a remainder of the order
$\mathcal{O}(\sqrt{\varepsilon})$. The refinement term of
\S\ref{subsec:exp8-ansatz} is designed to carry that remainder. The corrector
remains the solution of a linear equation under the slip
condition, since the velocity of the layer is of the order
$\mathcal{O}(\sqrt{\varepsilon})$ and the layer does not transport itself at
the leading order. Under the no-slip condition the corrector solves the
Prandtl equations instead. We are treating
the varying slip speed and the no-slip condition in a forthcoming work.

\section{Singular-layer Finite-Volume Neural Networks}
\label{sec:exp10}

In \S\ref{subsec:fv-residual} we replaced the Laplacian in the residual by a
finite volume flux difference, which reduced the cost of a training step at a moderate loss of accuracy on a
smooth problem (Table~\ref{tab:fv-stokes-rel}). In \S\ref{sec:slpinn} we
embedded the corrector in the trial function, which made the error uniform in
$\varepsilon$ at about twice the cost of an unenriched step. The two devices
have been used separately so far. In the present section we combine them into
the singular layer FVNN, and we ask whether the finite-volume residual keeps
the $\varepsilon$-uniform accuracy of the sl-PINN at a lower cost of the
derivatives.

\subsection{Consistency of the discrete Laplacian in the presence of a layer}
\label{subsec:exp10-why}

The finite-volume residual and the enrichment do not combine directly, since
the stencil is consistent only on a grid which resolves the field. The grid of
\S\ref{subsec:fv-residual} does not resolve the present layer at any viscosity
of the study. With $\Delta y=1/20$ and a layer of width
$\sqrt{\varepsilon}$, the ratio $\Delta y/\sqrt{\varepsilon}$ is $5$ at
$\varepsilon=10^{-4}$ and $5{,}000$ at $10^{-10}$, so that the layer is
thinner than one cell at every viscosity of the sweep.

The consequence can be measured before any network is trained, by applying the
discrete operator to the exact solution.

\begin{table}[h!]
\centering
\begin{tabular}{cccc}
\toprule
$\varepsilon$ & $\Delta y/\sqrt{\varepsilon}$ & whole field & split \\
\midrule
    $10^{-4}$ & $5$ & $2.7074\times10^{0}$ & $5.8186\times10^{-5}$ \\
    $10^{-6}$ & $50$ & $2.5357\times10^{2}$ & $4.2388\times10^{-4}$ \\
    $10^{-8}$ & $500$ & $2.5357\times10^{3}$ & $4.2388\times10^{-4}$ \\
    $10^{-10}$ & $5000$ & $2.5357\times10^{4}$ & $4.2388\times10^{-4}$ \\
\bottomrule
\end{tabular}
\caption{Relative error of the discrete Laplacian,
$\|\Delta_h\omega-\Delta\omega\|/\|\Delta\omega\|$, on the exact solution and
the $50\times21\times10$ grid of \S\ref{subsec:fv-residual}, where $\|\cdot\|$
is the root mean square over the interior nodes $j=1,\dots,N_y-2$ of the grid.  ``Whole field''
is the stencil
applied to $\omega$ as it stands; ``split'' is the stencil on the outer part with
the corrector's Laplacian taken exactly.  The Stokes and the Navier--Stokes
problems give the same numbers, since their exact solutions share the
construction of \S\ref{subsec:exp7-design}.}
\label{tab:fv-consistency}
\end{table}

Applied to the entire field, the stencil misrepresents the Laplacian by a
factor which grows like $\varepsilon^{-1/2}$ and reaches $2.5\times10^{4}$ at
$\varepsilon=10^{-10}$. This error is concentrated in the first row of cells
at each wall, where the flux difference \eqref{eq:fv-div} references the
wall-normal derivative at the wall, which is of the order
$\varepsilon^{-1/2}$. Minimizing a residual which the exact solution
fails to satisfy by four orders
of magnitude is of no use, since training against it moves the network away
from the exact solution. Refining the grid does not help either, because the
consistency would
require $\Delta y\lesssim\sqrt{\varepsilon}$, and hence $N_y\gtrsim10^{5}$ at
$\varepsilon=10^{-10}$.

The enrichment removes this obstruction.  The corrector-informed trial
function \eqref{eq:ansatz-B} is already written
as a sum of a regular part and a layer part, $\omega_\theta=\omega_{\theta,\rm
reg}+\Theta_\theta$. Here $\omega_{\theta,\rm reg}=a$ is given by the split
\eqref{eq:reg-split} and $\Theta_\theta$ is given by \eqref{eq:Theta-net}. We
discretize the two parts differently,
\begin{equation}
\Delta\omega_\theta \;\longrightarrow\;
\Delta_h\,\omega_{\theta,\rm reg} \;+\; \Delta\Theta_\theta ,
\label{eq:fv-split}
\end{equation}
and likewise for $\psi_\theta$. The stencil now differences the fields which
vary on the $\mathcal{O}(1)$
scale only, while the part which varies on the scale $\sqrt{\varepsilon}$ is a
closed-form expression whose Laplacian is taken exactly. The regular part is a
network of the unstretched variables, and nothing in the residual rewards a
structure on the scale $\sqrt{\varepsilon}$ in it, since the layer is already
carried by $\Theta_\theta$. The ratio of at most $1.04$ of the grid~B to the grid~A for
the sl-FVNN in Table~\ref{tab:fv-max} confirms that the trained regular part
contains no such structure. The right-hand column of
Table~\ref{tab:fv-consistency} is the
result, namely $4.2\times10^{-4}$, which is the truncation error of the order
$\mathcal{O}(\Delta y^2)$ of the outer solution, and which does not depend on
$\varepsilon$.

\begin{remark}
The Laplacian alone is discretized. The convection term carries no second
derivative. Replacing the convection term by a difference brings no benefit,
and it would
re-impose the resolution limit of the grid on a term which does not require
it. We evaluate it pointwise, as in
\S\ref{sec:slpinn}. This pointwise evaluation departs from the conservative
treatment of the
fluxes in \S\ref{subsubsec:fv-nonlinear}, for the reason just given.
\end{remark}

\begin{remark}
We apply the split to the corrector variant only. The regular part of the
corrector variant is smooth on the scale of the grid by construction, and the
Laplacian of its layer part is available in closed form. The layer parts
of the two scale variants are free sub-networks of
$Y=y/\sqrt{\varepsilon}$, whose Laplacians would have to be evaluated by
automatic differentiation. The split would then remove the saving in the cost
which is the purpose of the finite-volume residual.
\end{remark}

\subsection{Numerical results}
\label{subsec:exp10-results}

We compare four methods, namely the two trial spaces of \S\ref{sec:slpinn},
each with the pointwise and with the finite-volume residual.
All of the settings are those of \S\ref{subsec:exp8-setup}, and the
finite-volume residual uses the grid of \S\ref{subsec:fv-residual}. The column
FVNN is the whole-field stencil of
Table~\ref{tab:fv-consistency},
which we include in order to confirm by computation the failure predicted in
\S\ref{subsec:exp10-why}.

\begin{table}[h!]
\centering
\small
\begin{tabular}{ccccc}
\toprule
$\varepsilon$ & vanilla PINN & FVNN & sl-PINN & sl-FVNN \\
\midrule
\multicolumn{5}{l}{\emph{Stokes}}\\
    $10^{-4}$ & $1.469593\times10^{-2}$ & $1.484684\times10^{-2}$ & $3.870849\times10^{-3}$ & $2.076824\times10^{-3}$ \\
    $10^{-6}$ & $6.627423\times10^{-2}$ & $2.118641\times10^{-2}$ & $3.798262\times10^{-3}$ & $2.240233\times10^{-3}$ \\
    $10^{-8}$ & $8.438753\times10^{-2}$ & $5.280938\times10^{-2}$ & $3.947600\times10^{-3}$ & $2.215629\times10^{-3}$ \\
    $10^{-10}$ & $8.957216\times10^{-2}$ & $4.607932\times10^{-2}$ & $4.063590\times10^{-3}$ & $2.131244\times10^{-3}$ \\
\midrule
\multicolumn{5}{l}{\emph{Navier--Stokes}}\\
    $10^{-4}$ & $1.712628\times10^{-2}$ & $1.755776\times10^{-2}$ & $5.038883\times10^{-3}$ & $3.402620\times10^{-3}$ \\
    $10^{-6}$ & $6.936433\times10^{-2}$ & $2.344544\times10^{-2}$ & $5.020663\times10^{-3}$ & $4.572535\times10^{-3}$ \\
    $10^{-8}$ & $8.499346\times10^{-2}$ & $3.944250\times10^{-2}$ & $5.161699\times10^{-3}$ & $4.887293\times10^{-3}$ \\
    $10^{-10}$ & $9.154314\times10^{-2}$ & $5.176555\times10^{-2}$ & $5.223168\times10^{-3}$ & $4.791415\times10^{-3}$ \\
\bottomrule
\end{tabular}
\caption{Global space--time relative $L^2$ vorticity errors, layer-resolved
grid~B. The first and third columns repeat Tables~\ref{tab:sl-rel}
and~\ref{tab:nse-rel}.} \label{tab:fv-rel} \end{table}

\begin{figure}[h!]
\centering
\includegraphics[width=\textwidth]{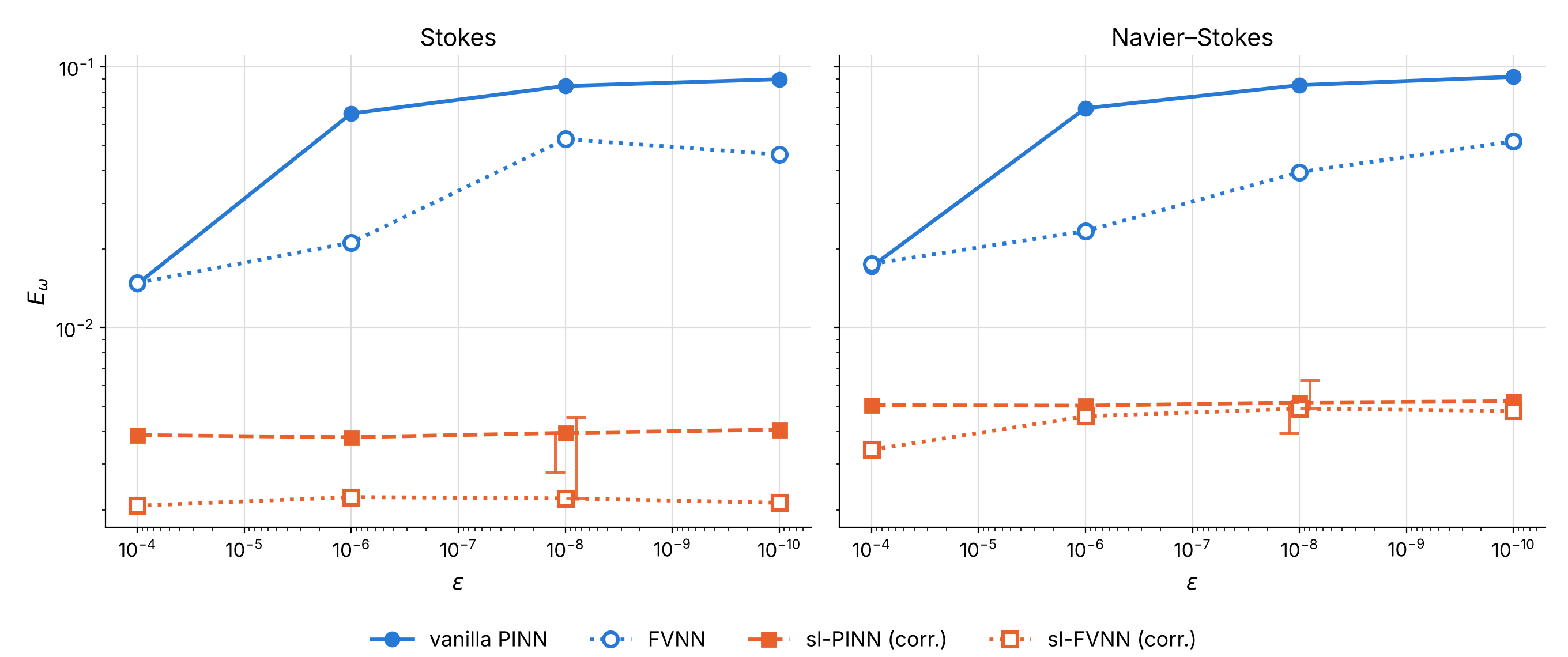}
\caption{Four methods across six decades.  Each trial space keeps its hue; the
finite-volume member of the pair is hollow and dotted.  Both residuals of the
enriched space give errors uniform in $\varepsilon$;
both residuals of the unenriched space give errors which grow. The bars
at $\varepsilon=10^{-8}$ are the full range over three random seeds
(\S\ref{subsec:exp9-seed}), drawn beside the markers of the sl-PINN and of the
sl-FVNN.}
\label{fig:fv-sweep}
\end{figure}

Table~\ref{tab:fv-rel} collects the global errors, and
Figure~\ref{fig:fv-sweep} plots the same four methods across the six decades.

\emph{Both of the residuals of the enriched trial space give errors uniform in
$\varepsilon$.} On the
Stokes problem the sl-FVNN gives $2.077$, $2.240$, $2.216$ and $2.131$
($\times10^{-3}$) across six decades, which is a drift of $8\%$ against the
$7\%$ of the sl-PINN, while being uniformly \emph{more} accurate at this
seed, by a factor
of $1.70$ to $1.91$. On the Navier--Stokes problem the finite-volume residual
is uniform in
$\varepsilon$ as well, with a smaller improvement, the values being $3.403$,
$4.573$, $4.887$ and
$4.791$ ($\times10^{-3}$), a drift of $44\%$, with an improvement over the
sl-PINN of $1.48$ at $\varepsilon=10^{-4}$ which falls to $1.09$ at
$10^{-10}$. The seed study of \S\ref{subsec:exp9-seed} puts these factors
in perspective. Over three seeds at $\varepsilon=10^{-8}$ the mean error of
the sl-FVNN is $1.09$ times that of the sl-PINN on the Stokes problem and
$1.23$ times on the Navier--Stokes problem. The two residuals are therefore
equally accurate to within a factor of $1.5$, and the finite-volume residual
is as accurate and cheaper on both problems. The drift of $44\%$ is of the same order as the seed spread of $27\%$ measured
in \S\ref{subsec:exp9-seed} for the sl-PINN, and with a single seed we cannot
decide whether it is a trend in $\varepsilon$ or scatter.

\emph{Both of the residuals of the unenriched trial space fail.} The FVNN
reaches $4.61\times10^{-2}$ on the Stokes problem and $5.18\times10^{-2}$ on
the Navier--Stokes problem at $\varepsilon=10^{-10}$, which is half the
value $9\times10^{-2}$ of the vanilla PINN, with
a maximum vorticity error of $2.95$, which is the full amplitude of the layer.
Its layer-mode error $E^{(1)}_\omega$ is $0.18$ and $0.24$, against
$0.30$ and $0.32$ for the vanilla PINN. The stencil misses the layer as the
pointwise residual does, and the smaller global error comes from the mode
$0$, whose error is half that of the vanilla PINN.
Table~\ref{tab:fv-consistency} predicted this failure.

\begin{table}[h!]
\centering
\small
\begin{tabular}{ccccccc}
\toprule
& \multicolumn{2}{c}{$\max|\omega_\theta-\omega|$, FVNN} &
\multicolumn{2}{c}{$\max|\omega_\theta-\omega|$, sl-FVNN} &
\multicolumn{2}{c}{$E_\omega$, grid B/A}\\
\cmidrule(lr){2-3}\cmidrule(lr){4-5}\cmidrule(lr){6-7}
$\varepsilon$ & Stokes & NSE & Stokes & NSE & FVNN & sl-FVNN \\
\midrule
    $10^{-4}$ & $2.5887\times10^{-1}$ & $4.3912\times10^{-1}$ & $1.8789\times10^{-2}$ & $6.5761\times10^{-2}$ & $1.02$ & $1.00$ \\
    $10^{-6}$ & $1.0190\times10^{0}$ & $1.6498\times10^{0}$ & $3.3705\times10^{-2}$ & $1.5395\times10^{-1}$ & $1.65$ & $1.03$ \\
    $10^{-8}$ & $2.6346\times10^{0}$ & $2.6066\times10^{0}$ & $3.9796\times10^{-2}$ & $1.6902\times10^{-1}$ & $3.22$ & $1.04$ \\
    $10^{-10}$ & $2.9536\times10^{0}$ & $2.9552\times10^{0}$ & $3.5400\times10^{-2}$ & $1.6383\times10^{-1}$ & $3.82$ & $1.04$ \\
\bottomrule
\end{tabular}
\caption{Maximum absolute vorticity errors against the layer amplitude
$W\in[1,3]$, and the ratio of $E_\omega$ on the layer-resolved grid~B to its
value on the uniform grid~A (Stokes problem). The ratio for the FVNN reaches
$3.8$, against
$3.2$ for the vanilla network, so that the uniform grid conceals the failure of
the two unenriched approximations alike.}
\label{tab:fv-max}
\end{table}

The last two columns of Table~\ref{tab:fv-max} repeat, for the
finite-volume residual, what \S\ref{subsec:exp8-grid} showed for the pointwise
one. The ratio of the
grid~B to the grid~A for the FVNN reaches $3.8$, against $3.2$ for the vanilla
PINN. The uniform evaluation grid therefore understates the failure of the
two unenriched approximations alike. For the enriched approximation the ratio
is at most $1.04$, and no error is hidden between the uniform nodes.

\subsection{Computational cost}
\label{subsec:exp10-cost}

\begin{table}[h!]
\centering
\small
\begin{tabular}{lccccc}
\toprule
& \multicolumn{2}{c}{cost} & & \multicolumn{2}{c}{$E_\omega$ at $\varepsilon=10^{-10}$}\\
\cmidrule(lr){2-3}\cmidrule(lr){5-6}
method & ms/step & min/run & parameters & Stokes & NSE \\
\midrule
vanilla PINN & $276$ & $\sim$19 & $7,952$ & $8.9572\times10^{-2}$ & $9.1543\times10^{-2}$ \\
FVNN & $75$ & 6.6--7.7 & $7,952$ & $4.6079\times10^{-2}$ & $5.1766\times10^{-2}$ \\
sl-PINN & $581$ & 43.7--50.0 & $8,894$ & $4.0636\times10^{-3}$ & $5.2232\times10^{-3}$ \\
sl-FVNN & $299$ & 21.2--23.3 & $8,894$ & $2.1312\times10^{-3}$ & $4.7914\times10^{-3}$ \\
\bottomrule
\end{tabular}
\caption{Cost against accuracy. All of the timings were obtained on the laptop
CPU of item (vi) in
\S\ref{subsubsec:settings} (Apple M3 Pro, 18 GB), in double precision, with
the same PyTorch installation for the four methods. Milliseconds per Adam step
are measured
over $400$ steps at $\varepsilon=10^{-8}$ under identical conditions; minutes
per run are the wall-clock times of the $5{,}000$ Adam plus $500$ L-BFGS runs
reported here, three runs at a time on that machine.}
\label{tab:fv-cost}
\end{table}

\begin{figure}[h!]
\centering
\includegraphics[width=\textwidth]{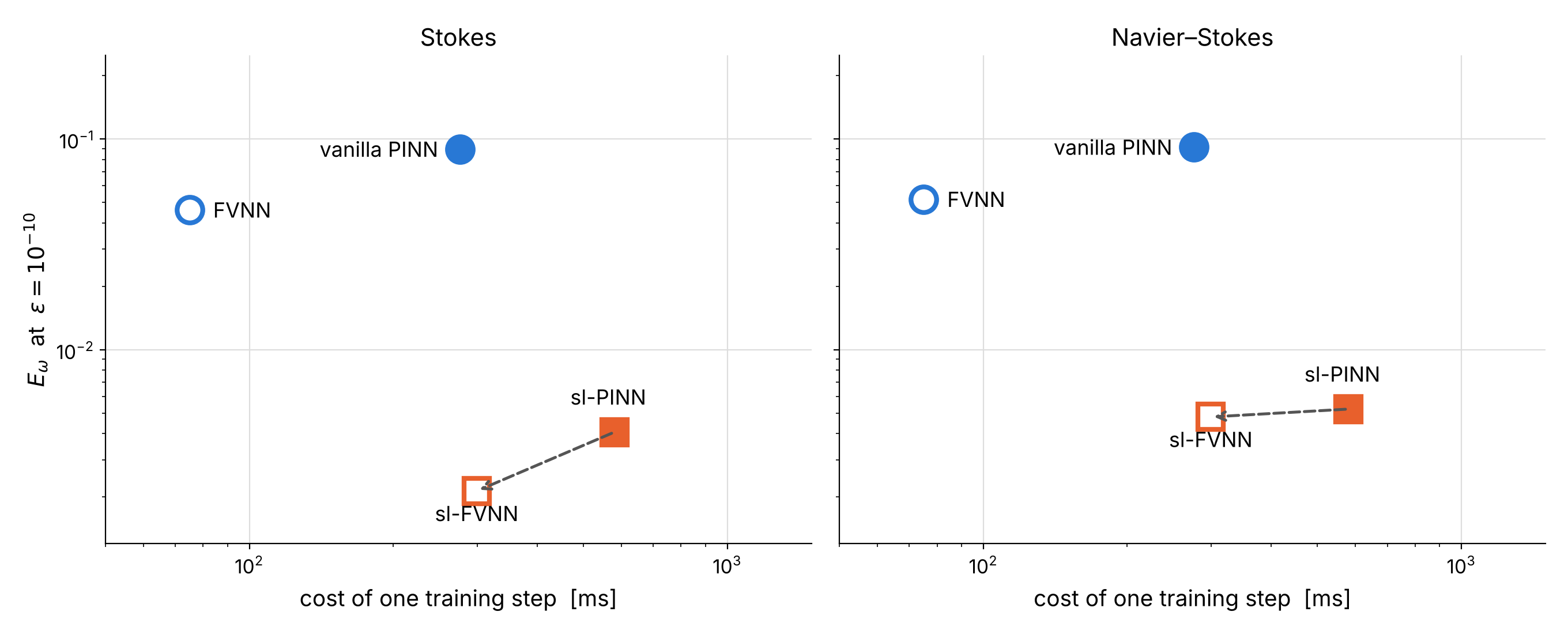}
\caption{Cost against accuracy at the smallest viscosity of the study.  Lower
error and lower cost lie down and to the left.  The dashed arrow marks the
change from the sl-PINN to the sl-FVNN: on the Stokes problem the arrow points
left and down, on the Navier--Stokes problem left and level.}
\label{fig:fv-cost}
\end{figure}

The enrichment of \S\ref{sec:slpinn} costs $2.1$ times a vanilla step, and
\S\ref{subsec:exp8-setup} gives the reason, namely that the layer factors and
the wall network are evaluated inside a forward pass which is then
differentiated twice. Replacing the Laplacian by the stencil removes the
second pass of the automatic differentiation through the $4\times50$ trunk,
which is the expensive half of it, and leaves a pass of the second order
through the $3\times20$ wall network inside $\Delta\Theta_\theta$, which is
not expensive. The measured result is $299$ ms per step against $581$ ms, a
factor of $1.94$, and $1.08$ times the cost of a step of the vanilla PINN.

The counts of the parameters are identical within each trial space, the
residual not being a change of the model, so that Table~\ref{tab:fv-cost} is a
statement about the arithmetic alone. Figure~\ref{fig:fv-cost} places the four
methods in the cost--accuracy plane. The enrichment moves a
method down in the error by a factor of $18$ for the Navier--Stokes problem
and $22$ for the Stokes problem, and the finite-volume residual then moves it
to the left by a factor of $1.94$, at no cost in the accuracy within the
seed spread of \S\ref{subsec:exp9-seed}.

The question which we asked in the present section has the following answer:
the finite-volume residual keeps the $\varepsilon$-uniform accuracy of the
corrector-informed trial space and halves the cost of a training step. It does
so in the decomposed form \eqref{eq:fv-split} only, since applied to the
enriched field as a whole the stencil is inconsistent, as
Table~\ref{tab:fv-consistency} shows before any training. Together
with \S\ref{sec:slpinn}, the results give the following sequence. The vanilla
PINN does not resolve the layer. The finite-volume residual, which is
inexpensive and accurate on a smooth problem, does not resolve it either, and
the uniform
evaluation grid understates that failure as it does for the vanilla PINN. The
corrector-informed trial function resolves the layer, at twice the cost of the
derivatives. The corrector-informed trial function combined with the
finite-volume residual resolves the layer at the cost of the unenriched
approximation. The enrichment and the discretization are therefore
complementary: the enrichment makes the coarse discretization consistent in
the presence of a layer, and the discretization reduces the cost of the
enrichment.

\begin{remark}
At the seed of the tables the sl-FVNN is more accurate than the
sl-PINN on the Stokes problem by a factor of $1.70$ to $1.91$. The three seeds
of \S\ref{subsec:exp9-seed} put the mean of the sl-FVNN at $1.09$ times the
mean of the sl-PINN on that problem, with overlapping ranges, so that we do
not read this factor as a property of the finite-volume residual.
\end{remark}

\section{Conclusion}
\label{sec:conclusion}

In this article, we developed singular layer Physics-Informed Neural Network
(sl-PINN) and singular layer Finite Volume Neural Network (sl-FVNN) methods
for the two-di\-men\-sional incompressible Navier--Stokes equations at a small
viscosity $\varepsilon$, written in the vorticity and stream function
formulation with the slip boundary conditions. Under the slip boundary
conditions the vorticity carries
a layer of width $\sqrt{\varepsilon}$ at the wall. Following the boundary layer analysis of Gie and Jung~\cite{GieJung2013}, we
found the corrector describing the singular behavior of the vorticity inside
the layer, and we embedded the corrector, together with the law that fixes its
amplitude at the wall, in the trial function of the sl-PINN. The amplitude is
tied to the wall trace of the network itself, so that the method requires no a
priori knowledge of the limit solution. We tested the methods on exact solutions consistent
with the vanishing viscosity limit (\S\ref{sec:exp7}).
Numerical experiments show that the corrector-informed sl-PINN keeps the
vorticity error uniform in $\varepsilon$ over six decades, at about
$4\times10^{-3}$, whereas the vanilla PINN loses a factor of six over the
same range. The same trial function transfers to the Navier--Stokes problem,
where the wall data travel at the slip speed, with an error again uniform in
$\varepsilon$ and about $1.4$ times the Stokes error. Since the enrichment
carries the layer analytically, the Laplacian in the residual can be
discretized by a finite volume flux difference on a coarse grid which does not
resolve the layer. The resulting sl-FVNN attains the same
$\varepsilon$-uniform accuracy, to within the scatter of the random seed,
at about half the cost of a training step of
the sl-PINN, and this combination of the asymptotic corrector with the
coarse-grid finite volume discretization is the main result of the article.
The scale-only enrichment, which uses the thickness $\sqrt{\varepsilon}$ of
the layer but leaves the profile and the amplitude free, loses a factor
of three over the six decades, whether the two walls share one layer network
or not. Its maximum pointwise error inside the layer reaches the amplitude of
the layer at $\varepsilon=10^{-10}$, so that the corrector, and not the
scale only, is the information that the analysis has to supply. Throughout the
article we measured the errors by the $x$-wavenumber and on a
grid graded toward the walls, which we refined until the errors of every
method were insensitive to it. A global relative $L^2$ error on a uniform grid
understates the error inside the layer by a factor of up to $3.8$ for the
unenriched methods, and by a factor of up to $8.5$ for the scale-only
enrichment.

The present construction has a flat channel, impermeable walls and a constant
slip speed, so that the corrector equation reduces to the heat equation in a
travelling frame. In future work, we aim to extend the methods to a slip speed
which varies along the wall, where the corrector is a Duhamel integral and the
refinement term of \S\ref{subsec:exp8-ansatz} carries the remainder, and to
curved walls. We also aim to treat the no-slip boundary condition, where the
Prandtl layer replaces the $\sqrt{\varepsilon}$ vorticity layer.

\section*{Declarations}

\textbf{Funding.} Gie was partially supported by the Fulbright U.S. Scholar Program, which is
sponsored by the U.S. Department of State and the Korean-American Educational
Commission.
Hong was supported by the Korea government (MSIT) through
the NRF (RS-2023-00219980 \& RS-2026-25491270); and by the Institute of
Information \& Communications Technology Planning \& Evaluation (IITP) grant
funded by the Korea government (MSIT) (RS-2021-II211343, Artificial
Intelligence Graduate School Program, Seoul National University).
Jung was supported by a grant from the National Research Foundation of Korea (NRF) funded by the Korea government (MSIT) (No. 2023R1A2C1003120) and by the 2025 Research Fund (1.250007.01) of UNIST (Ulsan National Institute of Science \& Technology).

\textbf{Competing interests.} The authors have no competing interests to declare
that are relevant to the content of this article.

\textbf{Data availability.} The code that generates the numerical results of
this article, together with the raw output of all of the runs reported in this
article, is available from the corresponding author on reasonable request.

\end{document}